\documentclass{aa}
\usepackage{graphicx}
\usepackage{txfonts}
\usepackage{longtable}
\usepackage{pdflscape}
\def \etal {\textit{et al. }}

\def \nNew {84 }
\def \nSystems {64 }
\def \nWereSingleUpper {Forty-one } 
\def \nWereSingleLower {forty-one } 
\def \nWereSingle {41 }

\def \nFoundOne {48 }
\def \nFoundTwo {13 }
\def \nFoundThree {two }
\def \nFoundFour {one }
\def \nReject {11 }
\def \nEphem {five }
\def \MultiFitFactor {nine }

\def \nHZupper {Five }

\def \nSmallupper {Forty-two }
\begin{document}
    \title{An independent planet search in the {\emph Kepler} dataset}
 	\subtitle{I. One hundred new candidates and revised KOIs}
    \author{Aviv Ofir \inst{1,2} \and Stefan Dreizler \inst{1}}

    \institute{Institut f\"ur Astrophysik, Georg-August-Universit\"at, 
Friedrich-Hund-Platz 1, 37077 G\"ottingen, Germany. \and 
\email{avivofir@astro.physik.uni-goettingen.de}}

    \date{Received XXX; accepted YYY}


   \abstract
    {}
    {We present first results of our efforts to re-analyze the \emph{Kepler} photometric dataset, searching for planetary transits using an alternative processing pipeline to the one used by the \emph{Kepler} Mission}
    {The SARS pipeline was tried and tested extensively by processing all available \emph{CoRoT} mission data. For this first paper of the series we used this pipeline to search for (additional) planetary transits only in a small subset of stars - the \emph{Kepler} objects of interest (KOIs), which are already known to include at least one promising planet candidate.}
    {Although less than 1\% of the \emph{Kepler} dataset are KOIs we are able to significantly update the overall statistics of planetary multiplicity: we find \nNew new transit signals on \nSystems systems on these light curves (LCs) only, nearly doubling the number of transit signals in these systems. \nWereSingleUpper of the systems were singly-transiting systems that are now multiply-transiting. This significantly reduces the chances of false positive in them. Notable among the new discoveries are KOI 435 as a new six-candidate system (of which kind only Kepler-11 was known before), KOI 277 (which includes two candidates in a 6:7 period commensurability that has anti-correlated transit timing variations) -- all but validating the system, KOIs 719, 1574, and 1871 that have small planet candidates ($1.15, 2.05$ and $1.71 R_\oplus$) in the habitable zone of their host star, and KOI 1843 that exhibits the shortest period (4.25hr) and among the smallest (0.63 $R_\oplus$) of all planet candidates. We are also able to reject \nReject KOIs as eclipsing binaries based on photometry alone, update the ephemeris for \nEphem KOIs and otherwise discuss a number of other objects, Which brings the total of new signals and revised KOIs in this study to more than one hundred. Interestingly a large fraction, about $\sim 1/3$, of the newly detected candidates participate in period commensurabilities. Finally, we discuss the possible overestimation of parameter errors in the current list of KOIs and point out apparent problems in at least two of the parameters.}
    {Our results strengthen previous analyses of the multi-transiting ensemble, and again highlight the great importance of this dataset. Nevertheless, we conclude that despite the phenomenal success of the \emph{Kepler} mission, parallel analysis of the data by multiple teams is required to make full use of the data.}

\titlerunning{A Search In The {\emph Kepler} Dataset Using SARS pipeline}

\maketitle
%

\section{Introduction}

The \emph{Kepler} mission goal is to determine the frequency of Earth-like planets around Sun-like stars, and in a wider sense the distribution of planets of all types around a variety of host stars (Borucki \etal 2010 and references therein). \emph{Kepler} has already produced a large number of milestone results (e.g. see introduction of Batalha \etal 2012 (hereafter B12) for a complete list), its importance to exoplanet studies is immense now and is expected to be even larger in the future.

The previous list of over 1200 Kepler candidates (Borucki \etal 2011b) was severely incomplete. Indeed, the \emph{Kepler} team recently almost doubled the number of candidates using improved detection tools (and only slightly more data) (B12). Even citizen scientists who just browsed the huge database were able to find several such missed signals (Fischer \etal 2012, Lintott \etal 2012) as an alternative approach to finding more candidates. We therefore checked the completeness of the updated \emph{Kepler} candidate list and report here that we are able to find a significant number of very good candidates that were missed by the \emph{Kepler} team.

The \emph{Kepler} pipeline meticulously processes the raw pixels transmitted from the spacecraft to produce the outstanding results achieved so far. This pipeline is progressively perfected, and together with the accumulating data has produced lists of growing length from 705 (Borucki \etal 2011a) to 1235 (Borucki \etal 2011b) to 2321 (B12) transiting-planet candidates. Still, the analysis of the raw data and subsequent transit searches that is done by the \emph{Kepler} team is to date almost the only analysis of this dataset (the very recent work by Huang \etal 2012, hereafter H12, is the only exception -- more below). This is in sharp contrast to  the \emph{CoRoT} space mission, where the calibrated light curves (LCs) are distributed to a number (up to 10) of different teams and each team tries its own tool set to detect the most and best candidate signals, which are then individually approved or rejected in a common discussion. This approach allows a more complete surveying of the dataset for transiting planets.

In this paper series we apply this approach to the \emph{Kepler} dataset and re-analyze it entirely with our own tool set. So far, very few attempts to do this are known: there is the Planet Hunters crowd-sourcing search (Fischer \etal 2012, Lintott \etal 2012) (which did not re-process the data) that found four, and the HATNet team (H12) that found 16 good periodic signals that were not identified by the \emph{Kepler} pipeline using Q0-Q6 data. In this first paper of the series we use a little-changed \emph{CoRoT}-oriented pipeline to remove systematic effects from the whole Q0-Q6 \emph{Kepler} data set. We then search for additional transit-like signals only on a subset of less than 1\% of all cleaned LCs - the \emph{Kepler} KOIs (Kepler Objects of Interest). All KOIs have already at least one promising transit candidate in them and passed a battery of tests for false positives (B12). Furthermore, the detection of a reliable second (or third, etc.) signal in the same LC dramatically reduces the remaining chances of false positives so that nearly all multi-transiting LCs are indeed of planetary origin (Lissauer \etal 2012, Fabrycky \etal 2012a). The true distribution of single- versus multi- transiting planets systems in itself is also of interest to understand the planet formation theory (Figueira \etal 2012). Other papers in this series are expected present the application of the same analysis to the entire \emph{Kepler} dataset (which is $>100$ times larger than just the KOIs) or to highlight particular systems.

The paper is organized in the following way: in \S \ref{SARSprocessing} we describe our pre-processing and detection pipeline and also discuss the fitting of multi-transiting systems. In \S \ref{NewCandidates} we present our new detections, in \S \ref{Notes} we discuss other KOIs that are determined to be eclipsing binaries (hereafter EBs) or are otherwise revised or noted, and we conclude in \S \ref{Discuss}. Appendix \ref{appA} includes the remaining graphical fits of systems with new detected candidates that were not presented earlier, and Appendix \ref{appB} has additional graphical content related to the other KOIs.

\section{SARS processing of {\emph Kepler} light curves}
\label{SARSprocessing}
\subsection{Algorithm}

We processed the raw Q0-Q6 public \emph{Kepler} photometry with a simple adaptation of the SARS pipeline (Ofir \etal 2010) that was designed for \emph{CoRoT} and is almost exactly identical to it. The SARS pipeline was tried and tested extensively by processing all available \emph{CoRoT} photometry with it, and participation on the detection of transiting exoplanets and brown dwarfs \emph{CoRoT} -12b, -13b, -14b, -15b, -16, -17b, -18b, -19b, -20b, -21b, -22b, -23b, and -24 b \& c (Gillon \etal 2010, Cabrera \etal 2010, Tingley \etal 2011, Bouchy \etal 2011, Ollivier \etal 2012, Csizmadia \etal 2011, H{\'e}brard \etal 2011, Guenther et al 2012, Deleuil \etal 2012, P\"atzold \etal (in prep.), Moutou \etal (in prep.), Rouan \etal 2012, and Alonso \etal (in prep.) respectively). In short, the SARS pipeline consists of these steps: the removal of all long-term variability with a median filter, the selection of a subset of stars that are intrinsically constant to serve as a "learning set" from which systematic effect (three pairs in this case) are deduced, and the removal of these effects from all stars. The SARS pipeline's core is a generalization of the SysRem algorithm (Tamuz \etal 2005) for the simultaneous identification and correction of both additive and relative systematic effects.

The Kepler version of the SARS pipeline has the ability to account for various discontinuities and anomalies in the data since they are pre-tabulated (such as the monthly Earth-point). This is done by calculating the long-term median filter separately for each continuous section (see also Ofir \etal 2012). Also, the very long LCs are naturally divided in to single-quarter lengths, so they are SARS-processed by quarter and channel and no division in to smaller blocks is required.

\subsection{Statistical comparison with PDC}
Batalha \etal (2012) presented two major improvements to the \emph{Kepler} pipeline: improved systematic errors correction (called PDC-MAP, Smith \etal 2012), and an ability to use multiple quarters in the detrending and transit search modules (Jenkins \etal 2010, Tenenbaum \etal 2012). These upgrades were the most significant factors in the near-doubling of the number of KOIs reported in B12. We compared the SARS results to this data set (at the time of writing of B12 thr PDC-MAP was not yet applied to all available quarters). The SARS pipeline aims squarely at detecting transiting planets and thus does not attempt to preserve intrinsic stellar variability. Still, it is possible to SARS-clean variable stars by iteratively applying SARS to the residuals of some model or smoothed data, as demonstrated by Ofir \etal (2009), but this should probably be done manually for each target.

The process described above is not (yet) optimized for \emph{Kepler} and is far from perfect in the simple sense that it has known bugs. Notably, the after-discontinuity transients are frequently poorly corrected and residuals remain. These are then usually modeled-out by the pipeline simply as the next transit signal, but since a transit is not a good model for these transients residuals sometimes remain and artifact are sometimes injected into the LC. When manifested, this particular bug significantly reduced our ability to detect long period signals. Promising solutions to this and other problems already exist and will be validated and then incorporated in the future.

\begin{figure}
\includegraphics[width=0.5\textwidth]{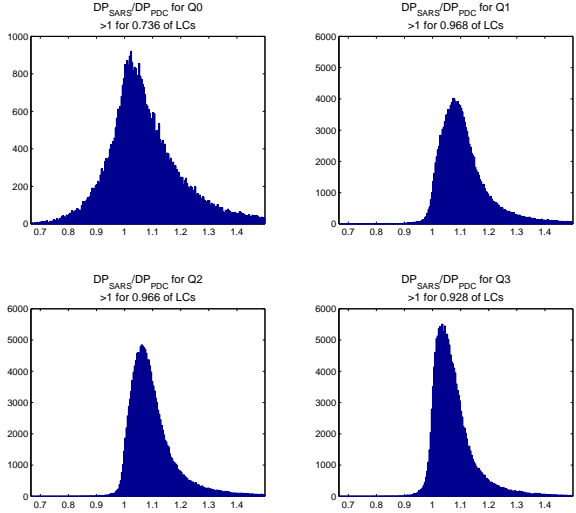}
\caption{Distribution of the ratio of detection power metrics (DP, see text for definition) $\frac{DP_{SARS}}{DP_{PDC}}$ for quarters 0 through 3 (quarters 4,5 and 6 are very similar to Q1 and Q2). The anomalously short Q0 is exceptionally underperforming but there too SARS has higher DP than PDC for $\sim 3/4$ of the data. For the sake of clarity we present only the range between 2/3 and 1.5 (representing a performance difference of 50\% in each direction).}
\label{DP}
\end{figure}

Still, we can show that the SARS-based pipeline compares well with the \emph{Kepler} processing. In Ofir \etal (2010) the detection power (DP) metric was defined as a function of standard deviation and the number of data points $DP=\frac{\sqrt(N)}{\sigma}$ (where $N$ is the number of surviving data points and $\sigma$ is measured after long-term variability is removed) so that differently processed LCs of the same object can be compared for their signal detectability potential even if the number of the data points in each is different due to different outlier rejection technique. We calculated this metric for each LC in each quarter of both the PDC-MAP and the SARS LCs after applying identical filtering to both. In figure \ref{DP} we plot the distribution of $\frac{DP_{SARS}}{DP_{PDC}}$: clearly for the vast majority of LCs (typically $\sim 95\%$ of them) the SARS processed LCs have a higher DP than PDC-MAP processed LCs (we believe, but have not yet checked, that the rest are mostly variable stars). The median ratio is in the range of 1.06 to 1.09, which may seem underwhelming, but as we pointed out in Ofir \etal 2010, the main difference is not the magnitude of the noise but its color (or lack thereof) (Pont \etal 2006), In Ofir \etal (2010) we observed a significant improvement in signal detectability even when both the above performance figures were lower. We note that these results were achieved using three pairs of effects (i.e., six effects) -- less than the eight vectors used by the \emph{Kepler} pipeline as cotrending basis vectors, therefore better results are obtained even though we use fewer decorrelation vectors.

\begin{figure}
\includegraphics[width=0.5\textwidth]{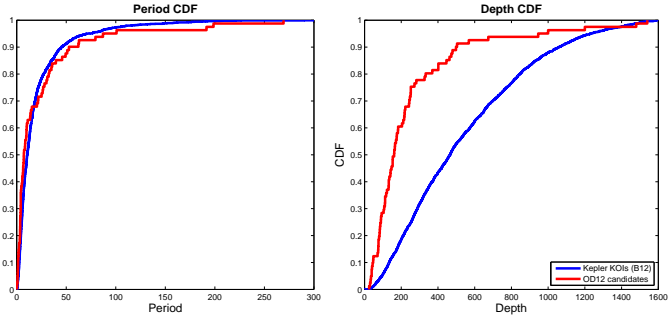}
\caption{LefT: period cumulative distribution function (CDF) of the B12 KOIs and of the new candidates described in this work (OD12). Right: similar CDFs for depth. Very deep B12 candidates are not included in these comparisons. See text for discussion.}
\label{PeriodDepthCDF}
\end{figure}

Importantly, it appears that the above statistical differences indeed translate into an improved ability to detect shallower transits. In the next section we describe in detail the detection and selection procedure, but Figure \ref{PeriodDepthCDF} shows that as a group the new candidates are drawn from the exact same population (identical period distribution) as the B12 KOIs, but they are typically shallower: 75\% of the new candidates have a depth of 280ppm or shallower, while only 24\% of the B12 KOIs are that shallow. This is a significant difference, even if one corrects for the roughly 14\% of the KOIs that are simply very deep and thus do not affect the new candidate sample \footnote{The threshold is at about 2000ppm, above which there is only one (anomalously deep) new detection, and below which there is approximately smooth distribution of new detections.} .

\subsection{Candidate detection and selection}
Once SARS has been run and the cleaned LCs are obtained, they are searched for transit-like features using the standard BLS technique (Kov{\'a}cs \etal 2002). The very first cut of significant signals is made based on the signal-to-noise ratio (S/N) of the highest periodogram peak, and not by the S/N of the proposed signal. This is because at this early stage one should not yet try to characterize the signal, but only to detect periodicity. Therefore, for detection purposes the correct question to ask is not about the significe of the signal (relative to the noise), but the significane of the folding at this particular period relative to other periods.

Detected signals (periodogram peak S/N$>20$) were first fitted with a trapezoidal model using a 5000 steps Markov Chain Monte Carlo (MCMC) procedure. The result of this quick fit was used to define a good starting point for the slower but more accurate fit using the Mandel \& Agol (2002) formalism, again using 5000 MCMC steps. The best fitting linear (i.e., no transit timing variations, hereafter TTVs) model was then subtracted from the data and a new BLS search initiated for possible other signal(s) in the data. All detected signals were manually inspected and had to pass some standard tests, such as: not being associated with a systematic error, no odd/even difference, generally plausible shape (when the S/N is high enough), transit duration not much longer than the maximal one expected form a central transit of F, G, or K dwarfs, individual transits must not be too different (especially if they are in a single quarter), no in-phase centroid motion, and a correct period (in case of detection of harmonics or subharmonics). We also added another test used by the \emph{CoRoT} science team \footnote{original idea by Roi Alonso} to help separate low-amplitude signals from noise that we dub the ``half-half test'': a true signal, even if barely detectable using the entire LC, should also produce a visible signature (i.e., a significant local peak in the periodogram) independently on the first and second half of the data. On the other hand, low-amplitude ``signals'' that arise from a single (or a few) ill-corrected systematic effect(s) would very likely not produce such coordinated local peaks in the two halves.

Lastly, once selected, the systems undergo more manual processing to which we sometimes added manual breaks in obvious discontinuities that are particular to each target. This in turn allows changing the long-term filter from a median filter (which is good at absorbing discontinuities) to a Savitzky Golay (SG) filter. The SG filter has a poor discontinuities behavior, but since at this stage these are already known and the filtering is thus performed on continuous sections only we can now enjoy the SG filter's superior smoothing over the median filter. This also means that all signals presented below -- including some very shallow signals -- were observed independently after applying two different filtering techniques, and are thus less likely to be an artifact of imperfect filtering (indeed, a few candidate signals not presented below failed this last test). As a final sanity check we checked for the presence of some (but not all) of the newly detected signals in the \emph{Kepler}-provided PDC data and nearly all of those checked were indeed found. The very shallowest new signals are probably where the SARS pipeline is absolutely required to extract the signals, and thus these are sometimes not found on the PDC data.

\subsection{Modeling multi-transiting systems}
\label{MultiFit}

The final model for each system was computed simultaneously for all detected components: an N-body Keplerian model gives the three-dimensional positions of all targets at the times of the data points. This information feeds a LC generator based on the Mandel \& Agol (2002) formalism, which itself is corrected for \emph{Kepler}'s finite integration time (Kipping 2010). The best-fitting LC is found using a 100,000-step MCMC chain, which also allows for the determination of all associated errors.

Importantly, the above procedure was born in the context of searching for circumbinary planets (Ofir 2008) and therefore our model assumes \emph{two} central bodies, around which other bodies may revolve. For a sufficiently low mass ratio $q=\frac{m_2}{m_1+m_2} \to 0$ these central bodies can just as well be the host star and the first (easiest) detected planet. At that point fitting the next planet in a circular orbit requires just $P$ and $T_{mid}$ - and does \emph{not} require the next planet's scaled semi-major axis $a/R_*$ because this can be deduced from Kepler's third law $P^2 \propto a^3$. Obviously, this assumes that the two (or more) planet candidates orbit the same star, which is a non-trivial assumption given \emph{Kepler}'s large pixels and point spread function. However, if this fit succeeds, and if none of the signals shows in-phase centroid motion, then the single-host assumption is more likely than invoking another host star at a special configuration to reproduce all the above. Moreover, since the star-planet separation during transit is no longer a free parameter, the transit duration is completely fixed by the period ratios. This also means that constraints on the orbital eccentricity can be made from photometry alone in multi-transiting systems (Ragozzine \& Holman 2010). To be precise, the fitted "$a/R_*$" are actually $d/R_*$ -- the instantaneous scaled distances during transit, which may be different from $a/R_*$ for eccentric orbits. Therefore, the above constraints are degenerate if the candidates have similar eccentricity \emph{and} have similar directions of the argument of periastron vectors.

\begin{figure}
\includegraphics[width=0.5\textwidth]{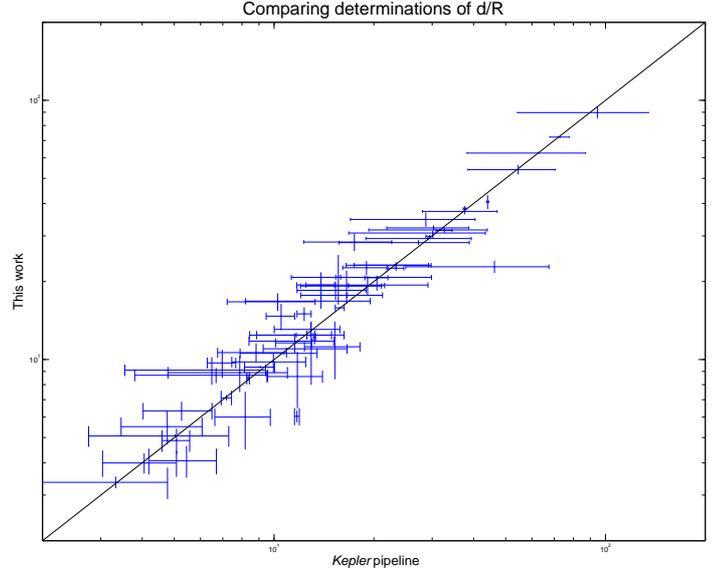}
\caption{Fitted values of $d/R_*$ for the sample of KOIs presented here, which are all part of multi-transiting systems by definition. The consistency of the $d/R_*$ fits between the \emph{Kepler} pipeline and this work is evident over about 2 orders of magnitude, but the associated error bars on each axis are significantly different -- by a factor \MultiFitFactor (mean factor)}
\label{d_over_R}
\end{figure}

\begin{figure}
\includegraphics[width=0.5\textwidth]{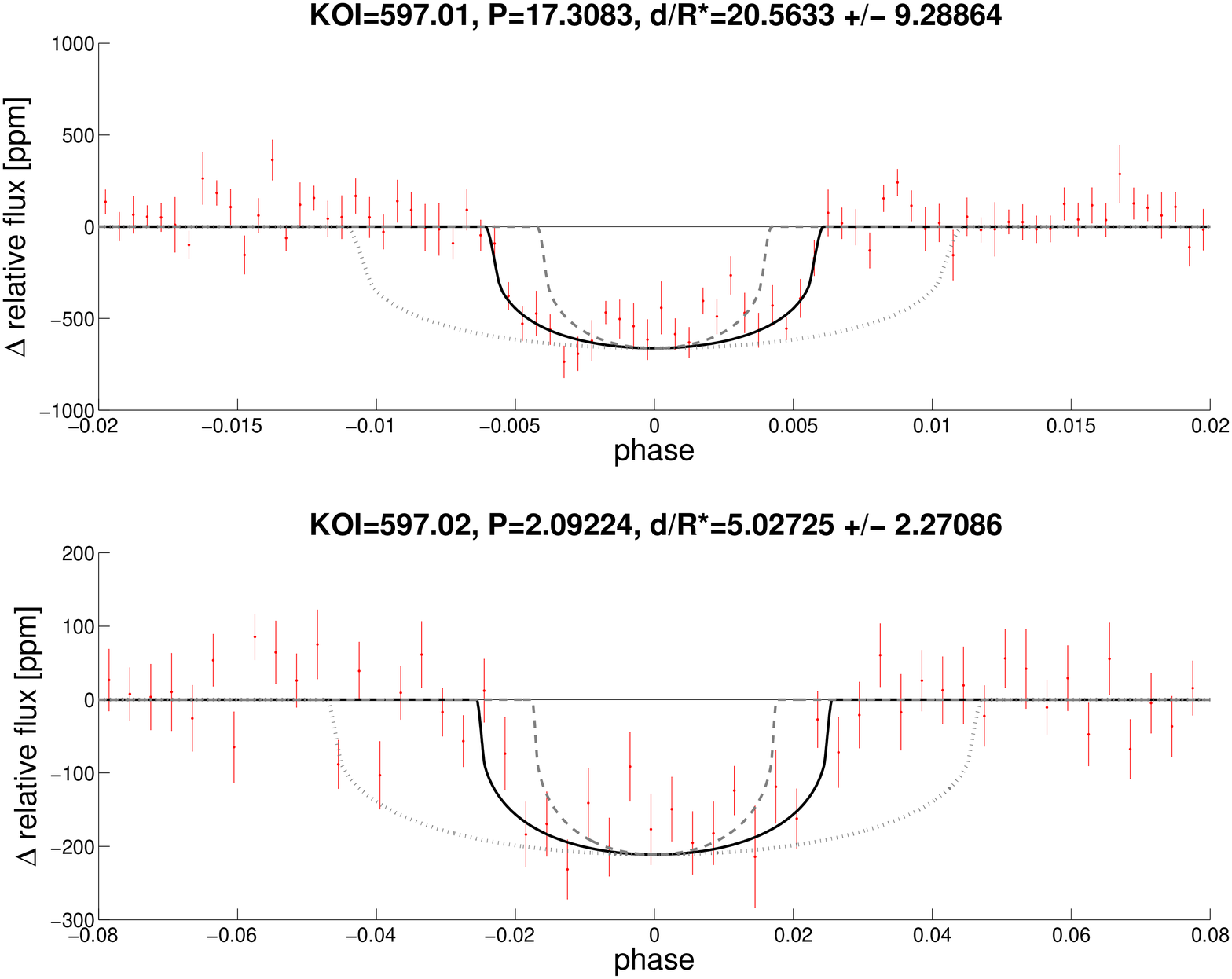}
\caption{SARS LC of KOI 597 phased to its two B12 candidates (top and bottom). Overplotted on each panel are the nominal (B12) model (solid black line) and the models corresponding to $\pm 1 \sigma$ on $d/R_*$ using the B12 values (dashed and dotted gray lines). It is evident that the errors are overestimated.}
\label{d_err_example}
\end{figure}

The above approach is equivalent to the one presented in B12. However, the B12 erros on $d/R_*$ are \MultiFitFactor times (mean factor) larger than our own, and sometimes much larger (see Figure \ref{d_over_R}). This is unexpected since the B12 fits also used quarters 7 and 8 for model fitting (i.e., about 30\% more data points). Moreover, the B12 error estimates are the diagonal elements of the model's covariance matrix, while our error estimates are derived from the MCMC analysis and thus the later should be more conservative (larger) than the former -- and the opposite is observed. This strange situation prompted us to conduct more in-depth checks, and we found the following:
\begin{itemize}
\item When simply plotting some of the KOIs with over-plotted models that show their B12 $d/R_*$ error ranges, frequently significant error overestimation is obvious (e.g. Figure \ref{d_err_example}). 
\item From the distribution of significances of the semimajor axis parameter $\big(  \frac{d}{\Delta d} \big)$ for the 2321 KOIs, one can see that half the KOIs have non-significant d parameters (less than 3 sigma). This is unlikely to be the true significance of $d/R_*$.
\item The planetary radius significances $\big( \frac{r}{\Delta r} \big)$ have weak correlation with the S/N parameter given in B12, despite the expected simple factor 2 relation between the two. In fact, $>11\%$ of the KOIs have significance lower than 3.05 (which corresponds to \emph{Kepler}'s $7.1 \sigma$ minimal S/N threshold value).
\end{itemize}
We conclude that it appears that the B12 error estimation of $d/R_*$ and $r_p/R_*$ -- on all KOIs, not just the ones presented here -- are unfortunately inaccurate, and we therefore did not try to reproduce them. 

In addition to the above subject of error estimation, we give the following notes on the fitting process and results:
\begin{itemize}
 
 \item MCMC procedures take some time to find to the region of global minimum and have to be truncated at their beginning (sometimes termed ``burn-in''). This point was determined as the first time each chain was within $5 \sigma$ of its global minimum $(\Delta \chi ^2 < 25)$. We also found that on rare occasions the chains become ``stuck'' in a particular location for the remainder of the chain, skewing the resulting distribution in the process. We therefore removed as ``burn-out'' all consecutive and final steps that had identical $\chi ^2$ to the very last $\chi ^2$.

 \item The limb-darkening parameters were computed for each star based on their KIC $T_{\mathrm{ eff}}$ and $log(g)$ values from Claret (2000) for the V band, and were not allowed to vary during the MCMC optimization. While these limb-darkening coefficients are not very accurate, this is of minor importance since the bulk of the objects is shallow and their fits are therefore almost insensitive to changes in these parameters.

 \item The derived radii for the previously known candidates were checked for consistency with the published B12 values. This is also a check that the SARS pipeline does not affect the signals more than the \emph{Kepler} pipeline. As a group, our values are typically slightly lower (7\% lower on median) than the B12 radii, however, this difference is frequently unimportant (less than $3 \sigma$ for 3/4 of the sample). This small difference arises even though (currently) we do not (yet) correct for the changing contamination levels across different quarters. We suspect that this effect can be attributed to the interaction of the long-term filter with the transit signal. In this case it is possible to follow Ofir \etal (2009) and effectively nullify this interaction by iterative filtering and fitting.

 \item In some cases the filtering was imperfect, and as a result, the out-of-transit LC directly adjacent to the transit may have an upward ``hump''. This is the same phenomenon as in the previous item, and the iterative approach of Ofir \etal (2009) will solve this.

 \item The mono-transits in our sample were given an initial period of twice the span of the data and a large initial period error ($\Delta P=1 $d) before the start of the fitting process.

\end{itemize}

\section{New candidates in known KOIs}
\label{NewCandidates}

\subsection{Overview}
In this section we give a case-by-case description of the newly detected signals. The models shown here are not taken from the serial identification during detection, but from a simultaneous solution of all identified signals (see \S \ref{MultiFit}). In total we found \nNew new transit-like signals. These signals are found in \nSystems systems, \nWereSingle of which were were considered singly-transiting systems that are now found to be multiple. In \nFoundOne systems a single new signal was found, \nFoundTwo systems now have two new signals, \nFoundThree have additional three signals, and in \nFoundFour system (KOI 435) four new transit signals were identified. Notable specific entries are (details below):
\begin{itemize}
\item KOI 435 is a new six-candidate system. The only other system with six transiting planets known to date was Kepler-11 (Lissauer \etal 2011). KOI 505 is a new five-candidate system, bringing the total to nine five-candidate systems.
\item The two signals in the KOI 277 system show anti-correlated TTVs, all but confirming these candidates as real planets.
\item KOI 1843 has a new extremely small (0.63 $R_\oplus$) and extremely short period (4.25hr) planet candidate.
\item In KOI 246 we detected a single transit event that may be compatible with the radial velocity (RV) signature of what the \emph{Kepler} team believed to be a non-transiting planet.
\item We found two additional transit signals in the LC of the confirmed planetary system Kepler-27 (KOI 841).
\end{itemize}

We note that our comments on period commensurability are given when the periods ratio is within 5\% of this value. The objects are ordered by KOI number and the complete list is given in Table \ref{NewCandsTable}. Physical sizes for the new candidates, as well as their physical semi-major axes and equilibrium temperatures, were calculated using the B12 technique and assumed values for flux redistribution factor and Bond albedo (i.e., $f=1$ and $A_\mathrm{B}=0.3$, respectively). These are tabulated in Table \ref{PhysProp}. Notable on this list are:
\begin{itemize}
 \item \nHZupper candidates have newly detected planets in the habitable zone (HZ) ($273 \le T_{eq} \le 373$) (as defined in B12).
 \item \nSmallupper candiates have sizes similar to Earth or smaller ($r_p \le 1.25 R_\oplus$).
 \item Of particular interest are the new candidates in KOIs 719, 1574, and 1871 that are both quite small ($1.15, 2.05$ and $1.71 R_\oplus$) and in the HZ of its host star. To date, the only validated planet with measured radius in the HZ of another star was the $2.4 R_\oplus$ Kepler-22b (Borucki \etal 2012).
\end{itemize}

\subsection{KOI 179}
In this 20.7d single-candidate object we found a single transit-like event over 22hr long of with a depth exceedign 179.01 by $\sim 40\%$ (see Figure \ref{KOI179fig}).

\subsection{KOI 239}
A candidate previously considered single with a new 3.62d signal that is within a few percent of the 2:3 period commensurability of KOI 239.01. (see Figure \ref{KOI239fig}).

\subsection{KOI 241}
A candidate previously considered single with two new periodic signals (see Figure \ref{KOI241fig}). These are near to a period commensurability with KOI 241.01: a 1:4 interior commensurability (to 1.5\%) and a 9:4 exterior commensurability (to 0.5\%) for the inner and outer new candidates, creating a resonant chain.

\subsection{KOI 246}
\begin{figure}[tbp]\includegraphics[width=0.5\textwidth]{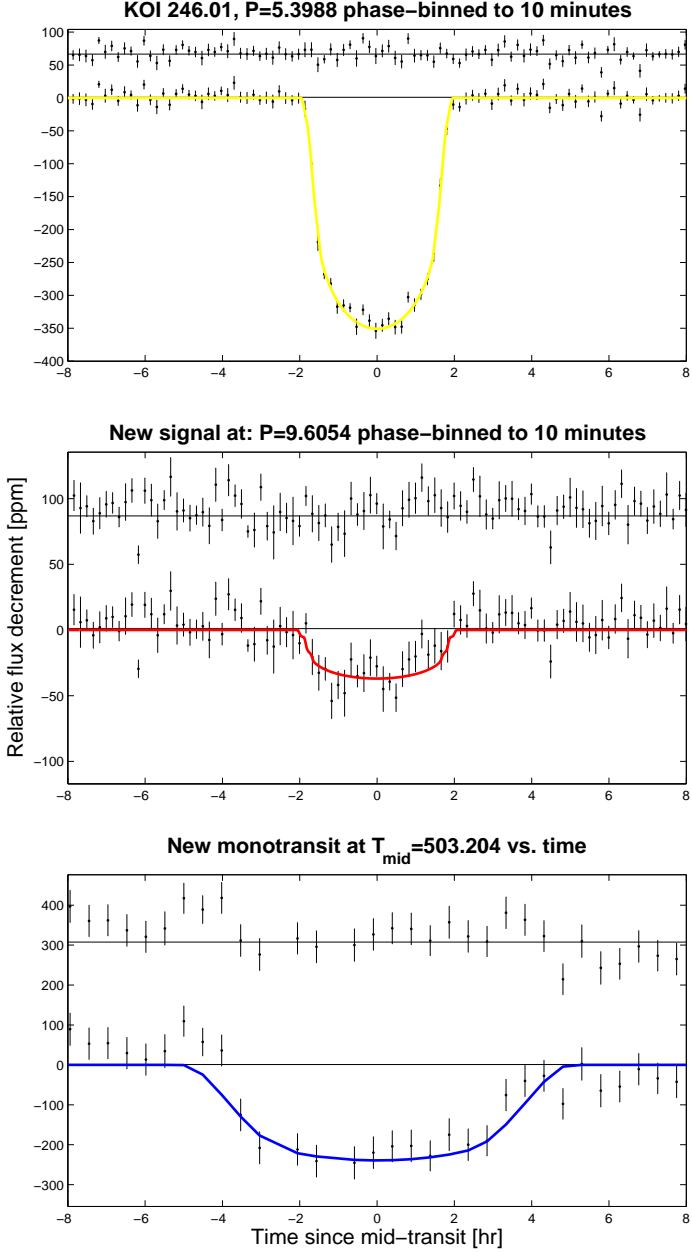}
\caption{Previously known (designated by their KOI number) and new signals in the KOI 246 LC. Periodic signals are shown binned and phased to the best-fit period, while mono-transits are given simply as relative flux vs. time (BJD-2454900). Because of the great variety of signals the time spanned by the shown window may be different for different candidates in the same system.}
\label{KOI246fig}
\end{figure}

Batalha \etal (2012) describes this object as having a single transiting planet candidate with a period of about 5.4d. We detected two new signals in this system: a periodic signal with $P\approx9.6$d and a mono-transit (see Figure \ref{KOI246fig}). Unpublished results \footnote[3]{Presented at the First Kepler Science Conference (December 2011) by G. Marcy and publicly available online.} indicate that the \emph{Kepler} team has not only detected the same second periodic signal in the extended photometry available to them, but also detected a third planet by its radial velocity signal. The association of this RV signal with the mono-transit we detected is non-trivial without the RV data, but on the one hand the duration of single transit ($<10$hr) is short compared to a central transit at the minimum period of this planet ($P>449$d) and the size of this star - and this implies that the orbit is probably eccentric -- as are the RV data of the additional planet. On the other hand, the mono-transit is significantly shallower than KOI 246.01, while the RV amplitude of the third RV planet is much higher than for KOI 246.01. Therefore, if the mono-transit and third RV signal come from the same object, its composition is radically different from that of KOI 246.01. This may be similar to what is observed in the KOI 277 system (below).

\subsection{KOI 260}
In this double-transiting system we detect another signal with a period between the two previously detected planet candidates (Figure \ref{KOI260fig}).

\subsection{KOI 271}
The two previously known planet candidates in this system are in a 5:3 period commensurability. We detect a new signal interior to both and within 2\% of the 1:2 period commensurability with the closer one, KOI 271.02 (see Figure \ref{KOI271fig}).

\subsection{KOI 277 - an exceptionally strong candidate}
\begin{figure}[tbp]\includegraphics[width=0.5\textwidth]{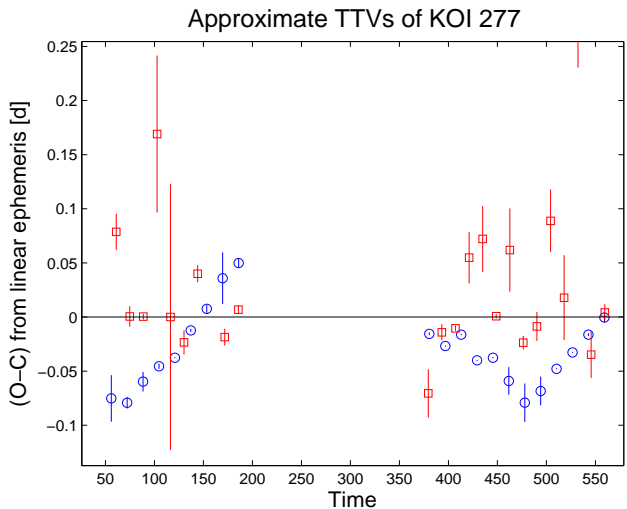}
\caption{The observed-calculated times of mid transit for KOI 277.01 (blue circles) and the newly detected 13.85d signal (red squares). We note two extremely imprecise (error$>0.1$d) timings of the new signal are slightly beyond (above) the shown region. While the anti-correlation of the TTVs is visible, we note that this timing analysis is preliminary (see text).}
\label{KOI277TTVs}
\end{figure}

We detect a second signal interior to the previously known singly-transiting planet candidate with a period of 13.85d, i.e., in a 6:7 inner period commensurability with KOI 277.01 (Figure \ref{KOI277fig}). The planets' periods are so close that they are actually interacting and show anti-correlated TTVs (Figure \ref{KOI277TTVs}), just like the confirmed systems Kepler-25 through -32 (Steffen \etal 2012, Fabrycky \etal 2012b).

We then checked that such a system can indeed be long-term stable. For a long-term dynamical stability in two-planet systems the separations of the semi-major axes of the two planets must exceed the mutual Hill radius for coplanar and circular orbits by about 3.5 times (Marchal \& Bozis 1982). Assuming planetary masses scaling with the planetary radius (see Table\,2) according to
$M_p=M_\oplus(R_p/R_\oplus)^{2.06}$, the separation is 4.3\,Hill radii and the system is therefore expected to be stable. We checked that using the hybrid symplectic integrator within the {\it Mercury} package (Chambers 1999), which we have run with a constant time step of 0.1 days, i.e., less than 1\% of the orbital period of the inner planet, for over more than $10^8$ orbital periods. During that time there was no close encounter, i.e., closer than 3 Hill radii. 

We also used a {\it Mercury} run over a shorter time (100 years) at a higher time resolution to estimate the TTV signal. While the period of the TTV modulation is mainly determined by the nearly 6:7 mean motion resonance, the amplitude depends on the planetary masses. From the mass ratio obtained with the scaling relation, the amplitude ratio should be quite different, which is not what is observed. The newly discovered inner planet is therefore suspected to have a significantly
higher density than the outer planet. A more detailed investigation of the TTV is possible, but beyond the scope of this paper.

The detection of anti-correlated TTVs in this system in a stable configuration make this system extremely likely to be a bona-fide system with two transiting planets. When one adds our multi-transit fit technique, false-positive scenarios become virtually impossible. An all-stellar false positive scenario should include a quadruple system of two EBs that are all very tightly bound and affect each other. Such a system, in just the right geometrical arrangement, can (maybe) produce the observed transit signals and the anti-correlated TTVs by some momentum exchange. However, such exchanges are accompanied by changes in the system's eccentricities and relative inclination, which should produce observable changes in the duration and/or depth of the observed transits. Such a detailed calculation is beyond the scope of this paper. We only note that this scenario is easily testable both by detailed LC analysis as above and by relatively low-precision RV to detect the motion of the two EBs around each other (they should be quite tightly bound, or they would not have caused the observed TTVs). Finally, we note that that the TTVs calculation is made from the serial detection pipeline, not the global N-body Keplerian fit, and therefore it is -- strictly speaking -- not fully consistent with it; we therefore label it ``preliminary''. Finally, we note that a manual break was added between cadences 23371 and 23372 for this object.

On the very last day of working on this paper, the KOI 277 system with its two planets and their associated TTVs, was published by the \emph{Kepler} team as confirmed and was designated Kepler-36 (Carter \etal 2012). While our analysis is an independent discovery and confirmation of this system, Carter \etal (2012) used a Newtonian model -- this system is obviously non-Keplerian -- and thus their analysis is more accurate.

\subsection{KOI 285}
We detect a second signal in this system with a period just below the outer 1:2 period commensurability with KOI 285.01 (see Fig. \ref{KOI285fig}). We also detect a third signal with a period of 49.3d. A manual break was added on cadence number 14601 for this object.

\subsection{KOI 289}
We detect an extremely shallow (27ppm) but significant second signal with a period of 1.47d (see Figure \ref{KOI289fig}).

\subsection{KOI 295}
A second 10.1d signal is detected in this system previously considered to be singly-transiting at 5.3d (Figure \ref{KOI295fig}). A manual break was added between cadences 17496 and 17497 for this object.

\subsection{KOI 316}
This was a double-transiting system with periods of 15.77d and 157.06d, i.e., within $5\%$ of the 1:10 period commensurability. We detect a new signal interior to these two (Figure \ref{KOI316fig}) with a period of 7.3d.

\subsection{KOI 321}
We detect a second planet outside this system previously considered to be singly-transiting (Figure \ref{KOI321fig}). The new planet has a radius of only $60\%$ of the first one, which itself is all but a confirmed planet as it was already RV followed-up by the \emph{Kepler} team $^3$ and its RV signal measured - which all but validates the second candidate without any additional measurement.

\subsection{KOI 330}
We detect a second planet candidate interior to the candidate previously considered to be singly-transiting (Figure \ref{KOI330fig}).

\subsection{KOI 339}
Two planet candidates with near-identical transit depths were known in this system, and we detect a third signal with slightly shallower transit and a longer (35.86d) period (Figure \ref{KOI339fig}).

\subsection{KOI 408}
We add a fourth candidate interior to the existing three (Figure \ref{KOI408fig}).

\subsection{KOI 435}
\begin{figure}[tbp]\includegraphics[width=0.5\textwidth, height=23.7cm]{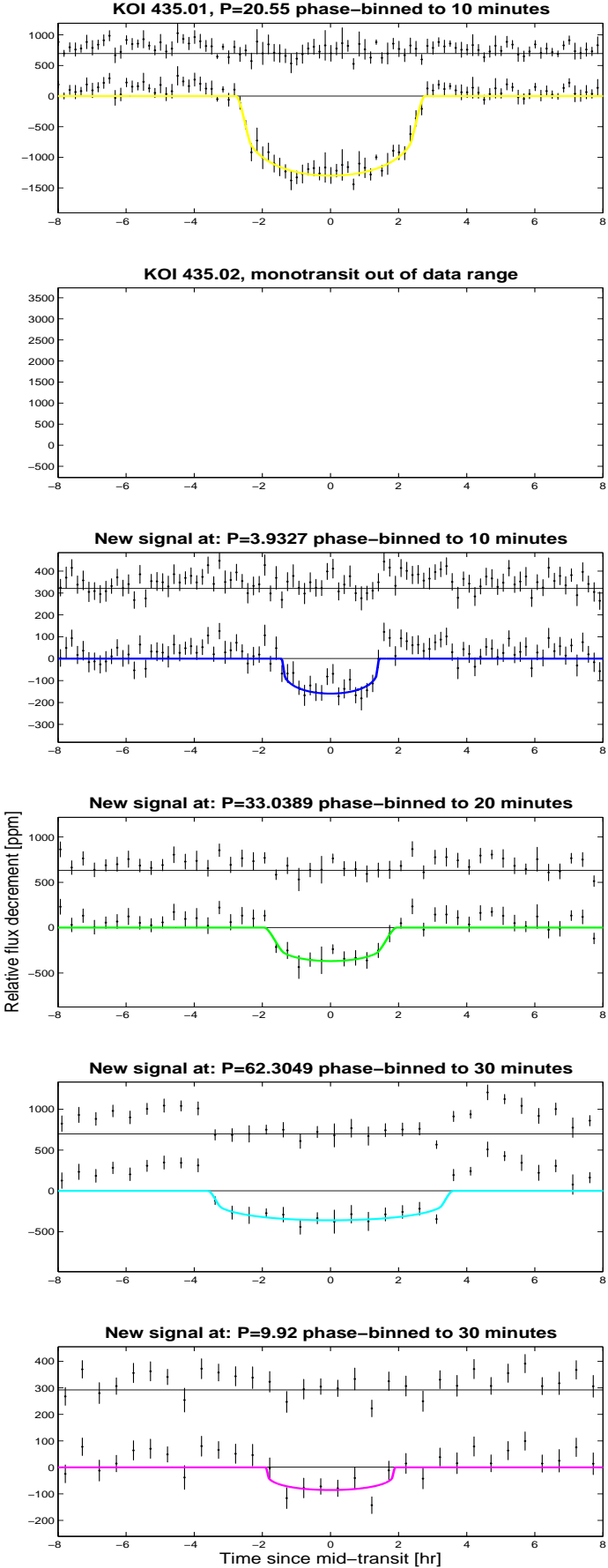}
\caption{Similar to Figure \ref{KOI246fig}. Note the panel of KOI 435.02 is intentionally empty (see text).}
\label{KOI435fig}
\end{figure}

This system was known to contain a single periodic signal with $P_1 \approx 20.55$d and a depth of about 1500ppm. The published candidate table (B12) also included a second signal - KOI 435.02 - a monotransit with a depth of $>8000$ppm that should have been quite easy to spot, but unfortunately the time of mid transit is $T_{mid,2}=2455490.27$, which is about a month after the end of the publicly available data. We therefore cannot model this candidate. On top of these two signals, we find four additional signals with periods of 3.93d, 33.04d, 62.30, and 9.92d (Figure \ref{KOI435fig}). This is only the second system with six (candidate) transiting planets, the first being the Kepler-11 system (Lissauer \etal 2011). We note that the 33.04d signal is within 0.5\% of the 8:5 outer period commensurability with KOI 435.01. Manual breaks were added in cadences 22442 and 23990 for this object.

\subsection{KOI 505}
In this double-transiting (13.76d and 6.19d) system we detect three more signals with periods of 3.25d, 87.09d, and 8.34d, where the latter is within 1\% of the 3:4 period commensurability with KOI 505.02 (see Figure \ref{KOI505fig}). We note that the 8.34d signal was independently found by HATNet (H12). A manual break was added between cadences 6712 and 6718 for this object.

\subsection{KOI 509}
We detect a third signal in this system previously considered to be double-transiting with a period of 39.59d, or within 1.5\% of the 7:2 period commensurability with KOI 509.02 (Figure \ref{KOI509fig}).

\subsection{KOI 510}
We detect a forth signal in this system previously considered to be triple-transiting with a period of 35.12d, or within $4\%$ of the 5:2 period commensurability with KOI 509.03 (Figure \ref{KOI510fig}).

\subsection{KOI 564}
A planet candidate interior to the previously known two is detected with a period of 6.21d (see Figure \ref{KOI564fig}). The larger-than-usual error bars on KOI 564.02 were caused by a particularly unfortunate timing of one of the three events shortly after a discontinuity in the data that affected the filtering.

\subsection{KOI 582}
We detect a 9.94d signal in this system, in addition to the previously known 5.94d and 17.74d signals. The new planet candidate appears to be in a double period commensurability: within 0.4\% of the 5:3 ratio exterior to KOI 582.01, and within 2\% of the 4:7 ratio interior to KOI 582.02 (see Figure \ref{KOI582fig}).

\subsection{KOI 584}
This shallow candidate (detected to $9.8 \sigma$) lies interior to the two previously known planet candidates in this system (Figure \ref{KOI584fig}). An even shallower transit-like signal about half as deep as the first one was detected at phase 0.59 of the new $6.47$d signal. This possible secondary eclipse is very close to the noise level -- if not below it -- and thus its very detection is much less robust. Still, we wish to point out that the proposed signal will need to be checked for binarity when more data become available. A manual break was added on cadence number 1386 for this object.

\subsection{KOI 593}
We detect a third planet candidate in this system with a period of $51.06$d, between the two previously known candidates (Figure \ref{KOI593fig}).

\subsection{KOI 597}
We detect a third planet candidate in this system within $2\%$ of the 3:1 period commensurability with KOI 597.01 (Figure \ref{KOI597fig}). Manual breaks were added on cadence numbers 1173 and 1796 for this object.

\subsection{KOI 623}
We detect a forth and outermost candidate in this very compact system 
(see Figure \ref{KOI623fig}).

\subsection{KOI 624}
We detect a candidate in a short period of 1.31d (Figure \ref{KOI624fig}) in addition to the previously known KOIs 624.01 and 624.02.

\subsection{KOI 627}
We detect a second candidate in this previously single-candidate system (Figure \ref{KOI627fig}).

\subsection{KOI 671}
We detect two additional planet candidates, both in between the two previously known planet candidates (Figure \ref{KOI671fig}) . The resulting compact system (four plant candidates within $P \approx 16$d) appears to be in a resonant chain of 4:7, 2:3 and 2:3 (from the inside out) period commensurabilities.

\subsection{KOI 710}
We detect a shallow signal ($8.8 \sigma$) in this system previously considered to be singly-transiting (Figure \ref{KOI710fig}).

\subsection{KOI 717}
A very short period planet candidate (P=0.9d) is detected in this system previously considered to be singly transiting (Figure \ref{KOI717fig}).

\subsection{KOI 719}
This system experienced a dramatic transformation from our re-analysis of the KOIs. We detect no less than three additional signals in this system previously considered to be singly-transiting with periods of 4.16d, 45.09d, and 28.12d (see Figure \ref{KOI719fig}). Moreover, the 45d planet candidate has very interesting physical parameters: at $1.15 R_\oplus$ and $T_{eq}=354$K this is one of the most Earth-like exoplanet candidates known -- both in size and in being inside the habitable zone.

\subsection{KOI 780}
We detect a second transit-like signal in the data that is very short: it lasts for about $0.5hr$ despite the relatively long period of 7.24d (see Figure \ref{KOI780fig}).

\subsection{KOI 841 = Kepler-27}
In this confirmed system of planets (Steffen \etal 2012) we detect, in addition to the previously known KOI 841.01 (P=15.3d) and KOI 841.02 (P=31.3d) two other signals with periods of 269d and 6.55d (Figure \ref{KOI841fig}). While these new signals are most likely the third and forth planets in this confirmed system, the photometric aperture of this target is known to contain significant contaminators.

\subsection{KOI 856}

We detect a shallow second marginal signal to $6.86 \sigma$. This significance level is lower than the $7.1 \sigma$ threshold used by the \emph{Kepler} team, but we chose to point it out as it looks quite problem-free, despite the low formal error. Note the previously known candidate KOI 856.01 also exhibits centroid motion, as noted in \S \ref{Notes}.

\subsection{KOI 1060}
We detect a forth signal in the system, exterior to the known three (Figure \ref{KOI1060fig}).

\subsection{KOI 1069}
The additional P=8.70d signal we detected in this system previously considered to be singly-transiting is very strong, with a depth $>3.5$ times the depth of KOI 1069.01 (Figure \ref{KOI1069fig}). We checked that this signal appears neither in the detected binaries nor in the list of false positives. This strong signal, together with the candidates detected by the Planet Hunters project, shows why parallel analyses of the same data are absolutely required.

\subsection{KOI 1082}
We detect two candidate planets in addition to the previously known candidate. They are both interior to and smaller than KOI 1082.01 (Figure \ref{KOI1082fig}).

\subsection{KOI 1108}
We detect two candidate planets with periods of 1.47d and 4.15d in addition to the previously known single candidate (Figure \ref{KOI1108fig}). The 4.15d signal is about 1\% from the 9:2 period commensurability with KOI 1108.01.

\subsection{KOI 1413}
We detect (Figure \ref{KOI1413fig}) a second signal in the system, marginally deeper than KOI 1413.01, and with a period about twice as long.

\subsection{KOI 1536}
This system is known to contain a 3.74d transit candidate. We detect a second signal about twice as deep and with a period of about 79.48d (see Figure \ref{KOI1536fig}).

\subsection{KOI 1574}
This is a particularly long-period system: in addition to the known periodic signal at P=114.7d, we detect two separate events separated by 191.5d. The pattern of data points seemed, to the eye, sufficiently different to believe that these are two separate mono-transits (and one of these signals was also identified by H12). However, when the signals were modeled in detail it turned out that they are statistically identical, so the suggested model now includes a second periodic signal (Figure \ref{KOI1574fig}) in an outer 5:3 period commensurability with 1574.01 to better than 0.2\%. The new planet candidate has a radius of just over two Earth radii and an equilibrium temperature within the habitable zone of $T_{eq}=281$K which makes it one of the most Earth-like planet candidates known.

\subsection{KOI 1593}
We detect an additional transit signal outside the known 1593.01 with a period of 15.38d (Figure \ref{KOI1593fig}).

\subsection{KOI 1599}
We detect a second transit candidate interior to KOI 1599.01 with a period of 13.61d, or within 0.01\% of the 2:3 resonance with KOI 1599.01 (Figure 
\ref{KOI1599fig}).

\subsection{KOI 1601}
We detect the second planet candidate in this system previously considered to be singly-transiting. The new signal is about 1\% off the 1:6 period commensurabilities with KOI 1601.01 (see Figure \ref{KOI1601fig}).

\subsection{KOI 1650}
We detect only two events of a second signal, suggesting a period of 100.8d in this system previously considered to be single-candidate (see Figure \ref{KOI1650fig}).

\subsection{KOI 1681}
This system was known to contain a planet candidate with a period of 6.94d. We detect two new transit signals with periods of 1.99d and 3.53d (Figure \ref{KOI1681fig}) -- the latter is within 2\% of the 1:2 period commensurabilities with KOI 1681.01.

\subsection{KOI 1830}
We detect a deep, long-period (198.7d) transit signal as the second signal in this system (see Figure \ref{KOI1830fig}).

\subsection{KOI 1831}
We detect an additional planet candidate in an orbit significantly interior to the previously known two (see Figure \ref{KOI1831fig}).

\subsection{KOI 1843}
\begin{figure}[tbp]\includegraphics[width=0.5\textwidth]{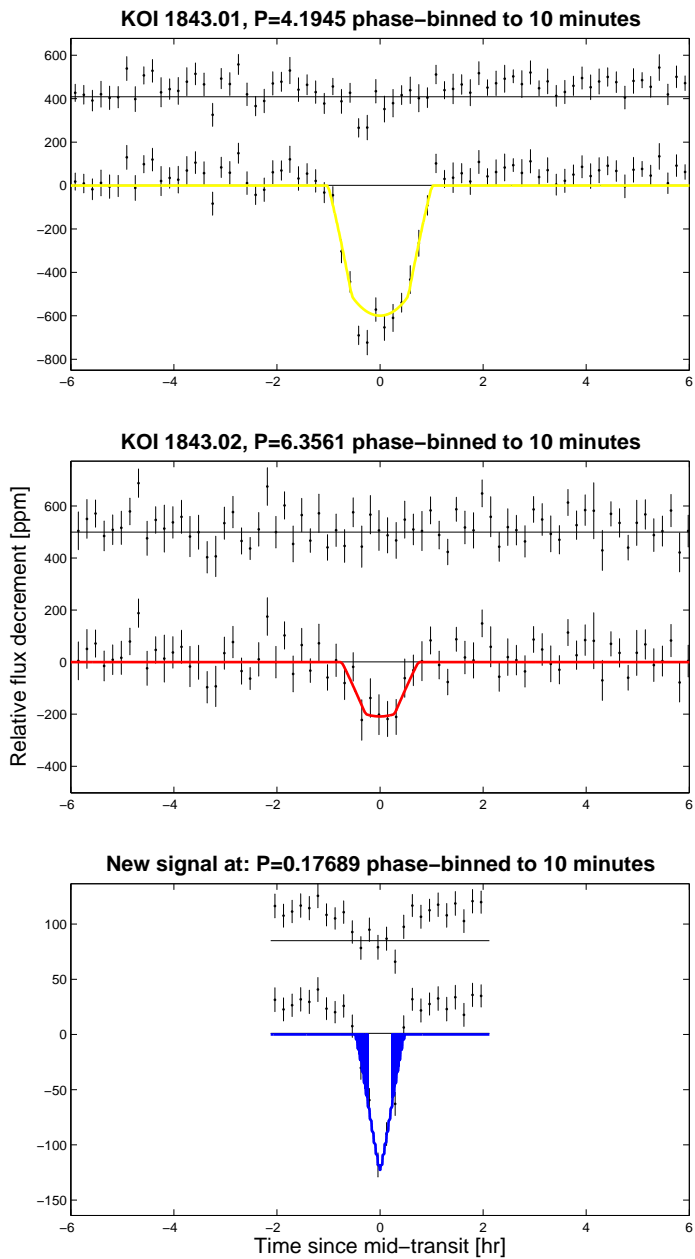}
\caption{Similar to Figure \ref{KOI246fig}. The large duty cycle of the new signal ($>22\%$) caused the mean of the LC to be significantly different from the mean of the out-of-transit points, despite the filtering of the data. The triangle, or thick line, near ingress/egress is the model line fluctuating between the correct value and zero due to numerical problems. These effects will be taken into account in the planned future detailed study of this exceptional object.}
\label{KOI1843fig}
\end{figure}

This system was known to host two planet candidates with periods of 4.19d and 6.36d. We detect an additional and extreme transit signal with a period of 0.176d (4.25hr) and depth of 120ppm (Figure \ref{KOI1843fig}). Interestingly, the short period couples with the long \emph{Kepler} cadence duration to create a seemingly unphysical duty cycle of $>22\%$. Importantly, the host star is relatively cool and small with $T_{eff}=3673$K and radius of 0.52 $R_\odot$ (KIC values). We plan to analyze this system in detail and attempt to fully validate it in a future paper, but this tantalizing signal implies that the newly detected signal is from a planet with the shortest known period, shorter even than post-common-envelope planets (Charpinet \etal 2011) and with a radius of 0.63 $R_\oplus$ -- among the smallest exoplanets known.

\subsection{KOI 1870}
We detect a 0.5\% deep monotransit about 12hr long with no associated centroid motion in addition to the known 7.96d KOI 1870.01 (Figure \ref{KOI1870fig}).

\subsection{KOI 1871}
We detect a periodic signal with a period of 32.38d in addition to the previously known KOI 1871.01 (Figure \ref{KOI1871fig}). The new planet candidate has a radius of about 1.7 Earth radii and an equilibrium temperature (barely) within the hot side of the habitable zone: $T_{eq}=370$K.

\subsection{KOI 1875}
We detect a very short period (0.54d) signal in addition to the P=9.92d previously known KOI 1875.01.

\subsection{KOI 1955}
We detect two new signals in this system previously considered to be double transiting. One is significantly shallower and has significantly shorter period (P=1.64d) than the known KOI 1955.01 and 1955.02. The second newly detected signal has a period of 26.23d - a 2:3 period commensurability with KOI 1955.02 to less than 0.5\% (Figure \ref{KOI1955fig}).

\subsection{KOI 1972}
We detect a second signal in this system previously considered to be singly transiting with 
a period of 1.23d (see Figure \ref{KOI1972fig}).

\subsection{KOI 1986}
A second signal is detected in this system previously considered to be singly transiting with a period of 7.13d (see Figure \ref{KOI1986fig}). We note that the local peak in the "half-half" test on the first half of the data exists, but is very small.

\subsection{KOI 2004}
A second signal is detected in this system previously considered to be singly transiting with a period of 3.19d (see Figure \ref{KOI2004fig}).

\subsection{KOI 2055}
We detect two more signals in this system with periods of 4.03d and 2.50d, or within 0.5\% of the 8:5 period commensurability between them (Figure \ref{KOI2055fig}), in addition to the existing 2055.01.

\subsection{KOI 2159}
A second signal is detected with a period of 2.39d (Figure \ref{KOI2159fig}).

\subsection{KOI 2193}
We detect a second signal outside of this system previously considered to be singly transiting with a period of 4.16d (see Figure \ref{KOI2193fig}). The two signals are within 1\% of the 7:4 period commensurability.

\subsection{KOI 2195}
We detect two signals with periods on both sides of the previously known KOI 2195.01: an inner P=6.85d within 3\% of the 1:3 period commensurability with 2195.01, and an outer candidate with P=30.0922 - or within 0.3\% of the 3:2 period commensurability with 2195.01 (see Figure 
\ref{KOI2195fig}) - creating a resonant chain.

\subsection{KOI 2243}
We detect a second signal with a period of 8.46d in this system previously considered to be singly transiting (Figure \ref{KOI2243fig}).

\subsection{KOI 2485}
We detect two planet candidates with periods of 3.60d and 5.73d, or about 0.5\% of the 5:8 period commensurability, in addition to the known 9.99d KOI 2485.01 (Figure \ref{KOI2485fig}).

\subsection{KOI 2579}
We detect two planet candidate outside the known KOI 2579.01. The first has a period of 3.60d, or about 1\% off the 3:4 period commensurability with KOI 2579.01. The second has a period of 10.30d (Figure \ref{KOI2579fig}).

\subsection{KOI 2597}
We detect an additional planet candidate, interior to the previously known two, with a period of 5.61d (Figure \ref{KOI2597fig}).

\section{Notes about other KOIs}
\label{Notes}

\subsection{Overview}
In this section we comment on a few of the other KOIs. Specifically, we 
reject \nReject KOIs as EBs based on close inspection of the LC,
correct the published ephemeris of for \nEphem KOIs,
suggest another interpretation of KOI 1401 as caused by stellar pulsations,
detect the secondary eclipse of KOIs 805 and 1468 and identify it as consistent with a planetary secondary eclipse in an eccentric orbit,
confirm the second transit signal identified by Nesvorn{\'y} \etal (2012) in the KOI 872 system.

\subsection{KOI 225.01}
\begin{figure}[tbp]\includegraphics[width=0.5\textwidth]{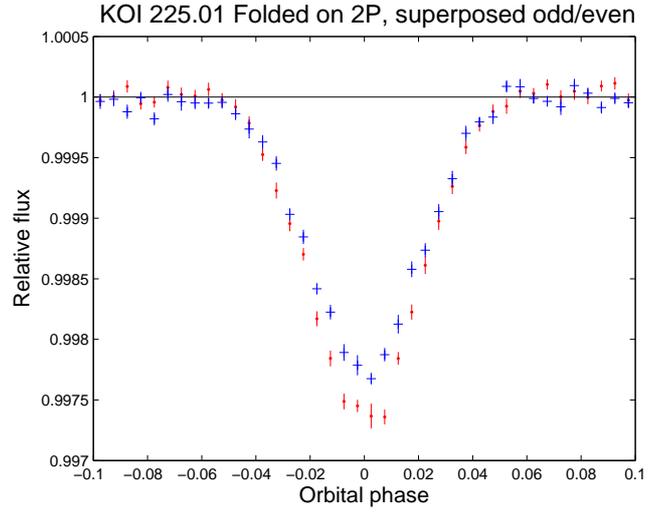}
\caption{Phase-folded odd (red dots) and even (blue '+' signs) eclipses of KOI 225.01 showing significant difference that is indicative of an EB with similar but not identical components.}
\label{KOI225oddeven}
\end{figure}

We change the interpretation of KOI 225.01 from a planet candidate to an EB with twice the period of KOI 225.01 since we detect significant differences between the odd and even events in this system (see Figure 
\ref{KOI225oddeven}).

\subsection{KOI 341.01}
The true period of this candidate is half of the published period, i.e., $P_{341.01}=7.1706171$d.

\subsection{KOI 536.01}
The true period of this candidate is half of the published period, i.e., $P_{536.01}=81.1697006$d.

\subsection{KOI 741.01}
\begin{figure}[tbp]\includegraphics[width=0.5\textwidth]{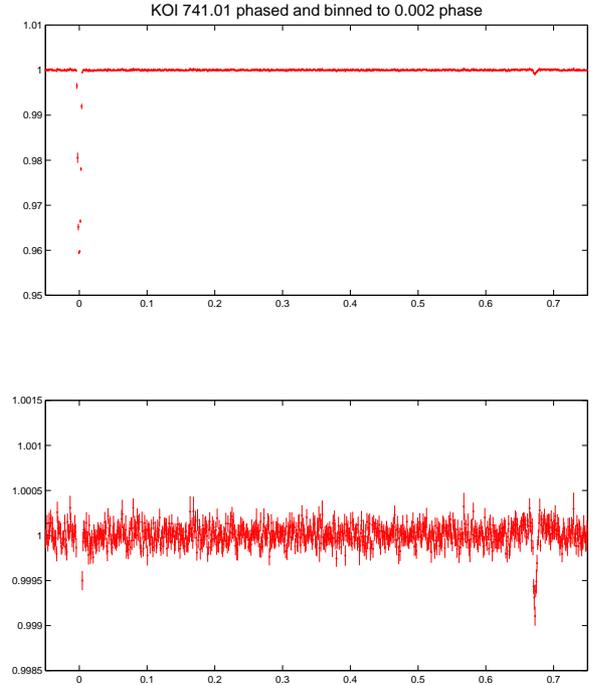}
\caption{Phase-folded and binned \emph{Kepler}'s PDC LC of KOI 741.01. Top: the 741.01 signal at phase 0.  Bottom: expanded scale of the same data, showing a secondary eclipse near phase 0.672.}
\label{KOI741secondary}
\end{figure}

This object has extremely deep eclipses ($\sim 4\%$ deep) causing our processing pipeline, which is built to handle low-amplitude signals, to reject some of the in-transit points. Still, we were able to detect a secondary signal with a depth of $\sim 1000$ ppm to KOI 741.01, far larger than the $  (\frac{R_p}{R_*} \frac{R_*}{a})^2 \sim 40$ ppm maximal expected signal for a planetary secondary eclipse (assuming $d/R_*=a/R_*$, or circular orbit, and geometrical albedo of 1), showing that this object is an eccentric EB. Figure \ref{KOI741secondary} shows \emph{Kepler}'s PDC LC since it clearly exhibits both the primary and secondary eclipses.

\subsection{KOI 743.01}
This object has deep eclipses ($\sim 1\%$ deep), but we were able to detect also a secondary eclipse to KOI 743.01 with a depth of $\sim 1000$ ppm, far larger than the $\sim 130 $ ppm expected (calculated similarly to KOI 741.01) for a planetary secondary eclipse, showing that this object is 
an eccentric EB (see Figure \ref{KOI743secondary}).

\subsection{KOI 805.01}
\begin{figure}[tbp]\includegraphics[width=0.5\textwidth]{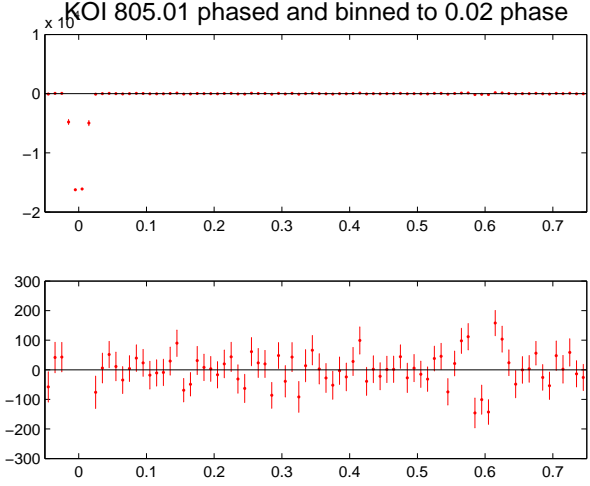}
\caption{Phase-folded and binned LC of KOI 805.01 showing the secondary eclipse of this object, with a depth that is consistent with a planetary interpretation of the transit signal.}
\label{KOI805secondary}
\end{figure}

This object has transit-like signals ($\sim 1.6\%$ deep), but we were able to detect also a secondary eclipse near phase 0.6 to KOI 805.01 with a depth of $\sim 140 $ ppm (see Figure \ref{KOI805secondary}). The expected depth for a planetary secondary eclipse (calculated similarly to KOI 741.01) is $141 \pm 5 $ ppm. This means that the observed signal is consistent with a secondary eclipse of KOI 805.01 as an eccentric planet with a very high geometric albedo.

\subsection{KOI 856.01}
\begin{figure}[tbp]\includegraphics[width=0.5\textwidth]{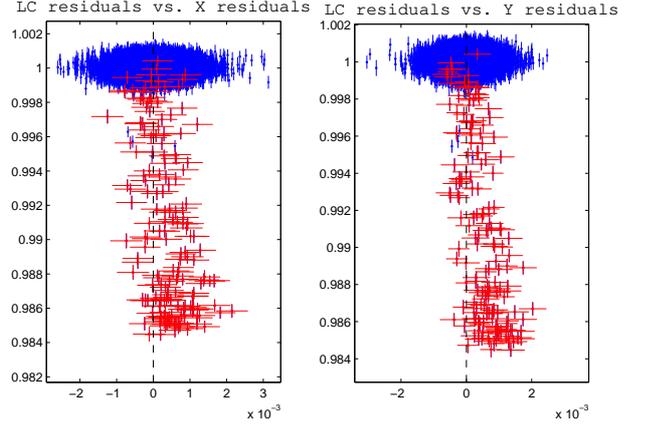}
\caption{Residuals of the LC model plotted vs. the residuals of the centroid of KOI 856.01, which were filtered in a similar way (i.e., segmented smooth). In-transit points (red) are given with full error bars, the other points (blue) are given with partial error bars for clarity.}
\label{KOI856centroid}
\end{figure}

This candidate shows $>1.3\%$ deep transit-like signals. We detect small but apparently in-phase centroid motion in the +X and +Y directions (Figure \ref{KOI856centroid}), which should be interpeted in a detailed analysis of the local stellar neighbors of KOI 856.

\subsection{KOI 872}
A second transit signal was detected by Nesvorn{\'y} \etal (2012). We confirm this detection; we can do so only in retrospect: for some yet unidentified reason the candidate was not promoted further in our pipeline.

\subsection{KOI 1003.01}
This object has deep eclipses ($\sim 1.9\%$ deep), but we were able to detect also a secondary eclipse to KOI 1003.01 with a depth of $\sim 900 $ ppm, much larger than the $\sim 230 $ ppm expected (calculated similarly to KOI 741.01) for a planetary secondary eclipse, showing that this object is an eccentric EB (see Figure \ref{KOI1003secondary}).

\subsection{KOI 1101.01}
The true period of this candidate is four times the published period, i.e., $P_{1101.01}=11.390958$d.

\subsection{KOI 1151.01}
The true period of this candidate is twice the published period, i.e., $P_{1151.01}=10.4356212$d.

\subsection{KOI 1152.01}
This object has extremely deep eclipses ($\sim 8.2\%$ deep), but we were able to detect also a secondary eclipse near phase 0.338 to KOI 1152.01 with a depth of $\sim 6000 $ ppm, far larger than the $\sim 55 $ ppm expected (calculated similarly to KOI 741.01) for a planetary secondary eclipse, showing that this object is an eccentric EB (see Figure \ref{KOI1152secondary}).

\subsection{KOI 1385.01}
This object exhibits V-shaped eclipses about 5\% deep. Fitting the data reveals a system with $b=0.9 \pm 0.005$ but $R_2/R_1=0.2972\pm0.0034$, i.e., a grazing eclipse (the published \emph{Kepler} fit is even poorer with $b>1$ directly). Such deep and V-shaped eclipses suggest a significantly larger secondary radius, along with the KIC estimates for the properties of host ($R_* =0.84 R_\odot ,  T_{eff}=5848 $K, log(g)=4.55) leave little doubt that the secondary cannot be a planet, but is likely to be an EB on an eccentric and inclined orbit (because there is no secondary eclipse).

\subsection{KOI 1401.01, KIC 9030447}
\begin{figure}[tbp]\includegraphics[width=0.5\textwidth]{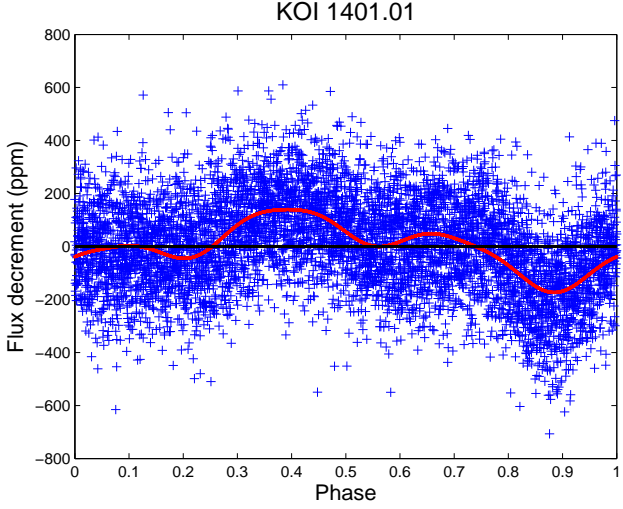}
\caption{KOI 1401.01 phased to the B12 orbital period. A transit-like feature appears near phase 0.88, and a similar-amplitude ``hump'' can be seen half a period offset from it. A two-sine fit with $P_{1401.01}$ and $P_{1401.01}/3$ is overplotted (solid red line).}
\label{KOI1401harm}
\end{figure}

This candidate appears in the B12 catalog and was therefore analyzed for this paper, but later we saw that it also appears on the list of \emph{Kepler} eclipsing binaries (Slawson \etal 2011). Stellar variability in this star is dominated by pulsations at the KOI 1401.01 period and its second harmonic (i.e., $P_1=0.56672$d and $P_2=P_1/3$) that together create a transit-like signal, but also a ``hump'' about half a period later -- which should not happen if the transit-like signal originates from either a planetary object or an EB. Practically all variability is explained by this two-period pulsation model (see Figure \ref{KOI1401harm}). We therefore believe that KOI 1401.01 is caused by pulsations, rather than transits or eclipses.

\subsection{KOI 1419.01}
We change the interpretation of KOI 1419.01 from a planet candidate to an EB with twice the period of KOI 1419.01 since we detect significant differences between the odd and even events in this system (see Figure \ref{KOI1419oddeven}). A manual break was added on cadence number 11108 for this object.

\subsection{KOI 1459.01}
\begin{figure}[tbp]\includegraphics[width=0.5\textwidth]{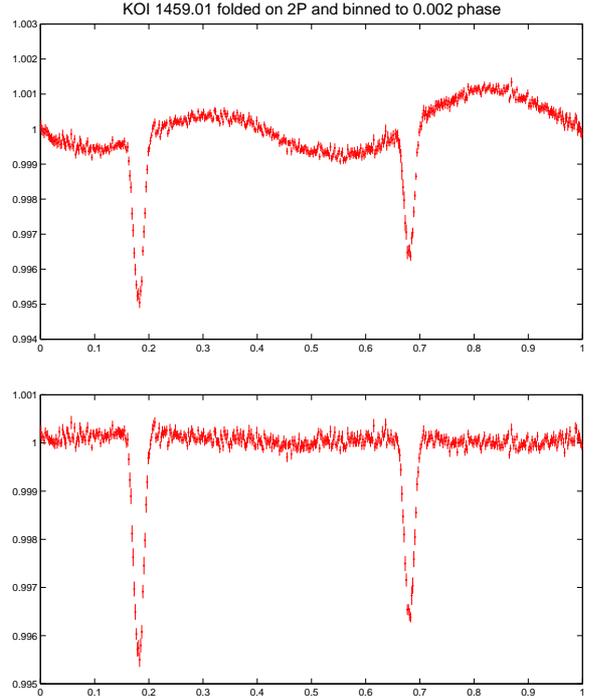}
\caption{Double-period phase-folded LC of KOI 1459 before (top) and after (buttom) background subtraction.}
\label{KOI1459oddeven}
\end{figure}

This object exhibits strong sine-like background variability that is almost, but not exactly, in phase with the transit-like signal. We iteratively computed the power spectrum for the next-strongest frequency and then simultaneously fitted sines with all previously found frequencies to the data with $3 \sigma$ iterative outlier rejection. It took only a single period of $P_1=0.691758$d and its near sub harmonic at $P_2=1.38248$d to all but remove the background and produce Figure \ref{KOI1459oddeven}, showing that this is an EB with twice the orbital period as suggested by B12.

\subsection{KOI 1468.01}
\begin{figure}[tbp]\includegraphics[width=0.5\textwidth]{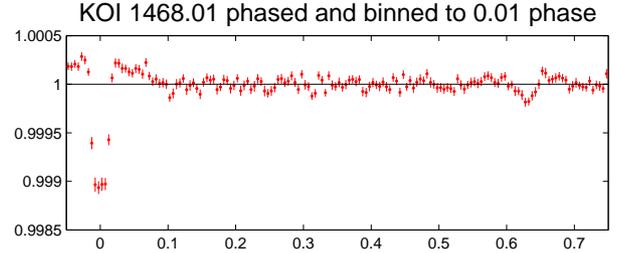}
\caption{Phase-folded and binned LC of KOI 1468.01 showing the secondary eclipse of this object, with depth that is consistent with a planetary interpretation of the transit signal.}
\label{KOI1468secondary}
\end{figure}

This object has transit-like signals ($\sim 1000 $ ppm deep), but we were able to detect also a secondary eclipse near phase 0.63 to KOI 1468.01 with a depth of $\sim 18 $ ppm (see Figure \ref{KOI1468secondary}). The expected depth for a planetary secondary eclipse (calculated similarly to KOI 741.01) has a fairly wide range: mostly in the $6-12 $ ppm range with $\sim 20\%$ chance of being higher, even as high as $40 $ ppm. This means that the observed signal is consistent with a secondary eclipse of KOI 1468.01 as an eccentric planet with a high geometric albedo.

\subsection{KOI 1540.01}
The long duration of this candidate signal caused us to examine this candidate in more datail. The background variability is difficult to remove and requires some fine-tuning. We managed to achieve good filtering by using a Savitzky Golay filter with a third degree polynomial and one day long window size with $4 \sigma$ iterative rejection applied separately for each continuous section. Once filtered, the odd and even transit events were clearly different (Figure \ref{KOI1540oddeven}), making this an EB at twice the published period for KOI 1540.01.

\subsection{KOI 1731.01}
\begin{figure}[tbp]\includegraphics[width=0.5\textwidth]{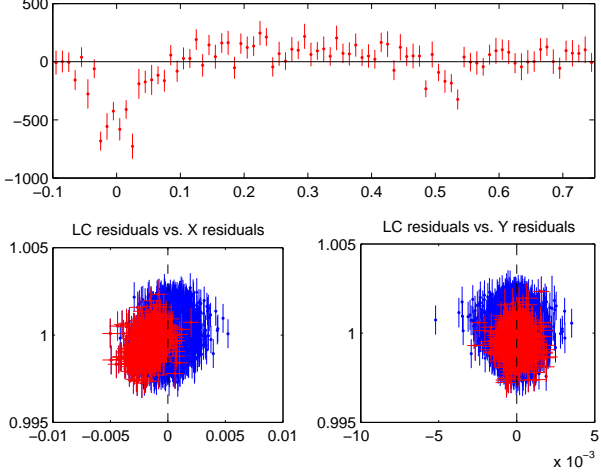}
\caption{Top: phase-folded to $P_2 \sim 0.84$d and heavily binned to 0.02 phase LC of KOI 1731. Bottom: centroid motion of KOI 1731 in phase with $P_2$, similar to Figure \ref{KOI856centroid}.}
\label{KOI1731contamination}
\end{figure}

This object was observed in quarters 3 and 4 only, and the contamination levels rose by a factor of 3 (from $\sim 8\%$ to $\sim 24\%$) between the two data sets. We note that in addition to the published $\sim 2.6$d transit-like signal we detect a second periodic transit-like signal with a period of $P_2=0.837894$d and depth of $500 $ ppm. This second signal also shows a very marginal secondary eclipse, slight out-of-transit variations and in-phase centroid motion (See Figure \ref{KOI1731contamination}), where the level of the latter is unchanged between the two quarters. We conclude that there is a contaminating EB in the KOI 1731 aperture with a short projected distance from the main target.

\subsection{KOI 2188.01}
\begin{figure}[tbp]\includegraphics[width=0.5\textwidth]{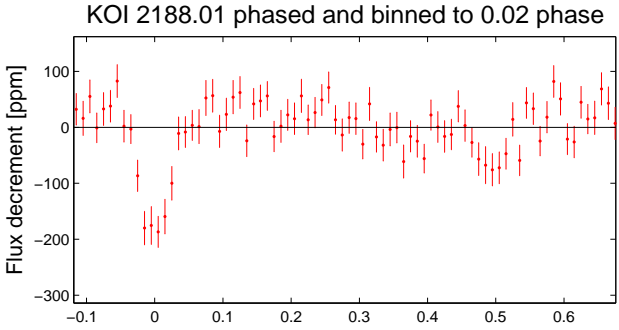}
\caption{Phase-folded to 2/5 of the published period of KOI 2188.01, heavily binned to 0.02 phase.}
\label{KOI2188secondary}
\end{figure}

The true period of this candidate is 2/5 of the published period, i.e. $P_{2188.01}=2.69615$d. When phased to this new period (see Figure \ref{KOI2188secondary}) the transit signal seems somewhat deeper than the B12 value of $157 $ ppm and a possible secondary eclipse is also marginally detected.

\subsection{KOI 2269.01}
We change the interpretation of KOI 2269.01 from a planet candidate to an EB with twice the period of KOI 2269.01 since we detect significant differences between the odd and even events in this system (Figure 
\ref{KOI2269oddeven}).

\subsection{KOI 2272.01}
This object has a scaled semi-major axis $a/R_*$ that is consistent with the minimal value of a=1. Additionally, the transit duration is too long: it is somewhat too long for a central transit of a $2.34 R_\odot$ star (the KIC value for KOI 2272), and definitely too long for a high impact parameter transit such as in this case ($b$ is nonzero with $>4 \sigma$ significance). This means that, as far as the we can tell from the data, the secondary object in this system is all but touching the host's surface, and it is not small compared to the primary (causing the long transit signal). We conclude that it is more likely to be a contact binary in a low-inclination orbit, with possibly some dilution, than a transiting planet.

\section{Discussion}
\label{Discuss}

We presented the discovery of \nNew transit signals found on a subset of less than 1\% of the \emph{Kepler} dataset - the Kepler objects of interest (KOIs). These signals were not detected by the \emph{Kepler}'s processing pipeline in the publicly available Q0-Q6 data. We were able to achieve that with the SARS pipeline that is was optimized for \emph{CoRoT}, and not for \emph{Kepler}, and was left almost unchanged from it. Still, the very basic fact that this pipeline is simply different from the one used by \emph{Kepler} team means that its sensitivities and biases are also different. Furthermore, we showed evidence that for the specific goal of transit searches the SARS pipeline may even be superior to \emph{Kepler}'s PDC-MAP -- helping to understand the significant number of new detections in such a small subsample of LCs. It is interesting to note that in ststems previously known to be multi-transiting often the \emph{order} in which signals are detected was different in the SARS pipeline compared to the published order -- this is a hallmark of the different sensitivities of each pipeline.

To boost the confidence level in the newly detected signals we employed a very large array of tests for candidate selection: to the regular binarity and contamination tests we added the ``half-half'' test for false-positives identification, required that signals are not effected by changing the long-term filter (and thus are even less likely to be an artifact of data processing), and sometimes confirmed that they are present in \emph{Kepler}'s PDC data as well. This suite of tests exceeds the one used by the \emph{Kepler} pipeline, helping to produce a clean catalog.


We also showed evidence that the errors given in B12 for the $d$ and $r_p$ parameters are overestimated for a significant fraction of the KOIs, to the point where for about half the KOIs the value of $d$ is not significantly different from zero ($\frac{d}{\Delta d} <3$), which is highly unlikely to be true.

The fact that additional transit candidates are found in the data of \nWereSingleLower systems that had only one previously known signal significantly boosts the chances that these systems are real planets since the false-alarm rate of multi-transiting planets is, intrinsically, very low. This is especially true for the KOI 277 system, which also exhibits anti-correlated TTVs in a dynamically stable configuration, all but validating the system. Also notable is the large fraction of planet candidates that are part of period commensurabilities, more than than presented in Fabrycky \etal (2012a). We do not have a good explanation for why this happened - but near-resonant systems are particularly useful because they may allow for the full confirmation of their planetary nature using photometry alone.

We advocate the parallel processing of the full data by other teams. However, in the near future the publicly available \emph{Kepler} data will significantly increase -- and with it the associated workload related to analyzing it, both in computer time and human time. We believe that the sheer volume of the work means that re-analyzing the data will be increasingly more difficult, and thus increasingly less possible for small groups of researches - but we hope that the results of this work will encourage others to work in this direction. For instance, the \emph{Kepler} pipeline has five primary modules: CAL, PA, PDC, TPS, and DV for the calibration, photometry, removal of systematic effects, transit search, and signal validation, respectively. In this paper we presented our own versions of the PDC, TPS, and DV modules - and these changes were productive. We therefore believe that an independent analysis matching the CAL and PA modules (which remain ``unchallenged'' to date) may also be productive.

Finally, since we were able to detect up to four additional transit signals in each of the presented systems, the chances of detecting more transiting systems -- and even more multi-transiting systems -- in the remaining $>99\%$ of the data are quite real.

\section{Acknowledgments}

A.O. acknowledges financial support from the Deutsche Forschungsgemeinschaft under DFG GRK 1351/2. We thank the anonymous referee for his/her report, which helped to improve the manuscript.

\onecolumn
\begin{landscape}
\begin{longtable}{|l|l|l|l|l|l|l|l|l|}
\caption{Results of the global fit for each system for both the previously known and new signals. Notes: (1) Negative period designates the best-fitting period that reproduces the observed transit duration/shape of a non-periodic mono-transit. (2) The $\frac{d}{R_*}$ is computed from the final $\frac{d_1}{R_*}$ for the first signal, times the Keplerian factor $(\frac{P_i}{P_1})^\frac{2}{3}$ - all directly from the MCMC distributions.} \\
\hline
KOI	&	New	&	Period$^{(1)}$	& Epoch 	& depth & $\frac{r}{R_*}$ & $\frac{d}{R_*} \quad ^{(2)}$ & i 	 &  $e \, sin \, \omega $ \\
	&	signal	&	[d]	& [BJD-2454900]	& [ppm]	&  		  & 		    & [degrees] &  \\
\hline
\endfirsthead
\hline
KOI	&	New	&	Period$^{(1)}$	& Epoch 	& depth & $\frac{r}{R_*}$ & $\frac{d}{R_*} \quad ^{(2)}$ & i 	 &  $e \, sin \, \omega $ \\
	&	signal	&	[d]	& [BJD-2454900]	& [ppm]	&  		  & 		    & [degrees] &  \\
\hline
\endhead
\hline
\multicolumn{9}{r}{\textit{Continued on next page.}} \\
\endfoot
\hline
\endlastfoot
KOI 179.01 &   & $20.74 \pm 6.1e-05$ & $75.8004 \pm 0.0012$ & $995$ & $0.02801 \pm 0.00014$ & $15.827 \pm 0.076$ & $89.9498 \pm 0.0022$ & 0 (fixed) \\ 
KOI 179 & + & $-603.7 \pm 2.2$ & $239.499 \pm 0.0018$ & $1783$ & $0.0407 \pm 0.00038$ & $149.77 \pm 0.82$ & $89.7432 \pm 0.0038$ & 0 (fixed) \\ 
\rule{-2pt}{2.5ex}

KOI 239.01 &   & $5.64068 \pm 1.2e-05$ & $71.5557 \pm 0.00069$ & $1263$ & $0.03533 \pm 0.00027$ & $10.569 \pm 0.075$ & $85.95 \pm 0.041$ & 0 (fixed) \\ 
KOI 239 & + & $3.62276 \pm 2.5e-06$ & $315.969 \pm 0.0017$ & $113$ & $0.00945 \pm 0.00045$ & $7.867 \pm 0.056$ & $89.9 \pm 0.06$ & 0 (fixed) \\ 
\rule{-2pt}{2.5ex}

KOI 241.01 &   & $13.8214 \pm 6.2e-05$ & $64.7923 \pm 0.0015$ & $738$ & $0.02408 \pm 0.00029$ & $29.29 \pm 0.43$ & $89.376 \pm 0.014$ & 0 (fixed) \\ 
KOI 241 & + & $3.41059 \pm 2.5e-06$ & $378.342 \pm 0.0009$ & $118$ & $0.00947 \pm 0.00041$ & $11.52 \pm 0.17$ & $89.764 \pm 0.078$ & 0 (fixed) \\ 
KOI 241 & + & $30.9489 \pm 3e-05$ & $396.344 \pm 0.0018$ & $207$ & $0.01253 \pm 0.0006$ & $50.12 \pm 0.74$ & $89.948 \pm 0.052$ & 0 (fixed) \\ 
\rule{-2pt}{2.5ex}

KOI 246.01 &   & $5.39876 \pm 7.5e-06$ & $106.856 \pm 0.00038$ & $351$ & $0.01932 \pm 0.00014$ & $7.12 \pm 0.17$ & $83.44 \pm 0.24$ & 0 (fixed) \\ 
KOI 246 & + & $9.6054 \pm 0.00026$ & $376.753 \pm 0.0029$ & $37$ & $0.00646 \pm 0.00024$ & $10.45 \pm 0.24$ & $85.32 \pm 0.15$ & 0 (fixed) \\ 
KOI 246 & + & $-740 \pm 160$ & $503.204 \pm 0.0059$ & $239$ & $0.01974 \pm 0.00093$ & $190 \pm 29$ & $89.708 \pm 0.042$ & 0 (fixed) \\ 
\rule{-2pt}{2.5ex}

KOI 260.01 &   & $10.4958 \pm 0.00011$ & $105.784 \pm 0.0031$ & $92$ & $0.01038 \pm 0.00021$ & $8.48 \pm 0.19$ & $84.08 \pm 0.17$ & 0 (fixed) \\ 
KOI 260.02 &   & $100.283 \pm 0.0011$ & $178.041 \pm 0.0019$ & $314$ & $0.01882 \pm 0.00021$ & $38.19 \pm 0.85$ & $88.717 \pm 0.039$ & 0 (fixed) \\ 
KOI 260 & + & $21.8701 \pm 0.00026$ & $379.472 \pm 0.0022$ & $93$ & $0.01083 \pm 0.00023$ & $13.84 \pm 0.31$ & $86.24 \pm 0.1$ & 0 (fixed) \\ 
\rule{-2pt}{2.5ex}

KOI 271.01 &   & $48.6309 \pm 0.00034$ & $105.545 \pm 0.0019$ & $332$ & $0.01878 \pm 0.00021$ & $31.5 \pm 0.7$ & $88.517 \pm 0.053$ & 0 (fixed) \\ 
KOI 271.02 &   & $29.3924 \pm 0.0002$ & $142.072 \pm 0.0016$ & $321$ & $0.01764 \pm 0.00017$ & $22.52 \pm 0.5$ & $88.145 \pm 0.081$ & 0 (fixed) \\ 
KOI 271 & + & $14.4364 \pm 0.00021$ & $305.487 \pm 0.0025$ & $81$ & $0.01018 \pm 0.00035$ & $14.02 \pm 0.31$ & $86.26 \pm 0.11$ & 0 (fixed) \\ 
\rule{-2pt}{2.5ex}

KOI 277.01 &   & $16.2295 \pm 0.00013$ & $104.661 \pm 0.0024$ & $460$ & $0.02521 \pm 0.00048$ & $6.04 \pm 0.31$ & $81.11 \pm 0.55$ & 0 (fixed) \\ 
KOI 277 & + & $13.8537 \pm 0.0016$ & $379.467 \pm 0.018$ & $34$ & $0.00703 \pm 0.0008$ & $5.44 \pm 0.28$ & $79.97 \pm 0.66$ & 0 (fixed) \\ 
\rule{-2pt}{2.5ex}

KOI 285.01 &   & $13.7487 \pm 9e-05$ & $112.281 \pm 0.0015$ & $404$ & $0.01871 \pm 0.00034$ & $15 \pm 0.91$ & $87.92 \pm 0.43$ & 0 (fixed) \\ 
KOI 285 & + & $26.7234 \pm 0.00071$ & $268.002 \pm 0.0035$ & $125$ & $0.01014 \pm 0.0003$ & $23.4 \pm 1.4$ & $89.06 \pm 0.45$ & 0 (fixed) \\ 
KOI 285 & + & $49.3616 \pm 0.0046$ & $264.715 \pm 0.01$ & $66$ & $0.00733 \pm 0.00042$ & $35.2 \pm 2.1$ & $89.44 \pm 0.36$ & 0 (fixed) \\ 
\rule{-2pt}{2.5ex}

KOI 289.01 &   & $26.6293 \pm 0.00024$ & $124.995 \pm 0.0021$ & $432$ & $0.0197 \pm 0.0012$ & $20.8 \pm 4.5$ & $88.3 \pm 1.1$ & 0 (fixed) \\ 
KOI 289 & + & $1.47762 \pm 5.3e-05$ & $377.688 \pm 0.0053$ & $26$ & $0.006 \pm 0.014$ & $3.03 \pm 0.66$ & $71.8 \pm 5.4$ & 0 (fixed) \\ 
\rule{-2pt}{2.5ex}

KOI 295.01 &   & $5.3174 \pm 1.9e-05$ & $104.875 \pm 0.00092$ & $275$ & $0.01511 \pm 0.00029$ & $13.07 \pm 0.86$ & $88.26 \pm 0.77$ & 0 (fixed) \\ 
KOI 295 & + & $10.1057 \pm 0.00015$ & $371.68 \pm 0.0025$ & $99$ & $0.01065 \pm 0.00059$ & $20.1 \pm 1.3$ & $87.54 \pm 0.22$ & 0 (fixed) \\ 
\rule{-2pt}{2.5ex}

KOI 316.01 &   & $15.7708 \pm 8.3e-05$ & $117.908 \pm 0.0014$ & $502$ & $0.02103 \pm 0.00032$ & $19.38 \pm 0.94$ & $88.27 \pm 0.25$ & 0 (fixed) \\ 
KOI 316.02 &   & $157.05 \pm 0.011$ & $414.269 \pm 0.0056$ & $447$ & $0.02176 \pm 0.00068$ & $89.7 \pm 4.4$ & $89.483 \pm 0.039$ & 0 (fixed) \\ 
KOI 316 & + & $7.30571 \pm 5.4e-05$ & $313.172 \pm 0.001$ & $194$ & $0.01771 \pm 0.00062$ & $11.6 \pm 0.56$ & $85.25 \pm 0.25$ & 0 (fixed) \\ 
\rule{-2pt}{2.5ex}

KOI 321.01 &   & $2.42632 \pm 1.1e-05$ & $103.454 \pm 0.0012$ & $162$ & $0.01339 \pm 0.00057$ & $4.07 \pm 0.57$ & $78.1 \pm 2.3$ & 0 (fixed) \\ 
KOI 321 & + & $4.62324 \pm 8.9e-05$ & $310.404 \pm 0.0037$ & $49$ & $0.00757 \pm 0.00049$ & $6.26 \pm 0.88$ & $82.1 \pm 1.4$ & 0 (fixed) \\ 
\rule{-2pt}{2.5ex}

KOI 330.01 &   & $7.974 \pm 7.1e-05$ & $107.53 \pm 0.0031$ & $231$ & $0.01447 \pm 0.0003$ & $9.35 \pm 0.11$ & $86.086 \pm 0.024$ & 0 (fixed) \\ 
KOI 330 & + & $1.88897 \pm 6e-06$ & $314.439 \pm 0.0014$ & $32$ & $0.00505 \pm 0.00038$ & $3.58 \pm 0.043$ & $89.67 \pm 0.2$ & 0 (fixed) \\ 
\rule{-2pt}{2.5ex}

KOI 339.01 &   & $1.98035 \pm 8.8e-06$ & $103.138 \pm 0.0012$ & $241$ & $0.01447 \pm 0.00058$ & $5.51 \pm 0.8$ & $84.5 \pm 3.1$ & 0 (fixed) \\ 
KOI 339.02 &   & $12.8339 \pm 0.00024$ & $71.3342 \pm 0.0065$ & $235$ & $0.01611 \pm 0.00076$ & $19.2 \pm 2.8$ & $87.49 \pm 0.53$ & 0 (fixed) \\ 
KOI 339 & + & $35.863 \pm 0.0013$ & $391.692 \pm 0.0066$ & $170$ & $0.01257 \pm 0.00086$ & $38 \pm 5.5$ & $88.99 \pm 0.35$ & 0 (fixed) \\ 
\rule{-2pt}{2.5ex}

KOI 408.01 &   & $7.38202 \pm 2.5e-05$ & $106.071 \pm 0.00084$ & $1403$ & $0.03723 \pm 0.00036$ & $12.42 \pm 0.27$ & $86.56 \pm 0.14$ & 0 (fixed) \\ 
KOI 408.02 &   & $12.5609 \pm 8.3e-05$ & $99.796 \pm 0.0017$ & $811$ & $0.02825 \pm 0.00048$ & $17.7 \pm 0.39$ & $87.6 \pm 0.1$ & 0 (fixed) \\ 
KOI 408.03 &   & $30.825 \pm 0.00047$ & $86.0151 \pm 0.005$ & $646$ & $0.02639 \pm 0.00074$ & $32.2 \pm 0.7$ & $88.541 \pm 0.052$ & 0 (fixed) \\ 
KOI 408 & + & $3.42818 \pm 6.1e-05$ & $312.184 \pm 0.0026$ & $164$ & $0.0134 \pm 0.00081$ & $7.45 \pm 0.16$ & $83.61 \pm 0.3$ & 0 (fixed) \\ 
\rule{-2pt}{2.5ex}

KOI 435.01 &   & $20.55 \pm 9.6e-05$ & $111.944 \pm 0.0011$ & $1301$ & $0.0374 \pm 0.0018$ & $20.68 \pm 0.37$ & $87.951 \pm 0.07$ & 0 (fixed) \\ 
KOI 435.02 (no data) &   & $-0 \pm 0$ & $0 \pm 0$ & $0$ & $0 \pm 0$ & $0 \pm 0$ & $0 \pm 0$ & 0 (fixed) \\ 
KOI 435 & + & $3.93268 \pm 7.1e-05$ & $311.48 \pm 0.0022$ & $159$ & $0.01335 \pm 0.00068$ & $6.87 \pm 0.12$ & $83.5 \pm 0.27$ & 0 (fixed) \\ 
KOI 435 & + & $33.0388 \pm 0.00069$ & $325.503 \pm 0.0039$ & $368$ & $0.0234 \pm 0.0011$ & $28.38 \pm 0.51$ & $88.128 \pm 0.036$ & 0 (fixed) \\ 
KOI 435 & + & $62.3049 \pm 0.0014$ & $361.311 \pm 0.004$ & $360$ & $0.0202 \pm 0.0011$ & $43.31 \pm 0.78$ & $88.962 \pm 0.027$ & 0 (fixed) \\ 
KOI 435 & + & $9.92003 \pm 0.00061$ & $317.853 \pm 0.011$ & $85$ & $0.00981 \pm 0.00082$ & $12.72 \pm 0.23$ & $86.47 \pm 0.36$ & 0 (fixed) \\ 
\rule{-2pt}{2.5ex}

KOI 505.01 &   & $13.7671 \pm 9.1e-05$ & $107.811 \pm 0.0018$ & $623$ & $0.02431 \pm 0.00081$ & $28.4 \pm 2.2$ & $88.59 \pm 0.23$ & 0 (fixed) \\ 
KOI 505.02 &   & $6.19551 \pm 7.7e-05$ & $67.7124 \pm 0.0032$ & $229$ & $0.01333 \pm 0.00047$ & $16.7 \pm 1.3$ & $89.02 \pm 0.85$ & 0 (fixed) \\ 
KOI 505 & + & $3.25059 \pm 2.8e-05$ & $313.737 \pm 0.0014$ & $207$ & $0.01312 \pm 0.00062$ & $10.86 \pm 0.82$ & $87.4 \pm 1.0$ & 0 (fixed) \\ 
KOI 505 & + & $87.09 \pm 0.0012$ & $294.498 \pm 0.0025$ & $1057$ & $0.02948 \pm 0.00087$ & $97.2 \pm 7.4$ & $89.72 \pm 0.095$ & 0 (fixed) \\ 
KOI 505 & + & $8.34808 \pm 0.00011$ & $311.937 \pm 0.0019$ & $246$ & $0.01462 \pm 0.00066$ & $20.4 \pm 1.5$ & $88.36 \pm 0.43$ & 0 (fixed) \\ 
\rule{-2pt}{2.5ex}

KOI 509.01 &   & $4.16694 \pm 2.4e-05$ & $102.715 \pm 0.0016$ & $698$ & $0.02331 \pm 0.00025$ & $11.78 \pm 0.1$ & $89.581 \pm 0.062$ & 0 (fixed) \\ 
KOI 509.02 &   & $11.4636 \pm 6.2e-05$ & $70.3806 \pm 0.0014$ & $889$ & $0.02962 \pm 0.00051$ & $23.12 \pm 0.2$ & $88.1599 \pm 0.0035$ & 0 (fixed) \\ 
KOI 509 & + & $39.587 \pm 3.8e-05$ & $299.126 \pm 0.00089$ & $402$ & $0.01767 \pm 0.00073$ & $52.82 \pm 0.46$ & $89.94 \pm 0.1$ & 0 (fixed) \\ 
\rule{-2pt}{2.5ex}

KOI 510.01 &   & $2.94033 \pm 2.2e-05$ & $102.902 \pm 0.0016$ & $407$ & $0.01967 \pm 0.00061$ & $6.34 \pm 0.53$ & $83.6 \pm 1.1$ & 0 (fixed) \\ 
KOI 510.02 &   & $6.38906 \pm 4.9e-05$ & $108.472 \pm 0.0021$ & $475$ & $0.02175 \pm 0.00073$ & $10.64 \pm 0.89$ & $85.95 \pm 0.61$ & 0 (fixed) \\ 
KOI 510.03 &   & $14.627 \pm 0.00028$ & $69.2403 \pm 0.0053$ & $325$ & $0.01901 \pm 0.00096$ & $18.5 \pm 1.5$ & $87.41 \pm 0.29$ & 0 (fixed) \\ 
KOI 510 & + & $35.1182 \pm 0.00088$ & $295.831 \pm 0.0034$ & $526$ & $0.0259 \pm 0.0014$ & $33.1 \pm 2.8$ & $88.44 \pm 0.17$ & 0 (fixed) \\ 
\rule{-2pt}{2.5ex}

KOI 564.01 &   & $21.0549 \pm 0.00047$ & $104.902 \pm 0.0062$ & $553$ & $0.02509 \pm 0.00059$ & $12.2 \pm 0.76$ & $85.98 \pm 0.34$ & 0 (fixed) \\ 
KOI 564.02 &   & $127.855 \pm 0.0038$ & $179.554 \pm 0.0067$ & $2415$ & $0.05293 \pm 0.00093$ & $40.6 \pm 2.5$ & $88.781 \pm 0.1$ & 0 (fixed) \\ 
KOI 564 & + & $6.21739 \pm 0.00018$ & $381.8 \pm 0.0047$ & $134$ & $0.01321 \pm 0.00086$ & $5.41 \pm 0.34$ & $80.27 \pm 0.71$ & 0 (fixed) \\ 
\rule{-2pt}{2.5ex}

KOI 582.01 &   & $5.94508 \pm 2.8e-05$ & $103.468 \pm 0.0012$ & $741$ & $0.02428 \pm 0.00057$ & $16.8 \pm 1.1$ & $88.8 \pm 0.65$ & 0 (fixed) \\ 
KOI 582.02 &   & $17.7388 \pm 0.00023$ & $81.9821 \pm 0.004$ & $477$ & $0.0222 \pm 0.0009$ & $34.7 \pm 2.2$ & $88.71 \pm 0.14$ & 0 (fixed) \\ 
KOI 582 & + & $9.93964 \pm 0.00018$ & $313.075 \pm 0.0026$ & $252$ & $0.0143 \pm 0.00065$ & $23.6 \pm 1.5$ & $88.98 \pm 0.46$ & 0 (fixed) \\ 
\rule{-2pt}{2.5ex}

KOI 584.01 &   & $9.92669 \pm 5.2e-05$ & $108.686 \pm 0.0012$ & $663$ & $0.02712 \pm 0.00074$ & $11.6 \pm 1$ & $85.88 \pm 0.54$ & 0 (fixed) \\ 
KOI 584.02 &   & $21.2233 \pm 0.00025$ & $103.374 \pm 0.0027$ & $510$ & $0.02428 \pm 0.00069$ & $19.3 \pm 1.7$ & $87.45 \pm 0.31$ & 0 (fixed) \\ 
KOI 584 & + & $6.47024 \pm 0.0002$ & $308.901 \pm 0.0048$ & $86$ & $0.00986 \pm 0.00077$ & $8.72 \pm 0.75$ & $84.44 \pm 0.69$ & 0 (fixed) \\ 
\rule{-2pt}{2.5ex}

KOI 593.01 &   & $9.99772 \pm 9.6e-05$ & $104.784 \pm 0.0024$ & $492$ & $0.02363 \pm 0.00061$ & $12.43 \pm 0.5$ & $86.06 \pm 0.21$ & 0 (fixed) \\ 
KOI 593.02 &   & $90.4108 \pm 0.0025$ & $141.502 \pm 0.0057$ & $1107$ & $0.0361 \pm 0.0011$ & $54 \pm 2.2$ & $89.072 \pm 0.051$ & 0 (fixed) \\ 
KOI 593 & + & $51.065 \pm 0.0013$ & $335.943 \pm 0.0046$ & $438$ & $0.0237 \pm 0.0012$ & $36.9 \pm 1.5$ & $88.588 \pm 0.071$ & 0 (fixed) \\ 
\rule{-2pt}{2.5ex}

KOI 597.01 &   & $17.3083 \pm 0.0003$ & $109.939 \pm 0.0034$ & $489$ & $0.02127 \pm 0.00054$ & $20.75 \pm 0.27$ & $88.1624 \pm 0.005$ & 0 (fixed) \\ 
KOI 597.02 &   & $2.09224 \pm 2.6e-05$ & $66.0648 \pm 0.0029$ & $163$ & $0.01209 \pm 0.00043$ & $5.072 \pm 0.067$ & $83.078 \pm 0.069$ & 0 (fixed) \\ 
KOI 597 & + & $52.8153 \pm 0.00013$ & $309.161 \pm 0.00089$ & $344$ & $0.0177 \pm 0.0015$ & $43.65 \pm 0.58$ & $89.15 \pm 0.19$ & 0 (fixed) \\ 
\rule{-2pt}{2.5ex}

KOI 623.01 &   & $10.3498 \pm 0.00014$ & $107.06 \pm 0.0031$ & $95$ & $0.00928 \pm 0.00037$ & $14.7 \pm 1.7$ & $87.52 \pm 0.79$ & 0 (fixed) \\ 
KOI 623.02 &   & $15.6776 \pm 0.00023$ & $112.467 \pm 0.0044$ & $84$ & $0.00853 \pm 0.00036$ & $19.4 \pm 2.3$ & $88.38 \pm 0.67$ & 0 (fixed) \\ 
KOI 623.03 &   & $5.59936 \pm 6.9e-05$ & $104.473 \pm 0.0032$ & $65$ & $0.00756 \pm 0.00032$ & $9.7 \pm 1.2$ & $86.5 \pm 1.2$ & 0 (fixed) \\ 
KOI 623 & + & $25.2112 \pm 0.0015$ & $314.306 \pm 0.0082$ & $41$ & $0.00591 \pm 0.00044$ & $26.6 \pm 3.2$ & $88.9 \pm 0.62$ & 0 (fixed) \\ 
\rule{-2pt}{2.5ex}

KOI 624.01 &   & $17.7898 \pm 3.7e-05$ & $115.439 \pm 0.00056$ & $861$ & $0.02591 \pm 0.00021$ & $31.62 \pm 0.24$ & $89.9904 \pm 0.0065$ & 0 (fixed) \\ 
KOI 624.02 &   & $49.5668 \pm 0.0002$ & $102.542 \pm 0.0018$ & $604$ & $0.02549 \pm 0.00065$ & $62.61 \pm 0.47$ & $89.2521 \pm 6.7e-05$ & 0 (fixed) \\ 
KOI 624 & + & $1.31184 \pm 9.2e-07$ & $312.733 \pm 0.00062$ & $210$ & $0.01283 \pm 0.00018$ & $5.56 \pm 0.042$ & $89.927 \pm 0.039$ & 0 (fixed) \\ 
\rule{-2pt}{2.5ex}

KOI 627.01 &   & $7.75196 \pm 4.9e-05$ & $109.169 \pm 0.0016$ & $311$ & $0.0179 \pm 0.0012$ & $11 \pm 3$ & $85.9 \pm 1.6$ & 0 (fixed) \\ 
KOI 627 & + & $4.16521 \pm 8.7e-05$ & $314.963 \pm 0.0022$ & $99$ & $0.00979 \pm 0.0007$ & $7.3 \pm 2$ & $84.3 \pm 3$ & 0 (fixed) \\ 
\rule{-2pt}{2.5ex}

KOI 671.01 &   & $4.22865 \pm 4.9e-05$ & $103.741 \pm 0.0028$ & $133$ & $0.0134 \pm 0.00045$ & $3.99 \pm 0.35$ & $76.6 \pm 1.5$ & 0 (fixed) \\ 
KOI 671.03 &   & $16.2601 \pm 0.00054$ & $77.5275 \pm 0.0095$ & $97$ & $0.01203 \pm 0.00072$ & $9.79 \pm 0.86$ & $84.43 \pm 0.57$ & 0 (fixed) \\ 
KOI 671 & + & $7.46647 \pm 0.00029$ & $312.176 \pm 0.0044$ & $84$ & $0.01065 \pm 0.00069$ & $5.83 \pm 0.51$ & $80.8 \pm 1.1$ & 0 (fixed) \\ 
KOI 671 & + & $11.1317 \pm 0.00056$ & $312.328 \pm 0.0068$ & $69$ & $0.00984 \pm 0.00078$ & $7.61 \pm 0.66$ & $82.91 \pm 0.76$ & 0 (fixed) \\ 
\rule{-2pt}{2.5ex}

KOI 710.01 &   & $5.37515 \pm 0.00014$ & $103.928 \pm 0.0067$ & $115$ & $0.01184 \pm 0.00074$ & $4.87 \pm 0.54$ & $79.3 \pm 1.5$ & 0 (fixed) \\ 
KOI 710 & + & $8.58573 \pm 0.00037$ & $371.47 \pm 0.0067$ & $88$ & $0.01088 \pm 0.00098$ & $6.65 \pm 0.74$ & $81.9 \pm 1$ & 0 (fixed) \\ 
\rule{-2pt}{2.5ex}

KOI 717.01 &   & $14.7074 \pm 0.00026$ & $108.79 \pm 0.0035$ & $231$ & $0.01469 \pm 0.00047$ & $28.27 \pm 0.6$ & $88.621 \pm 0.014$ & 0 (fixed) \\ 
KOI 717 & + & $0.900351 \pm 4.3e-06$ & $376.749 \pm 0.0015$ & $33$ & $0.00511 \pm 0.00033$ & $4.391 \pm 0.093$ & $89.71 \pm 0.25$ & 0 (fixed) \\ 
\rule{-2pt}{2.5ex}

KOI 719.01 &   & $9.03423 \pm 2.3e-05$ & $104.013 \pm 0.00066$ & $537$ & $0.02566 \pm 0.00052$ & $22.8 \pm 1.2$ & $87.79 \pm 0.16$ & 0 (fixed) \\ 
KOI 719 & + & $4.15978 \pm 2.7e-05$ & $312.214 \pm 0.00093$ & $138$ & $0.01183 \pm 0.00043$ & $13.61 \pm 0.74$ & $86.81 \pm 0.33$ & 0 (fixed) \\ 
KOI 719 & + & $45.9041 \pm 0.0007$ & $329.074 \pm 0.0022$ & $321$ & $0.01602 \pm 0.00037$ & $67.5 \pm 3.7$ & $89.73 \pm 0.16$ & 0 (fixed) \\ 
KOI 719 & + & $28.1225 \pm 0.00048$ & $304.893 \pm 0.0027$ & $176$ & $0.01186 \pm 0.00038$ & $48.7 \pm 2.6$ & $89.62 \pm 0.21$ & 0 (fixed) \\ 
\rule{-2pt}{2.5ex}

KOI 780.01 &   & $2.33745 \pm 9.9e-06$ & $104.76 \pm 0.001$ & $857$ & $0.02594 \pm 0.00064$ & $8.7 \pm 0.58$ & $87.7 \pm 1.4$ & 0 (fixed) \\ 
KOI 780 & + & $7.24078 \pm 7.7e-05$ & $309.698 \pm 0.0015$ & $398$ & $0.06 \pm 0.033$ & $18.5 \pm 1.2$ & $86.84 \pm 0.31$ & 0 (fixed) \\ 
\rule{-2pt}{2.5ex}

KOI 841.01 &   & $15.3349 \pm 7.2e-05$ & $107.006 \pm 0.0014$ & $2850$ & $0.05338 \pm 0.0007$ & $22.99 \pm 0.64$ & $88.12 \pm 0.092$ & 0 (fixed) \\ 
KOI 841.02 &   & $31.3309 \pm 0.00014$ & $86.4256 \pm 0.0012$ & $5005$ & $0.07035 \pm 0.0008$ & $37 \pm 1$ & $88.85 \pm 0.058$ & 0 (fixed) \\ 
KOI 841 & + & $269.321 \pm 0.031$ & $487.939 \pm 0.022$ & $1160$ & $0.0344 \pm 0.0021$ & $155.4 \pm 4.3$ & $89.715 \pm 0.018$ & 0 (fixed) \\ 
KOI 841 & + & $6.54629 \pm 6.5e-05$ & $315.62 \pm 0.0012$ & $670$ & $0.02678 \pm 0.00083$ & $13.04 \pm 0.36$ & $86.44 \pm 0.16$ & 0 (fixed) \\ 
\rule{-2pt}{2.5ex}

KOI 856.01 &   & $39.749 \pm 6.3e-05$ & $105.853 \pm 0.0004$ & $13731$ & $0.1285 \pm 0.0013$ & $29.89 \pm 0.29$ & $88.334 \pm 0.024$ & 0 (fixed) \\ 
KOI 856 & + & $0.870588 \pm 1.5e-05$ & $380.218 \pm 0.0023$ & $94$ & $0.036 \pm 0.031$ & $2.34 \pm 0.023$ & $64.2 \pm 1.2$ & 0 (fixed) \\ 
\rule{-2pt}{2.5ex}

KOI 1060.01 &   & $12.1097 \pm 0.00033$ & $73.2107 \pm 0.0066$ & $193$ & $0.0139 \pm 0.00088$ & $11.2 \pm 2.8$ & $86.1 \pm 1.6$ & 0 (fixed) \\ 
KOI 1060.02 &   & $4.75768 \pm 0.00019$ & $70.713 \pm 0.01$ & $101$ & $0.0098 \pm 0.00062$ & $6 \pm 1.5$ & $83.3 \pm 3.3$ & 0 (fixed) \\ 
KOI 1060.04 &   & $8.19414 \pm 0.00054$ & $71.972 \pm 0.029$ & $63$ & $0.0092 \pm 0.0034$ & $8.6 \pm 2.1$ & $83.8 \pm 1.9$ & 0 (fixed) \\ 
KOI 1060 & + & $20.4973 \pm 0.0013$ & $305.914 \pm 0.0089$ & $119$ & $0.0112 \pm 0.001$ & $15.9 \pm 3.9$ & $87.1 \pm 1$ & 0 (fixed) \\ 
\rule{-2pt}{2.5ex}

KOI 1069.02 &   & $2.46693 \pm 5.6e-05$ & $287.933 \pm 0.0039$ & $284$ & $0.0158 \pm 0.0011$ & $9.1 \pm 1.1$ & $86.3 \pm 1.6$ & 0 (fixed) \\ 
KOI 1069 & + & $8.70386 \pm 0.00012$ & $421.123 \pm 0.0011$ & $995$ & $0.0285 \pm 0.0012$ & $21.1 \pm 2.5$ & $88.82 \pm 0.81$ & 0 (fixed) \\ 
\rule{-2pt}{2.5ex}

KOI 1082.01 &   & $6.50329 \pm 9.7e-05$ & $68.271 \pm 0.0052$ & $384$ & $0.01713 \pm 0.00088$ & $22.6 \pm 1.4$ & $89.76 \pm 0.11$ & 0 (fixed) \\ 
KOI 1082 & + & $1.19659 \pm 1.9e-06$ & $313.218 \pm 0.0026$ & $175$ & $0.0116 \pm 0.00078$ & $7.3 \pm 0.47$ & $89.51 \pm 0.21$ & 0 (fixed) \\ 
KOI 1082 & + & $4.09657 \pm 9e-06$ & $311.893 \pm 0.0021$ & $234$ & $0.0134 \pm 0.001$ & $16.6 \pm 1.1$ & $89.917 \pm 0.059$ & 0 (fixed) \\ 
\rule{-2pt}{2.5ex}

KOI 1108.01 &   & $18.9251 \pm 0.00011$ & $82.4951 \pm 0.0025$ & $540$ & $0.02215 \pm 0.0006$ & $30.77 \pm 0.58$ & $88.802 \pm 0.035$ & 0 (fixed) \\ 
KOI 1108 & + & $1.47534 \pm 2.1e-06$ & $312.687 \pm 0.00088$ & $121$ & $0.00968 \pm 0.00041$ & $5.62 \pm 0.11$ & $89.77 \pm 0.24$ & 0 (fixed) \\ 
KOI 1108 & + & $4.15245 \pm 6.6e-06$ & $312.575 \pm 0.0015$ & $166$ & $0.01132 \pm 0.00049$ & $11.19 \pm 0.21$ & $89.77 \pm 0.13$ & 0 (fixed) \\ 
\rule{-2pt}{2.5ex}

KOI 1413.01 &   & $12.645 \pm 0.0006$ & $71.637 \pm 0.012$ & $116$ & $0.01046 \pm 0.00074$ & $8.9 \pm 1.4$ & $85.6 \pm 1.5$ & 0 (fixed) \\ 
KOI 1413 & + & $21.5274 \pm 0.0011$ & $302.07 \pm 0.0073$ & $132$ & $0.01173 \pm 0.0008$ & $12.7 \pm 2$ & $86.42 \pm 0.87$ & 0 (fixed) \\ 
\rule{-2pt}{2.5ex}

KOI 1536.01 &   & $3.74446 \pm 6.3e-05$ & $66.5566 \pm 0.0054$ & $59$ & $0.00936 \pm 0.0005$ & $3.36 \pm 0.17$ & $73.55 \pm 0.98$ & 0 (fixed) \\ 
KOI 1536 & + & $79.4942 \pm 0.0054$ & $355.839 \pm 0.014$ & $137$ & $0.01354 \pm 0.00089$ & $25.8 \pm 1.3$ & $87.95 \pm 0.13$ & 0 (fixed) \\ 
\rule{-2pt}{2.5ex}

KOI 1574.01 &   & $114.737 \pm 0.00032$ & $98.1481 \pm 0.001$ & $4750$ & $0.06276 \pm 0.00024$ & $72.18 \pm 0.25$ & $89.6501 \pm 0.0031$ & 0 (fixed) \\ 
KOI 1574 & + & $191.549 \pm 0.00045$ & $219.072 \pm 0.0029$ & $651$ & $0.0233 \pm 0.0015$ & $101.58 \pm 0.35$ & $89.744 \pm 0.065$ & 0 (fixed) \\ 
\rule{-2pt}{2.5ex}

KOI 1593.01 &   & $9.69455 \pm 0.00016$ & $115.466 \pm 0.0039$ & $509$ & $0.0223 \pm 0.0013$ & $21.58 \pm 0.73$ & $88.054 \pm 0.048$ & 0 (fixed) \\ 
KOI 1593 & + & $15.3834 \pm 2.3e-05$ & $287.431 \pm 0.0023$ & $461$ & $0.019 \pm 0.0011$ & $29.35 \pm 0.99$ & $89.9 \pm 0.11$ & 0 (fixed) \\ 
\rule{-2pt}{2.5ex}

KOI 1599.01 &   & $20.4201 \pm 0.0007$ & $73.0055 \pm 0.0079$ & $369$ & $0.0192 \pm 0.0013$ & $21.1 \pm 3.7$ & $87.94 \pm 0.69$ & 0 (fixed) \\ 
KOI 1599 & + & $13.6129 \pm 0.00026$ & $305.419 \pm 0.0027$ & $331$ & $0.0196 \pm 0.0013$ & $16.1 \pm 2.9$ & $86.91 \pm 0.79$ & 0 (fixed) \\ 
\rule{-2pt}{2.5ex}

KOI 1601.01 &   & $10.3515 \pm 0.00031$ & $66.7014 \pm 0.0085$ & $205$ & $0.0148 \pm 0.0011$ & $8.9 \pm 2.4$ & $84.8 \pm 2.1$ & 0 (fixed) \\ 
KOI 1601 & + & $62.9133 \pm 0.0041$ & $336.376 \pm 0.0086$ & $230$ & $0.0153 \pm 0.0013$ & $29.7 \pm 7.9$ & $88.5 \pm 0.66$ & 0 (fixed) \\ 
\rule{-2pt}{2.5ex}

KOI 1650.01 &   & $6.53191 \pm 0.00026$ & $69.032 \pm 0.016$ & $593$ & $0.0275 \pm 0.0012$ & $7.63 \pm 0.078$ & $83.2177 \pm 0.0021$ & 0 (fixed) \\ 
KOI 1650 & + & $100.82 \pm 9.5e-05$ & $351.054 \pm 0.0012$ & $1210$ & $0.0387 \pm 0.0019$ & $47.3 \pm 0.48$ & $88.923 \pm 0.019$ & 0 (fixed) \\ 
\rule{-2pt}{2.5ex}

KOI 1681.01 &   & $6.93933 \pm 9.1e-05$ & $71.3278 \pm 0.0038$ & $597$ & $0.02173 \pm 0.00094$ & $27.1 \pm 2$ & $89.276 \pm 0.083$ & 0 (fixed) \\ 
KOI 1681 & + & $1.99273 \pm 2e-06$ & $311.811 \pm 0.00096$ & $275$ & $0.01449 \pm 0.00079$ & $11.8 \pm 0.88$ & $89.929 \pm 0.071$ & 0 (fixed) \\ 
KOI 1681 & + & $3.53103 \pm 3.4e-06$ & $310.956 \pm 0.0016$ & $279$ & $0.01458 \pm 0.00095$ & $17.3 \pm 1.3$ & $89.83 \pm 0.13$ & 0 (fixed) \\ 
\rule{-2pt}{2.5ex}

KOI 1830.01 &   & $13.2268 \pm 8.5e-05$ & $75.2071 \pm 0.0016$ & $742$ & $0.02676 \pm 0.00086$ & $22.4 \pm 2$ & $88.17 \pm 0.31$ & 0 (fixed) \\ 
KOI 1830 & + & $198.705 \pm 0.0025$ & $288.227 \pm 0.0022$ & $1997$ & $0.0429 \pm 0.0014$ & $137 \pm 12$ & $89.718 \pm 0.055$ & 0 (fixed) \\ 
\rule{-2pt}{2.5ex}

KOI 1831.01 &   & $51.804 \pm 0.00086$ & $116.011 \pm 0.0038$ & $832$ & $0.0298 \pm 0.0018$ & $42.3 \pm 8.7$ & $88.91 \pm 0.32$ & 0 (fixed) \\ 
KOI 1831.03 &   & $34.2154 \pm 0.0017$ & $81.72 \pm 0.013$ & $116$ & $0.0131 \pm 0.0017$ & $32.1 \pm 6.6$ & $88.32 \pm 0.35$ & 0 (fixed) \\ 
KOI 1831 & + & $4.38497 \pm 0.00012$ & $379.194 \pm 0.0046$ & $81$ & $0.0103 \pm 0.0013$ & $8.2 \pm 1.7$ & $83.6 \pm 1.4$ & 0 (fixed) \\ 
\rule{-2pt}{2.5ex}

KOI 1843.01 &   & $4.19454 \pm 2.2e-05$ & $67.1311 \pm 0.0013$ & $599$ & $0.02278 \pm 0.0008$ & $19.98 \pm 0.79$ & $89.32 \pm 0.45$ & 0 (fixed) \\ 
KOI 1843.02 &   & $6.35611 \pm 0.00015$ & $70.4902 \pm 0.0047$ & $209$ & $0.015 \pm 0.0017$ & $26.4 \pm 1$ & $88.19 \pm 0.17$ & 0 (fixed) \\ 
KOI 1843 & + & $0.176893 \pm 4.1e-07$ & $314.145 \pm 0.00039$ & $123$ & $0.01116 \pm 0.00068$ & $2.421 \pm 0.096$ & $76.2 \pm 2.4$ & 0 (fixed) \\ 
\rule{-2pt}{2.5ex}

KOI 1870.01 &   & $7.96431 \pm 3.5e-05$ & $69.8153 \pm 0.0011$ & $667$ & $0.025 \pm 0.0011$ & $29 \pm 3.4$ & $88.63 \pm 0.35$ & 0 (fixed) \\ 
KOI 1870 & + & $-1340 \pm 430$ & $537.107 \pm 0.0011$ & $6407$ & $0.07042 \pm 0.001$ & $890 \pm 280$ & $89.983 \pm 0.012$ & 0 (fixed) \\ 
\rule{-2pt}{2.5ex}

KOI 1871.01 &   & $92.7272 \pm 0.00071$ & $110.446 \pm 0.0023$ & $1071$ & $0.03173 \pm 0.00068$ & $85.5 \pm 1.1$ & $89.5355 \pm 0.002$ & 0 (fixed) \\ 
KOI 1871 & + & $32.3755 \pm 2.1e-05$ & $310.049 \pm 0.0007$ & $652$ & $0.0258 \pm 0.00092$ & $42.37 \pm 0.53$ & $88.972 \pm 0.031$ & 0 (fixed) \\ 
\rule{-2pt}{2.5ex}

KOI 1875.01 &   & $9.91701 \pm 0.00015$ & $68.1447 \pm 0.0042$ & $357$ & $0.01724 \pm 0.00045$ & $21.65 \pm 0.49$ & $88.854 \pm 0.049$ & 0 (fixed) \\ 
KOI 1875 & + & $0.538347 \pm 2.5e-06$ & $313.369 \pm 0.00093$ & $164$ & $0.01139 \pm 0.00023$ & $3.104 \pm 0.07$ & $89.936 \pm 0.066$ & 0 (fixed) \\ 
\rule{-2pt}{2.5ex}

KOI 1955.01 &   & $15.1701 \pm 0.0002$ & $65.6677 \pm 0.0035$ & $199$ & $0.01457 \pm 0.00075$ & $12.85 \pm 0.27$ & $85.93 \pm 0.1$ & 0 (fixed) \\ 
KOI 1955.02 &   & $39.4627 \pm 0.0016$ & $69.438 \pm 0.013$ & $162$ & $0.01279 \pm 0.00048$ & $24.3 \pm 0.5$ & $88.072 \pm 0.071$ & 0 (fixed) \\ 
KOI 1955 & + & $1.64423 \pm 3.6e-05$ & $313.028 \pm 0.0029$ & $35$ & $0.00597 \pm 0.00033$ & $2.92 \pm 0.06$ & $73.96 \pm 0.75$ & 0 (fixed) \\ 
KOI 1955 & + & $26.2335 \pm 0.00059$ & $311.103 \pm 0.0031$ & $226$ & $0.0173 \pm 0.0056$ & $18.51 \pm 0.38$ & $86.942 \pm 0.079$ & 0 (fixed) \\ 
\rule{-2pt}{2.5ex}

KOI 1972.01 &   & $17.7908 \pm 0.00011$ & $82.2599 \pm 0.0019$ & $358$ & $0.01734 \pm 0.00049$ & $49.6 \pm 1.6$ & $89.481 \pm 0.021$ & 0 (fixed) \\ 
KOI 1972 & + & $1.22625 \pm 1.6e-06$ & $313.389 \pm 0.0015$ & $63$ & $0.00707 \pm 0.00038$ & $8.34 \pm 0.28$ & $89.81 \pm 0.18$ & 0 (fixed) \\ 
\rule{-2pt}{2.5ex}

KOI 1986.01 &   & $148.457 \pm 0.0022$ & $84.3886 \pm 0.0046$ & $1437$ & $0.0379 \pm 0.0019$ & $140 \pm 19$ & $89.69 \pm 0.071$ & 0 (fixed) \\ 
KOI 1986 & + & $7.12775 \pm 0.00023$ & $311.347 \pm 0.0065$ & $197$ & $0.0131 \pm 0.0018$ & $18.5 \pm 2.5$ & $88.2 \pm 1$ & 0 (fixed) \\ 
\rule{-2pt}{2.5ex}

KOI 2004.01 &   & $56.1882 \pm 0.0014$ & $79.3777 \pm 0.007$ & $274$ & $0.01924 \pm 0.00082$ & $27.5 \pm 1.6$ & $88.08 \pm 0.13$ & 0 (fixed) \\ 
KOI 2004 & + & $3.18895 \pm 8.2e-05$ & $379.045 \pm 0.0046$ & $68$ & $0.00983 \pm 0.00088$ & $4.06 \pm 0.23$ & $76.67 \pm 0.9$ & 0 (fixed) \\ 
\rule{-2pt}{2.5ex}

KOI 2055.01 &   & $8.67887 \pm 6.6e-05$ & $66.037 \pm 0.0026$ & $431$ & $0.01834 \pm 0.00051$ & $19.35 \pm 0.61$ & $89.774 \pm 0.04$ & 0 (fixed) \\ 
KOI 2055 & + & $4.02586 \pm 5.1e-06$ & $269.717 \pm 0.0014$ & $204$ & $0.01263 \pm 0.00055$ & $11.6 \pm 0.37$ & $89.8 \pm 0.12$ & 0 (fixed) \\ 
KOI 2055 & + & $2.50463 \pm 3.6e-06$ & $267.007 \pm 0.0012$ & $143$ & $0.01057 \pm 0.00059$ & $8.45 \pm 0.27$ & $89.87 \pm 0.1$ & 0 (fixed) \\ 
\rule{-2pt}{2.5ex}

KOI 2159.01 &   & $7.59685 \pm 0.00012$ & $64.9473 \pm 0.0048$ & $104$ & $0.0093 \pm 0.00048$ & $15.9 \pm 1.9$ & $88.4 \pm 1.1$ & 0 (fixed) \\ 
KOI 2159 & + & $2.39266 \pm 3.2e-05$ & $313.226 \pm 0.0019$ & $77$ & $0.00929 \pm 0.00067$ & $7.35 \pm 0.88$ & $83.4 \pm 1.2$ & 0 (fixed) \\ 
\rule{-2pt}{2.5ex}

KOI 2193.01 &   & $2.36172 \pm 1.6e-05$ & $66.2383 \pm 0.0018$ & $506$ & $0.0264 \pm 0.0011$ & $6.65 \pm 0.22$ & $82.27 \pm 0.36$ & 0 (fixed) \\ 
KOI 2193 & + & $4.1629 \pm 2.6e-05$ & $314.452 \pm 0.00087$ & $473$ & $0.125 \pm 0.037$ & $9.7 \pm 0.33$ & $83.56 \pm 0.23$ & 0 (fixed) \\ 
\rule{-2pt}{2.5ex}

KOI 2195.01 &   & $20.0541 \pm 0.00074$ & $69.5792 \pm 0.0091$ & $236$ & $0.0172 \pm 0.0013$ & $11.5 \pm 2.9$ & $85.5 \pm 1.4$ & 0 (fixed) \\ 
KOI 2195 & + & $30.0921 \pm 0.0014$ & $322.777 \pm 0.0071$ & $191$ & $0.0157 \pm 0.0013$ & $15.1 \pm 3.8$ & $86.5 \pm 1$ & 0 (fixed) \\ 
KOI 2195 & + & $6.85034 \pm 0.00025$ & $314.253 \pm 0.0053$ & $101$ & $0.0109 \pm 0.0011$ & $5.6 \pm 1.4$ & $81 \pm 2.9$ & 0 (fixed) \\ 
\rule{-2pt}{2.5ex}

KOI 2243.01 &   & $5.18553 \pm 0.00016$ & $65.3294 \pm 0.0086$ & $184$ & $0.01242 \pm 0.00063$ & $9.2 \pm 1.1$ & $87.3 \pm 1.7$ & 0 (fixed) \\ 
KOI 2243 & + & $8.45812 \pm 0.00023$ & $306.887 \pm 0.0034$ & $208$ & $0.015 \pm 0.001$ & $12.8 \pm 1.5$ & $86.31 \pm 0.68$ & 0 (fixed) \\ 
\rule{-2pt}{2.5ex}

KOI 2485.01 &   & $9.9912 \pm 0.00022$ & $67.3913 \pm 0.0069$ & $384$ & $0.0235 \pm 0.0016$ & $11.4 \pm 1.7$ & $85.35 \pm 0.69$ & 0 (fixed) \\ 
KOI 2485 & + & $3.60087 \pm 6.2e-05$ & $376.705 \pm 0.0026$ & $244$ & $0.0186 \pm 0.0014$ & $5.79 \pm 0.85$ & $80.8 \pm 1.4$ & 0 (fixed) \\ 
KOI 2485 & + & $5.72689 \pm 0.00012$ & $376.226 \pm 0.0031$ & $265$ & $0.02 \pm 0.0017$ & $7.9 \pm 1.2$ & $83.17 \pm 0.99$ & 0 (fixed) \\ 
\rule{-2pt}{2.5ex}

KOI 2579.01 &   & $2.72969 \pm 7.3e-05$ & $66.2767 \pm 0.0071$ & $110$ & $0.01042 \pm 0.00085$ & $5.24 \pm 0.92$ & $81.8 \pm 2.7$ & 0 (fixed) \\ 
KOI 2579 & + & $3.59658 \pm 0.00017$ & $313.923 \pm 0.0063$ & $102$ & $0.0105 \pm 0.0012$ & $6.3 \pm 1.1$ & $82.5 \pm 1.9$ & 0 (fixed) \\ 
KOI 2579 & + & $10.3007 \pm 0.0005$ & $304.826 \pm 0.0074$ & $152$ & $0.0132 \pm 0.0015$ & $12.7 \pm 2.2$ & $86.12 \pm 0.87$ & 0 (fixed) \\ 
\rule{-2pt}{2.5ex}

KOI 2597.01 &   & $8.00514 \pm 0.00027$ & $71.711 \pm 0.009$ & $135$ & $0.011 \pm 0.0011$ & $15.1 \pm 2.8$ & $87.7 \pm 1.4$ & 0 (fixed) \\ 
KOI 2597.02 &   & $12.1311 \pm 0.0014$ & $72.879 \pm 0.031$ & $158$ & $0.012 \pm 0.0023$ & $19.9 \pm 3.7$ & $88.2 \pm 1.3$ & 0 (fixed) \\ 
KOI 2597 & + & $5.61367 \pm 0.00018$ & $311.815 \pm 0.0041$ & $127$ & $0.0109 \pm 0.001$ & $11.9 \pm 2.2$ & $86.7 \pm 1.4$ & 0 (fixed) 
\label{NewCandsTable}
\end{longtable}
\end{landscape}

\onecolumn
\begin{longtable}{|l|l|l|l|l|}
\caption{Physical properties of the newly detected planet candidate. Notes: (1) Negative period designate the best-fitting period that reproduces the observed transit duration/shape of a non-periodic mono-transit. (2) The $\frac{d}{R_*}$ is computed from the final $\frac{d_1}{R_*}$ for the first signal, times the Keplerian factor $(\frac{P_i}{P_1})^\frac{2}{3}$ - all directly from the MCMC distributions.} \\
\hline
Host star&	Period$^{(1)}$	& $r$ 		& $a^{(2)}$	& $T_{eq}$  \\
	 &	(d)		& [$R_\oplus$] 	& [AU]		& [K]	    \\
\hline
\endfirsthead
\hline
Host star&	Period$^{(1)}$	& $r$ 		& $a^{(2)}$	& $T_{eq}$  \\
	 &	(d)		& [$R_\oplus$] 	& [AU]		& [K]	    \\
\hline
\endhead
\hline
\multicolumn{5}{r}{\textit{Continued on next page.}} \\
\endfoot
\hline
\endlastfoot
KOI 179 & -603.7 & 4.26 & 1.358 & 216 \\ 
\rule{-2pt}{2.5ex}

KOI 239 & 3.62276 & 0.96 & 0.047 & 1175 \\ 
\rule{-2pt}{2.5ex}

KOI 241 & 3.41059 & 0.62 & 0.039 & 873 \\ 
KOI 241 & 30.9489 & 0.82 & 0.170 & 418 \\ 
\rule{-2pt}{2.5ex}

KOI 246 & 9.6054 & 0.87 & 0.090 & 946 \\ 
KOI 246 & -740 & 2.67 & 1.631 & 222 \\ 
\rule{-2pt}{2.5ex}

KOI 260 & 21.8701 & 1.49 & 0.158 & 767 \\ 
\rule{-2pt}{2.5ex}

KOI 271 & 14.4364 & 1.41 & 0.125 & 866 \\ 
\rule{-2pt}{2.5ex}

KOI 277 & 13.8537 & 1.23 & 0.116 & 981 \\ 
\rule{-2pt}{2.5ex}

KOI 285 & 26.7234 & 1.69 & 0.184 & 746 \\ 
KOI 285 & 49.3616 & 1.22 & 0.277 & 608 \\ 
\rule{-2pt}{2.5ex}

KOI 289 & 1.47762 & 0.66 & 0.025 & 1616 \\ 
\rule{-2pt}{2.5ex}

KOI 295 & 10.1057 & 1.23 & 0.093 & 880 \\ 
\rule{-2pt}{2.5ex}

KOI 316 & 7.30571 & 2.28 & 0.075 & 979 \\ 
\rule{-2pt}{2.5ex}

KOI 321 & 4.62324 & 0.92 & 0.054 & 1107 \\ 
\rule{-2pt}{2.5ex}

KOI 330 & 1.88897 & 0.68 & 0.030 & 1639 \\ 
\rule{-2pt}{2.5ex}

KOI 339 & 35.863 & 1.19 & 0.207 & 542 \\ 
\rule{-2pt}{2.5ex}

KOI 408 & 3.42818 & 1.27 & 0.043 & 1110 \\ 
\rule{-2pt}{2.5ex}

KOI 435 & 3.93268 & 1.15 & 0.047 & 1035 \\ 
KOI 435 & 33.0388 & 2.01 & 0.193 & 509 \\ 
KOI 435 & 62.3049 & 1.74 & 0.294 & 412 \\ 
KOI 435 & 9.92003 & 0.84 & 0.086 & 760 \\ 
\rule{-2pt}{2.5ex}

KOI 505 & 3.25059 & 1.30 & 0.041 & 1035 \\ 
KOI 505 & 87.09 & 2.92 & 0.366 & 346 \\ 
KOI 505 & 8.34808 & 1.45 & 0.077 & 756 \\ 
\rule{-2pt}{2.5ex}

KOI 509 & 39.587 & 1.56 & 0.218 & 462 \\ 
\rule{-2pt}{2.5ex}

KOI 510 & 35.1182 & 2.77 & 0.199 & 523 \\ 
\rule{-2pt}{2.5ex}

KOI 564 & 6.21739 & 1.22 & 0.065 & 906 \\ 
\rule{-2pt}{2.5ex}

KOI 582 & 9.93964 & 1.18 & 0.085 & 671 \\ 
\rule{-2pt}{2.5ex}

KOI 584 & 6.47024 & 0.81 & 0.064 & 805 \\ 
\rule{-2pt}{2.5ex}

KOI 593 & 51.065 & 2.40 & 0.270 & 468 \\ 
\rule{-2pt}{2.5ex}

KOI 597 & 52.8153 & 1.87 & 0.265 & 491 \\ 
\rule{-2pt}{2.5ex}

KOI 623 & 25.2112 & 0.76 & 0.160 & 703 \\ 
\rule{-2pt}{2.5ex}

KOI 624 & 1.31184 & 1.16 & 0.023 & 1469 \\ 
\rule{-2pt}{2.5ex}

KOI 627 & 4.16521 & 1.28 & 0.054 & 1270 \\ 
\rule{-2pt}{2.5ex}

KOI 671 & 7.46647 & 1.15 & 0.075 & 933 \\ 
KOI 671 & 11.1317 & 1.06 & 0.098 & 817 \\ 
\rule{-2pt}{2.5ex}

KOI 710 & 8.58573 & 1.60 & 0.090 & 1135 \\ 
\rule{-2pt}{2.5ex}

KOI 717 & 0.900351 & 0.60 & 0.019 & 1901 \\ 
\rule{-2pt}{2.5ex}

KOI 719 & 4.15978 & 0.85 & 0.045 & 788 \\ 
KOI 719 & 45.9041 & 1.15 & 0.225 & 354 \\ 
KOI 719 & 28.1225 & 0.85 & 0.162 & 416 \\ 
\rule{-2pt}{2.5ex}

KOI 780 & 7.24078 & 4.38 & 0.067 & 674 \\ 
\rule{-2pt}{2.5ex}

KOI 841 & 269.321 & 3.45 & 0.813 & 252 \\ 
KOI 841 & 6.54629 & 2.68 & 0.068 & 873 \\ 
\rule{-2pt}{2.5ex}

KOI 856 & 0.870588 & 3.49 & 0.018 & 1824 \\ 
\rule{-2pt}{2.5ex}

KOI 1060 & 20.4973 & 1.40 & 0.156 & 766 \\ 
\rule{-2pt}{2.5ex}

KOI 1069 & 8.70386 & 3.45 & 0.080 & 855 \\ 
\rule{-2pt}{2.5ex}

KOI 1082 & 1.19659 & 0.97 & 0.021 & 1359 \\ 
KOI 1082 & 4.09657 & 1.12 & 0.048 & 902 \\ 
\rule{-2pt}{2.5ex}

KOI 1108 & 1.47534 & 0.74 & 0.023 & 1287 \\ 
KOI 1108 & 4.15245 & 0.86 & 0.047 & 911 \\ 
\rule{-2pt}{2.5ex}

KOI 1413 & 21.5274 & 1.06 & 0.139 & 581 \\ 
\rule{-2pt}{2.5ex}

KOI 1536 & 79.4942 & 1.70 & 0.379 & 450 \\ 
\rule{-2pt}{2.5ex}

KOI 1574 & 191.549 & 2.13 & 0.634 & 281 \\ 
\rule{-2pt}{2.5ex}

KOI 1593 & 15.3834 & 1.84 & 0.121 & 679 \\ 
\rule{-2pt}{2.5ex}

KOI 1599 & 13.6129 & 2.09 & 0.111 & 737 \\ 
\rule{-2pt}{2.5ex}

KOI 1601 & 62.9133 & 1.37 & 0.299 & 401 \\ 
\rule{-2pt}{2.5ex}

KOI 1650 & 100.82 & 3.71 & 0.423 & 345 \\ 
\rule{-2pt}{2.5ex}

KOI 1681 & 1.99273 & 0.82 & 0.025 & 748 \\ 
KOI 1681 & 3.53103 & 0.83 & 0.037 & 618 \\ 
\rule{-2pt}{2.5ex}

KOI 1830 & 198.705 & 2.71 & 0.565 & 220 \\ 
\rule{-2pt}{2.5ex}

KOI 1831 & 4.38497 & 0.85 & 0.050 & 890 \\ 
\rule{-2pt}{2.5ex}

KOI 1843 & 0.176893 & 0.63 & 0.005 & 1654 \\ 
\rule{-2pt}{2.5ex}

KOI 1870 & -1340 & 6.98 & 2.247 & 140 \\ 
\rule{-2pt}{2.5ex}

KOI 1871 & 32.3755 & 1.71 & 0.172 & 370 \\ 
\rule{-2pt}{2.5ex}

KOI 1875 & 0.538347 & 1.14 & 0.013 & 2157 \\ 
\rule{-2pt}{2.5ex}

KOI 1955 & 1.64423 & 0.68 & 0.028 & 1684 \\ 
KOI 1955 & 26.2335 & 1.98 & 0.180 & 669 \\ 
\rule{-2pt}{2.5ex}

KOI 1972 & 1.22625 & 0.73 & 0.022 & 1671 \\ 
\rule{-2pt}{2.5ex}

KOI 1986 & 7.12775 & 2.34 & 0.071 & 1080 \\ 
\rule{-2pt}{2.5ex}

KOI 2004 & 3.18895 & 0.99 & 0.041 & 1153 \\ 
\rule{-2pt}{2.5ex}

KOI 2055 & 4.02586 & 1.09 & 0.047 & 994 \\ 
KOI 2055 & 2.50463 & 0.91 & 0.035 & 1164 \\ 
\rule{-2pt}{2.5ex}

KOI 2159 & 2.39266 & 0.85 & 0.034 & 1229 \\ 
\rule{-2pt}{2.5ex}

KOI 2193 & 4.1629 & 7.76 & 0.043 & 695 \\ 
\rule{-2pt}{2.5ex}

KOI 2195 & 30.0921 & 1.83 & 0.198 & 644 \\ 
KOI 2195 & 6.85034 & 1.27 & 0.074 & 1055 \\ 
\rule{-2pt}{2.5ex}

KOI 2243 & 8.45812 & 1.54 & 0.082 & 885 \\ 
\rule{-2pt}{2.5ex}

KOI 2485 & 3.60087 & 1.52 & 0.042 & 929 \\ 
KOI 2485 & 5.72689 & 1.63 & 0.057 & 796 \\ 
\rule{-2pt}{2.5ex}

KOI 2579 & 3.59658 & 1.25 & 0.046 & 1277 \\ 
KOI 2579 & 10.3007 & 1.57 & 0.092 & 899 \\ 
\rule{-2pt}{2.5ex}

KOI 2597 & 5.61367 & 1.25 & 0.064 & 1107
\label{PhysProp}
\end{longtable}

\appendix
\clearpage
\twocolumn
\section{Appendix A}
\label{appA}
This appendix contains the figures representing the global fits for ``regular'' systems with new detections, to which we have little/no special comments. The figures are given primarily to show the reality of the signals.

\begin{figure}[tbp]\includegraphics[width=0.5\textwidth]{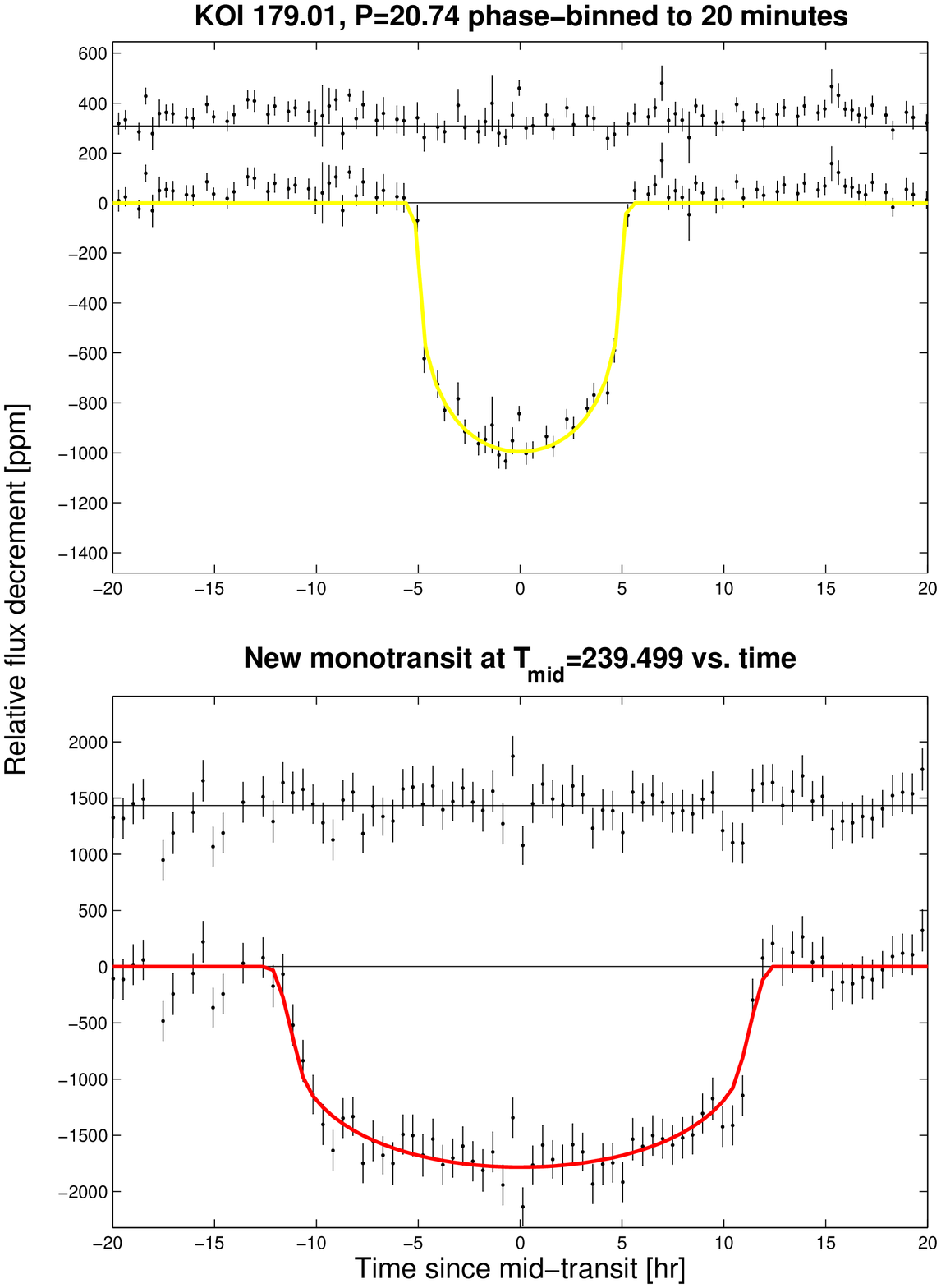}
\caption{Similar to Figure \ref{KOI246fig}.}
\label{KOI179fig}
\end{figure}

\begin{figure}[tbp]\includegraphics[width=0.5\textwidth]{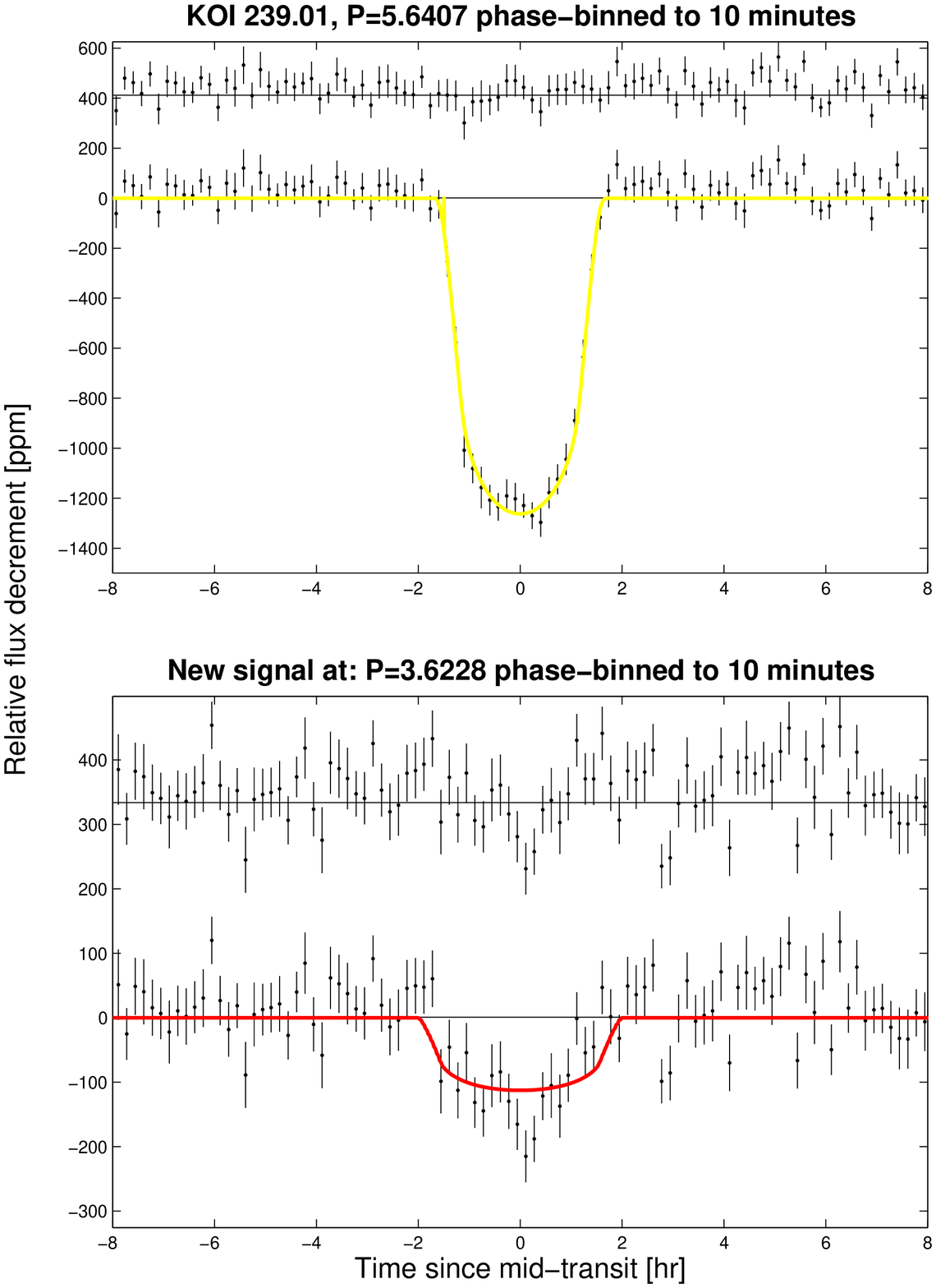}
\caption{Similar to Figure \ref{KOI246fig}.}
\label{KOI239fig}
\end{figure}

\begin{figure}[tbp]\includegraphics[width=0.5\textwidth]{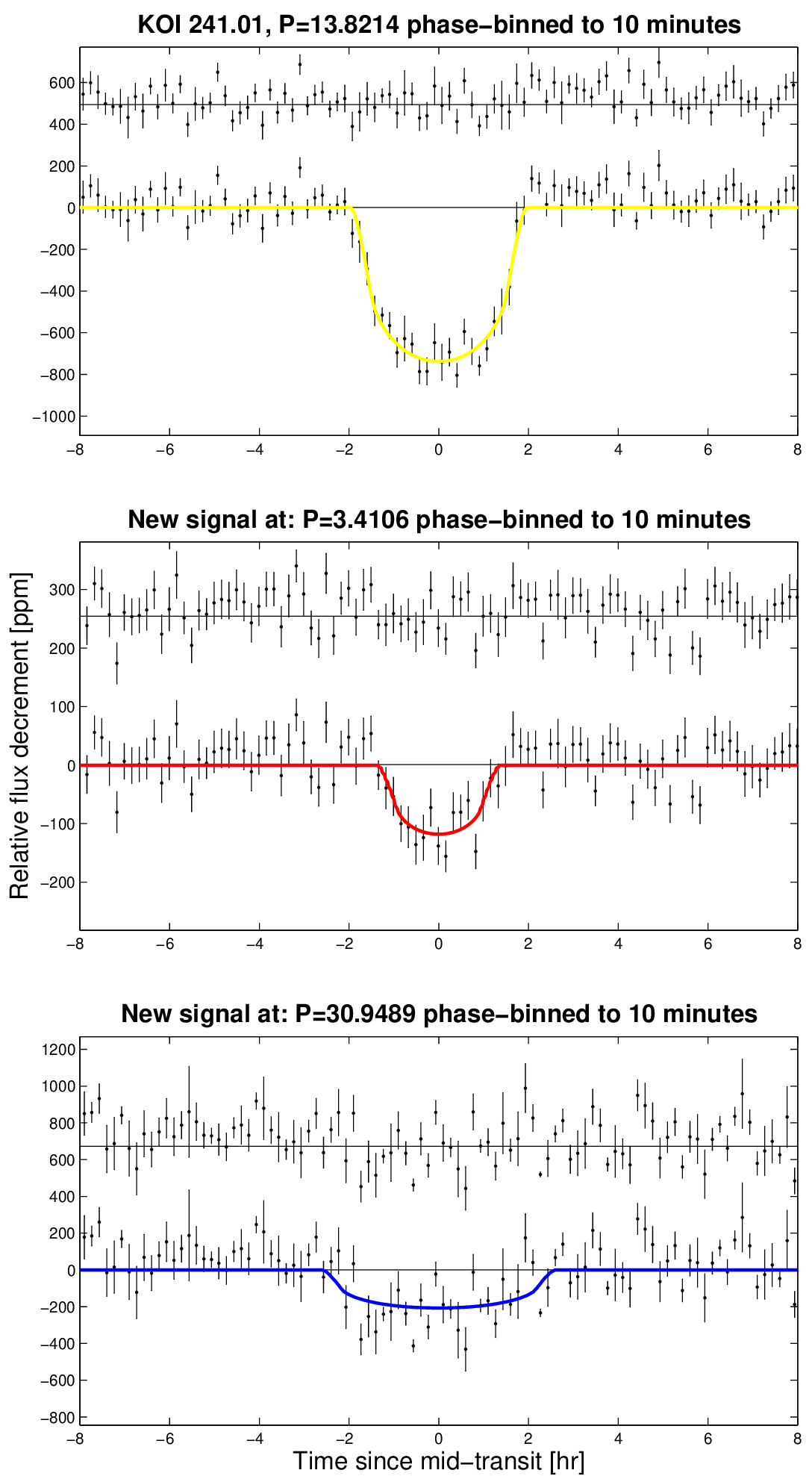}
\caption{Similar to Figure \ref{KOI246fig}.}
\label{KOI241fig}
\end{figure}

\begin{figure}[tbp]\includegraphics[width=0.5\textwidth]{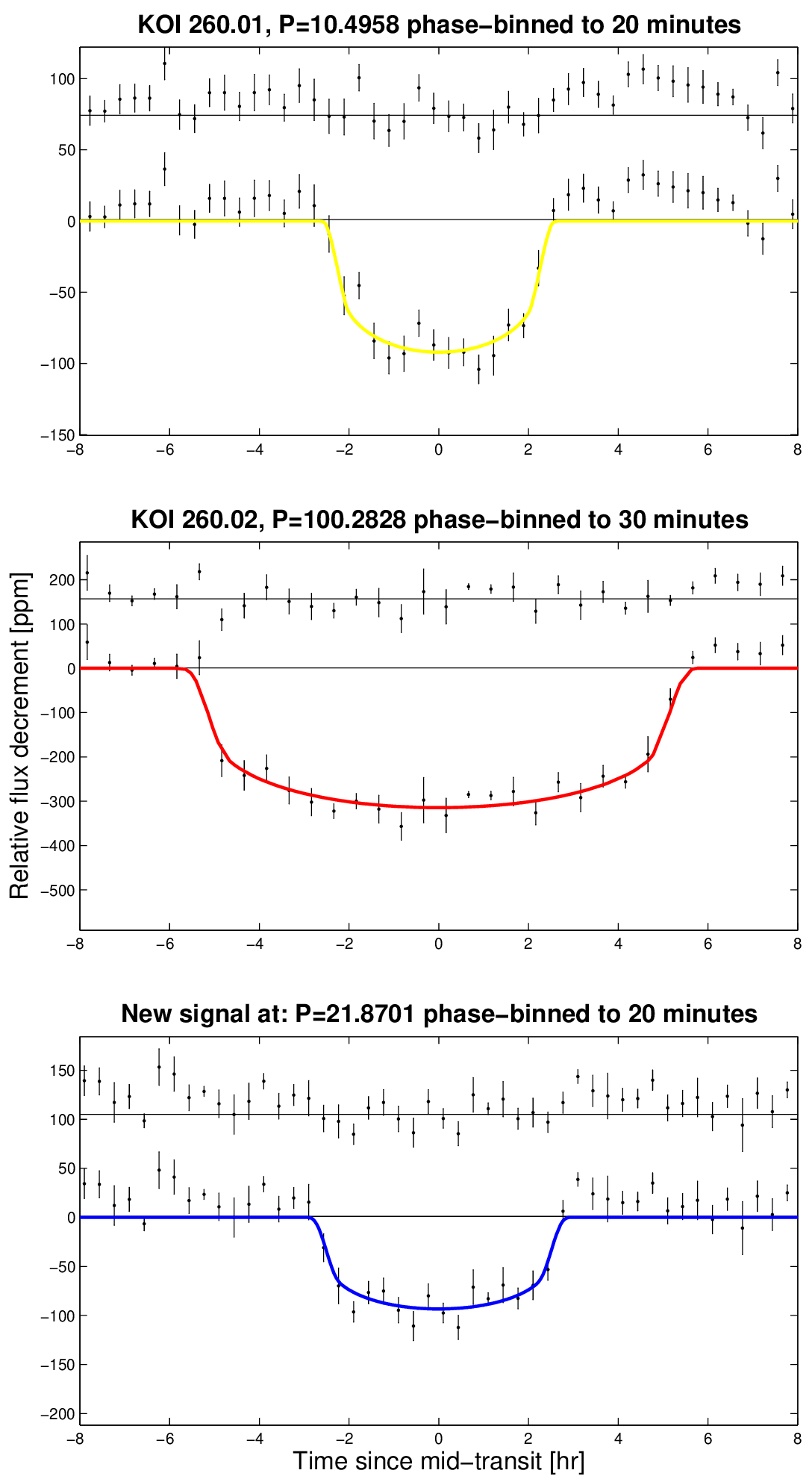}
\caption{Similar to Figure \ref{KOI246fig}.}
\label{KOI260fig}
\end{figure}

\begin{figure}[tbp]\includegraphics[width=0.5\textwidth]{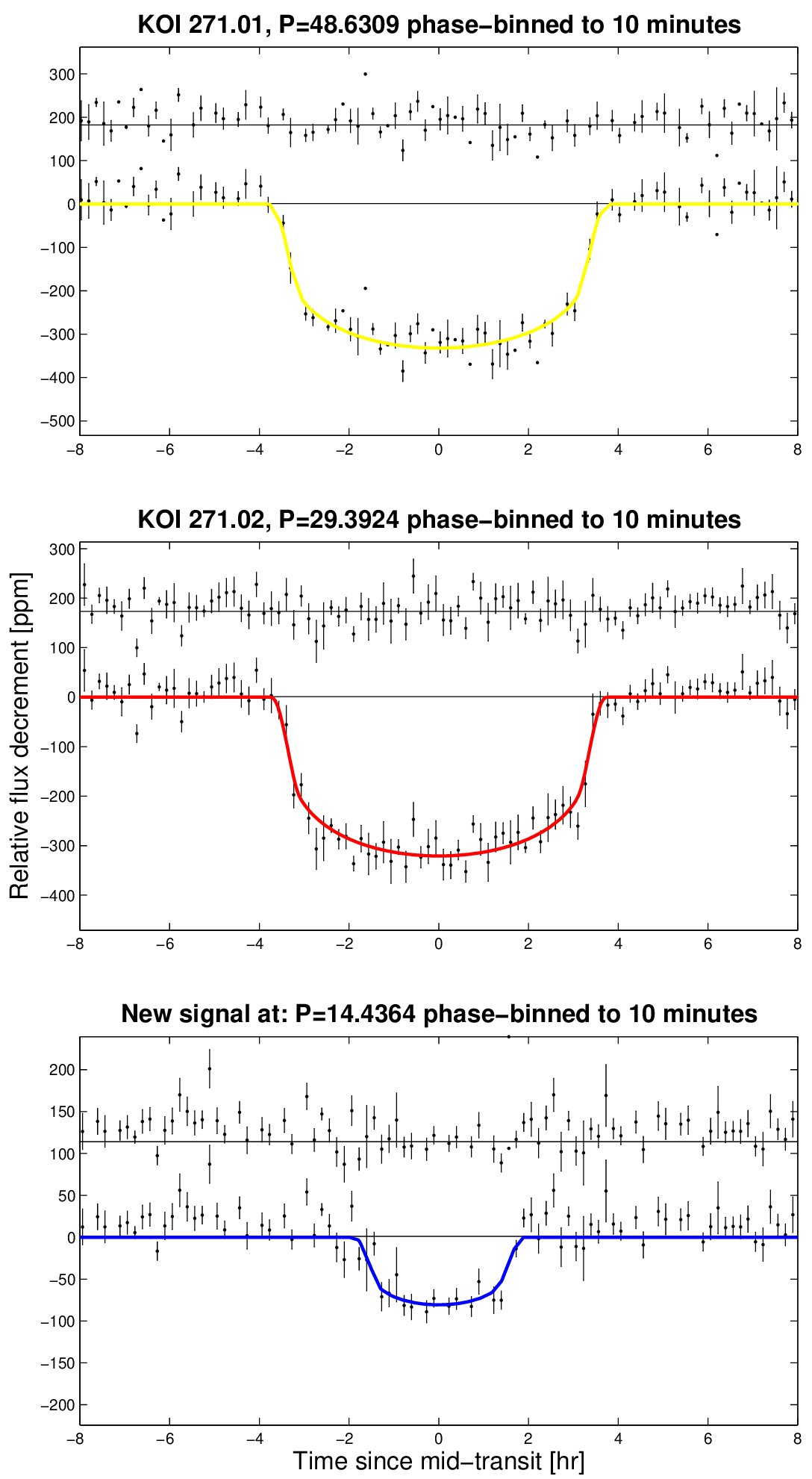}
\caption{Similar to Figure \ref{KOI246fig}.}
\label{KOI271fig}
\end{figure}

\begin{figure}[tbp]\includegraphics[width=0.5\textwidth]{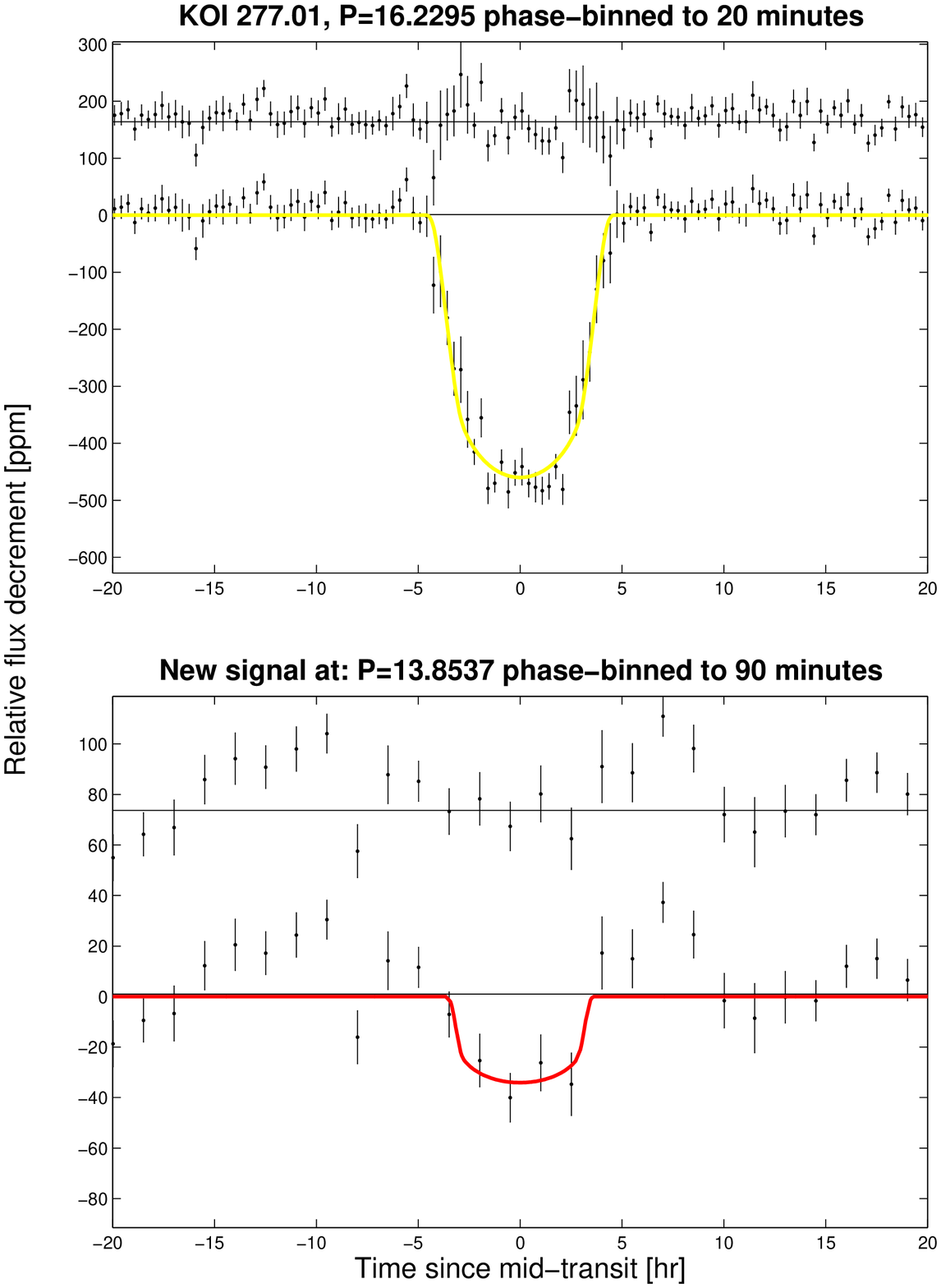}
\caption{Similar to Figure \ref{KOI246fig}.}
\label{KOI277fig}
\end{figure}

\begin{figure}[tbp]\includegraphics[width=0.5\textwidth]{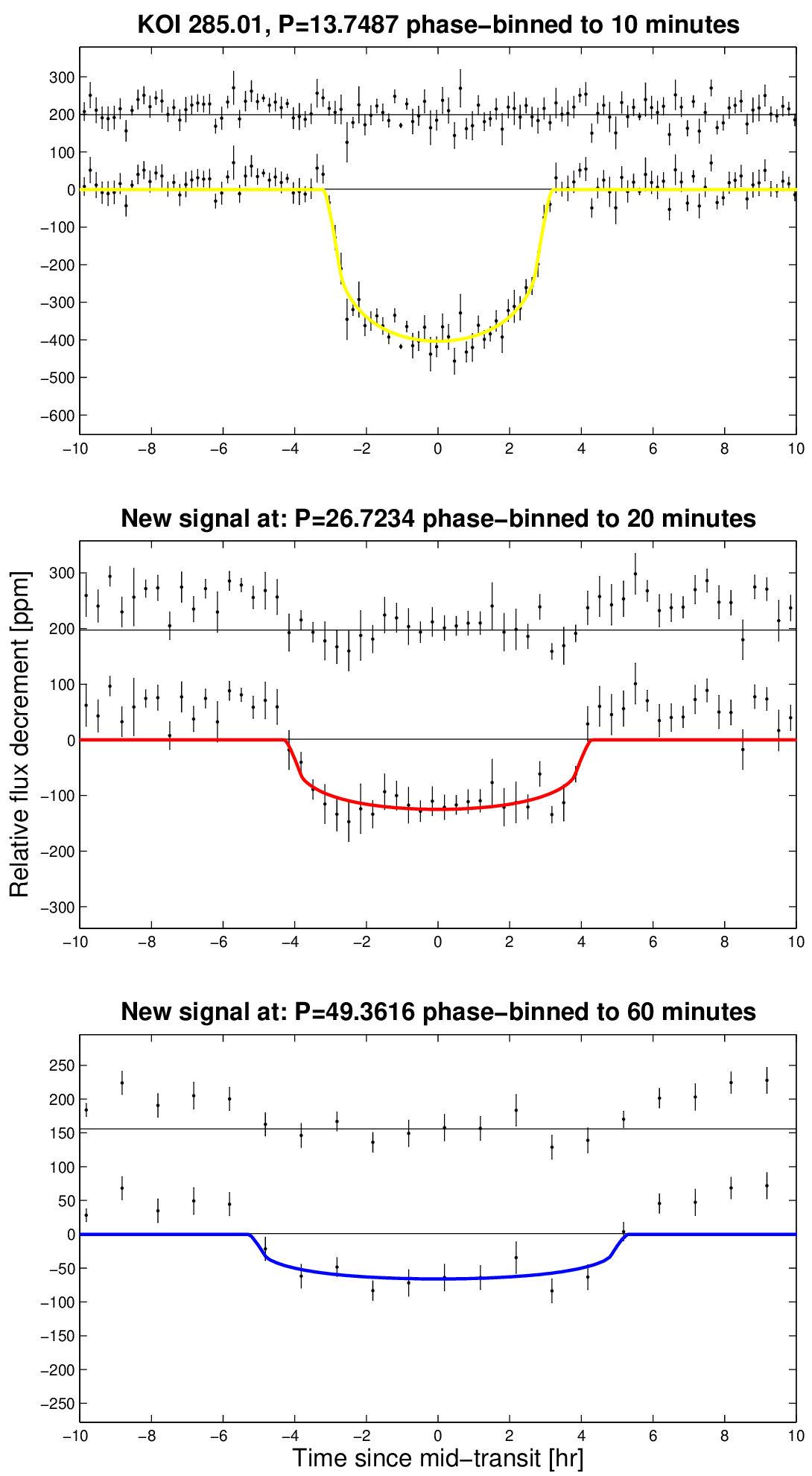}
\caption{Similar to Figure \ref{KOI246fig}.}
\label{KOI285fig}
\end{figure}

\begin{figure}[tbp]\includegraphics[width=0.5\textwidth]{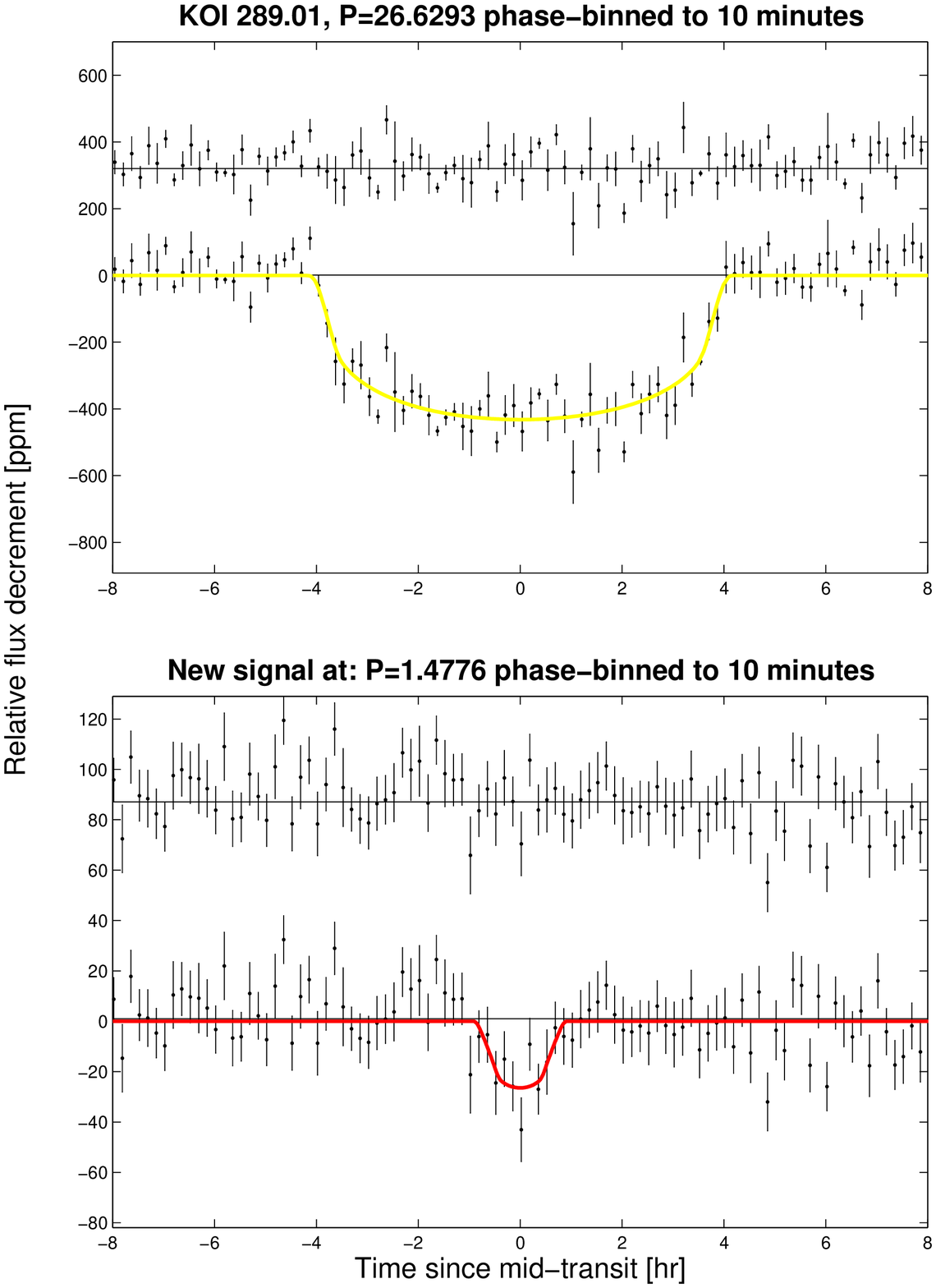}
\caption{Similar to Figure \ref{KOI246fig}.}
\label{KOI289fig}
\end{figure}

\begin{figure}[tbp]\includegraphics[width=0.5\textwidth]{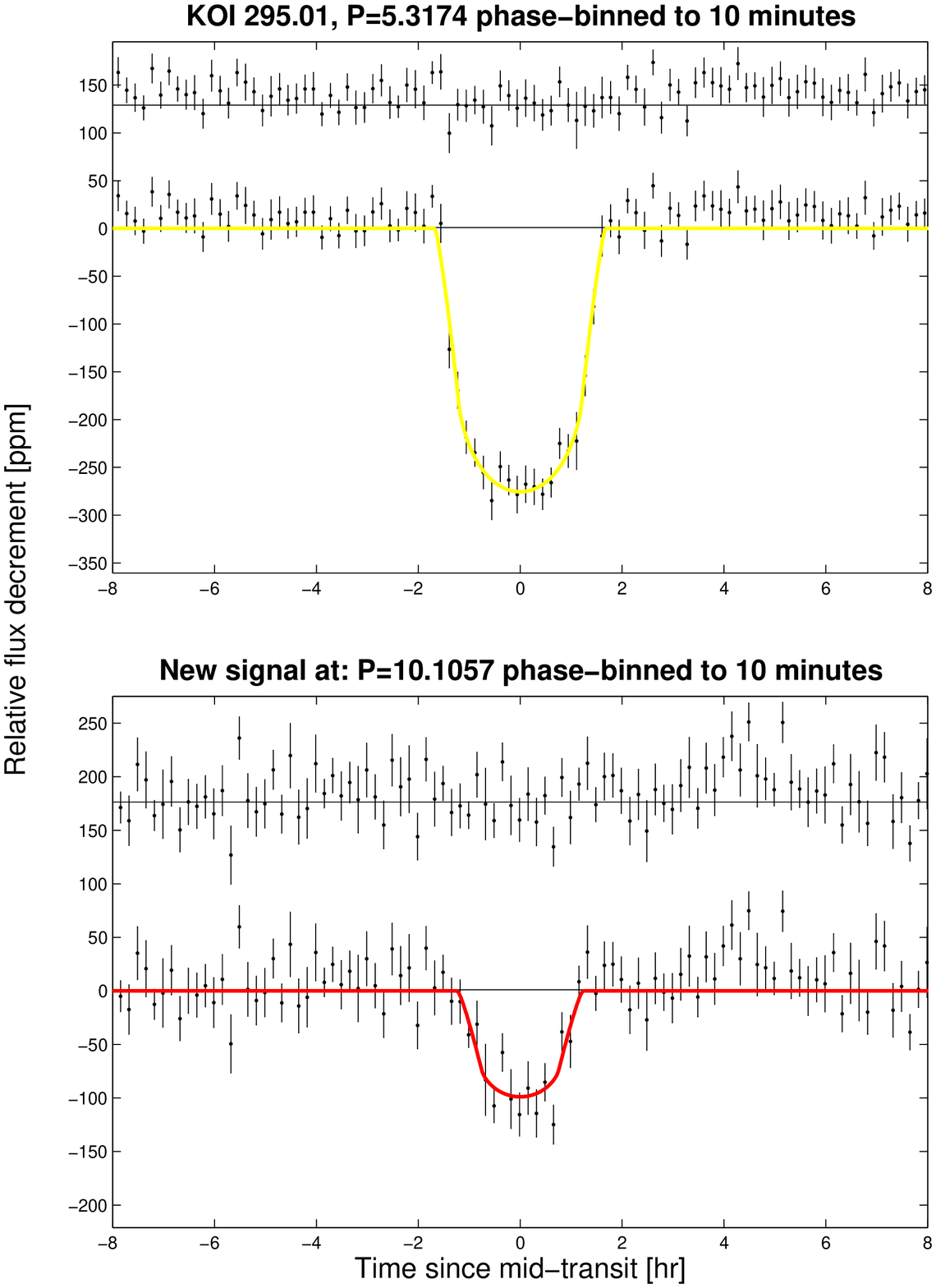}
\caption{Similar to Figure \ref{KOI246fig}.}
\label{KOI295fig}
\end{figure}

\begin{figure}[tbp]\includegraphics[width=0.5\textwidth]{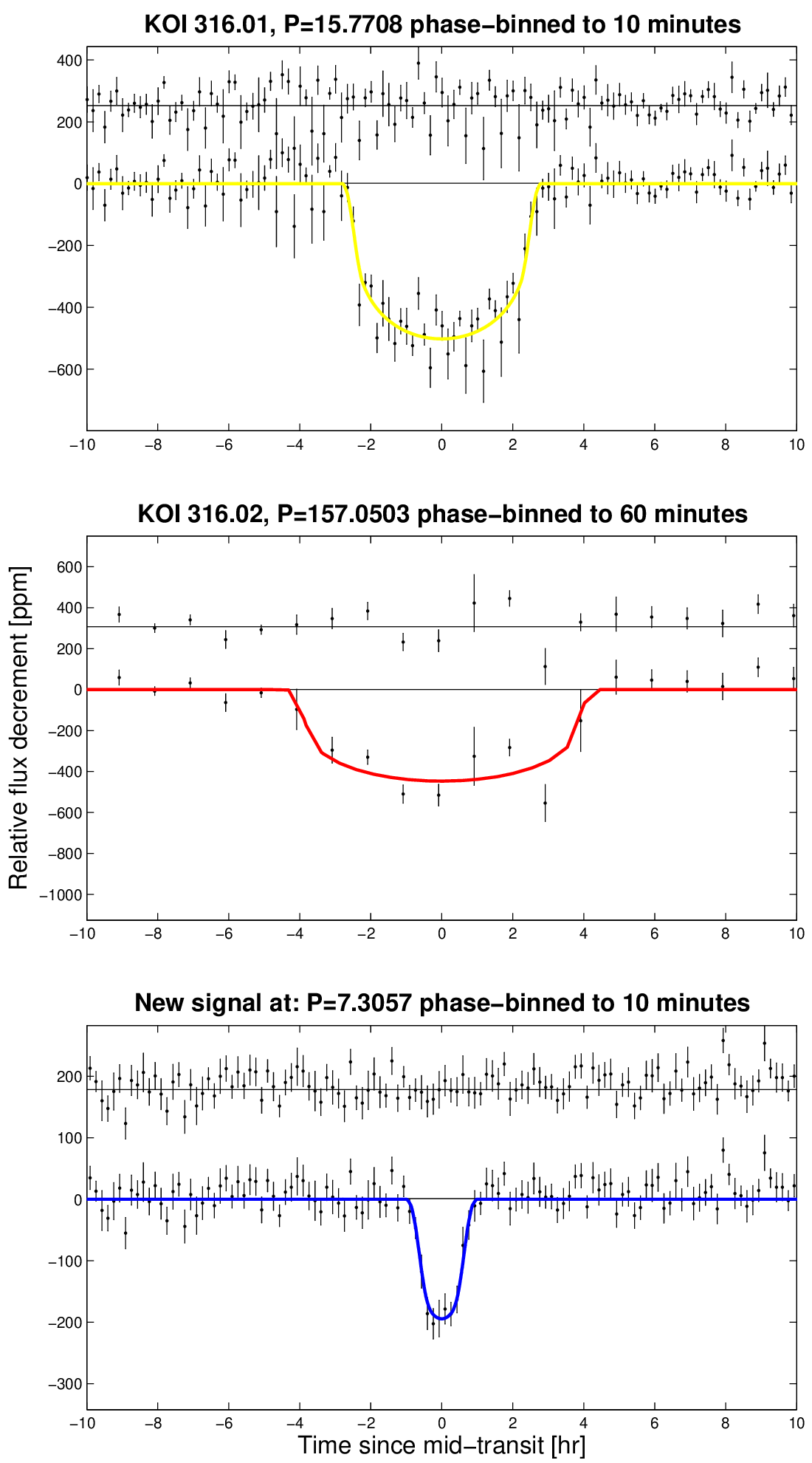}
\caption{Similar to Figure \ref{KOI246fig}.}
\label{KOI316fig}
\end{figure}

\begin{figure}[tbp]\includegraphics[width=0.5\textwidth]{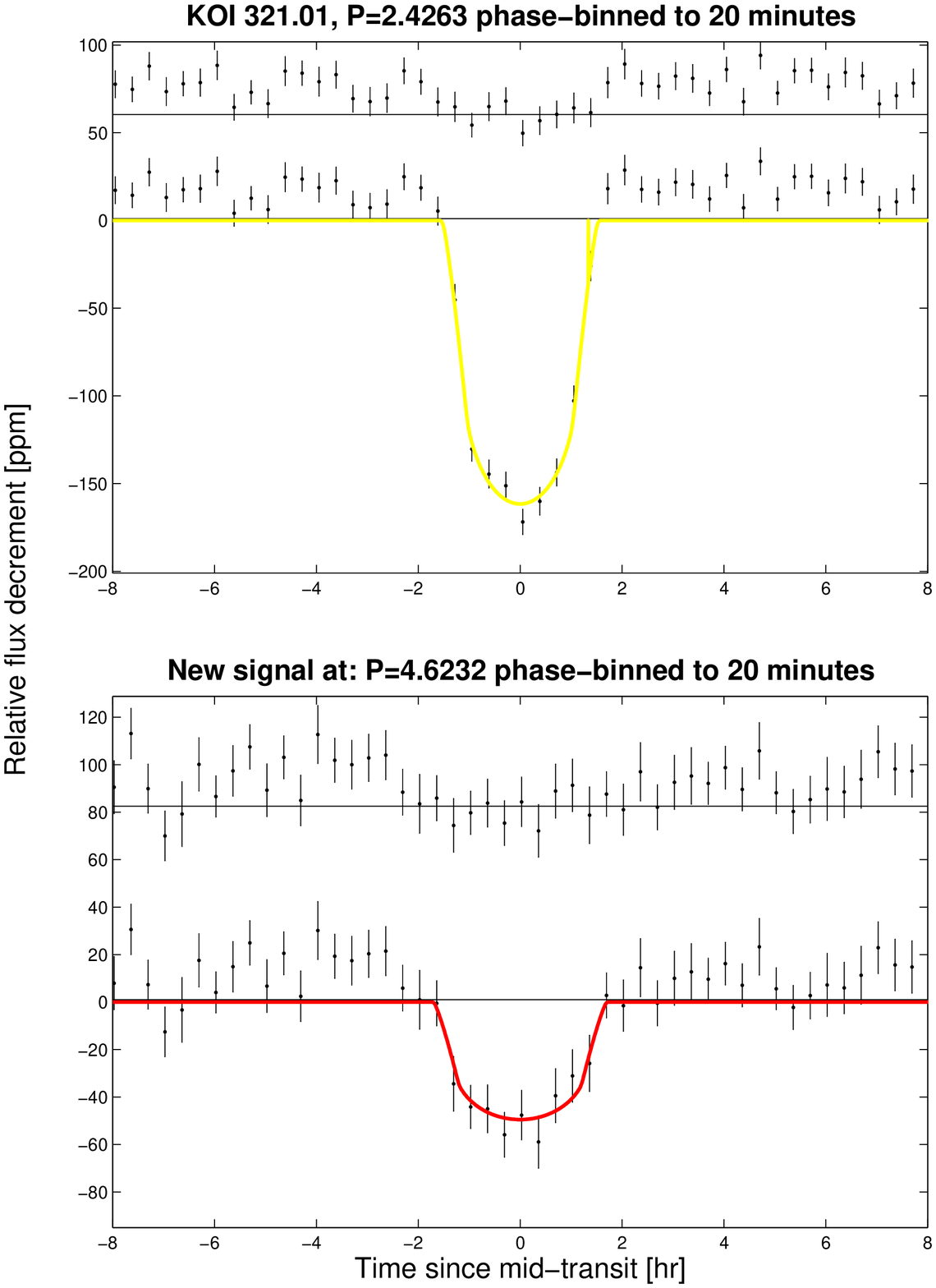}
\caption{Similar to Figure \ref{KOI246fig}.}
\label{KOI321fig}
\end{figure}

\begin{figure}[tbp]\includegraphics[width=0.5\textwidth]{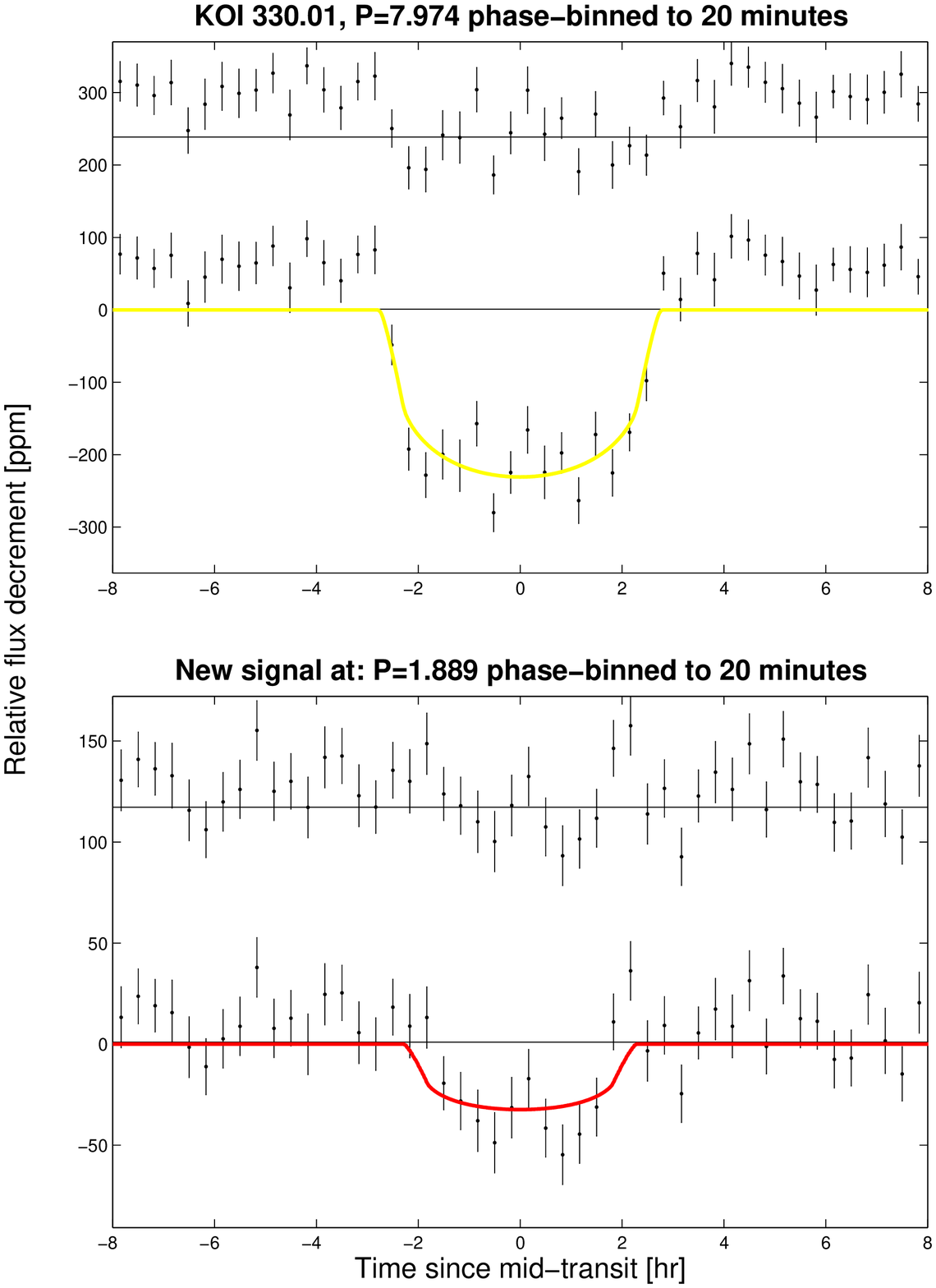}
\caption{Similar to Figure \ref{KOI246fig}.}
\label{KOI330fig}
\end{figure}

\begin{figure}[tbp]\includegraphics[width=0.5\textwidth]{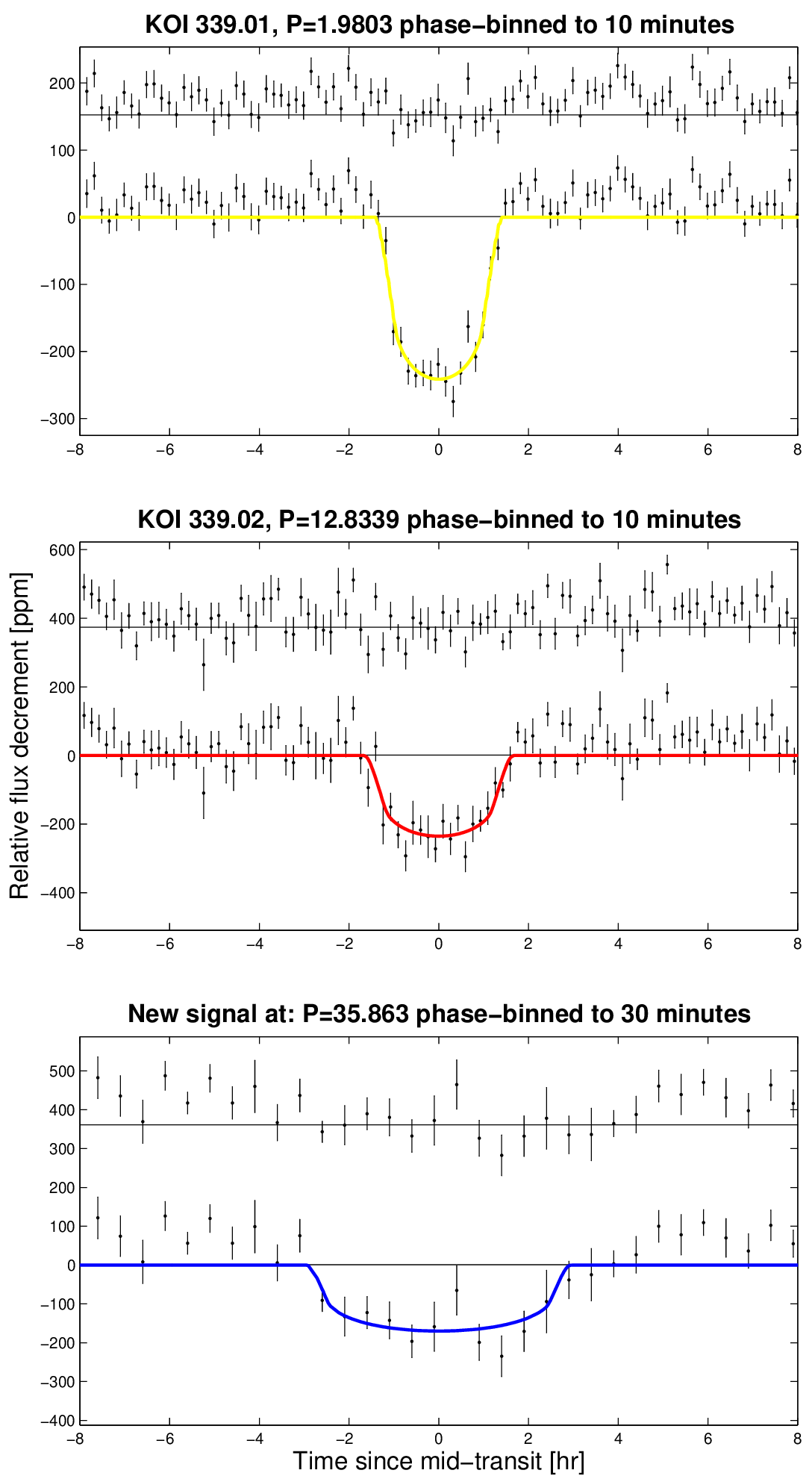}
\caption{Similar to Figure \ref{KOI246fig}.}
\label{KOI339fig}
\end{figure}

\begin{figure}[tbp]\includegraphics[width=0.5\textwidth]{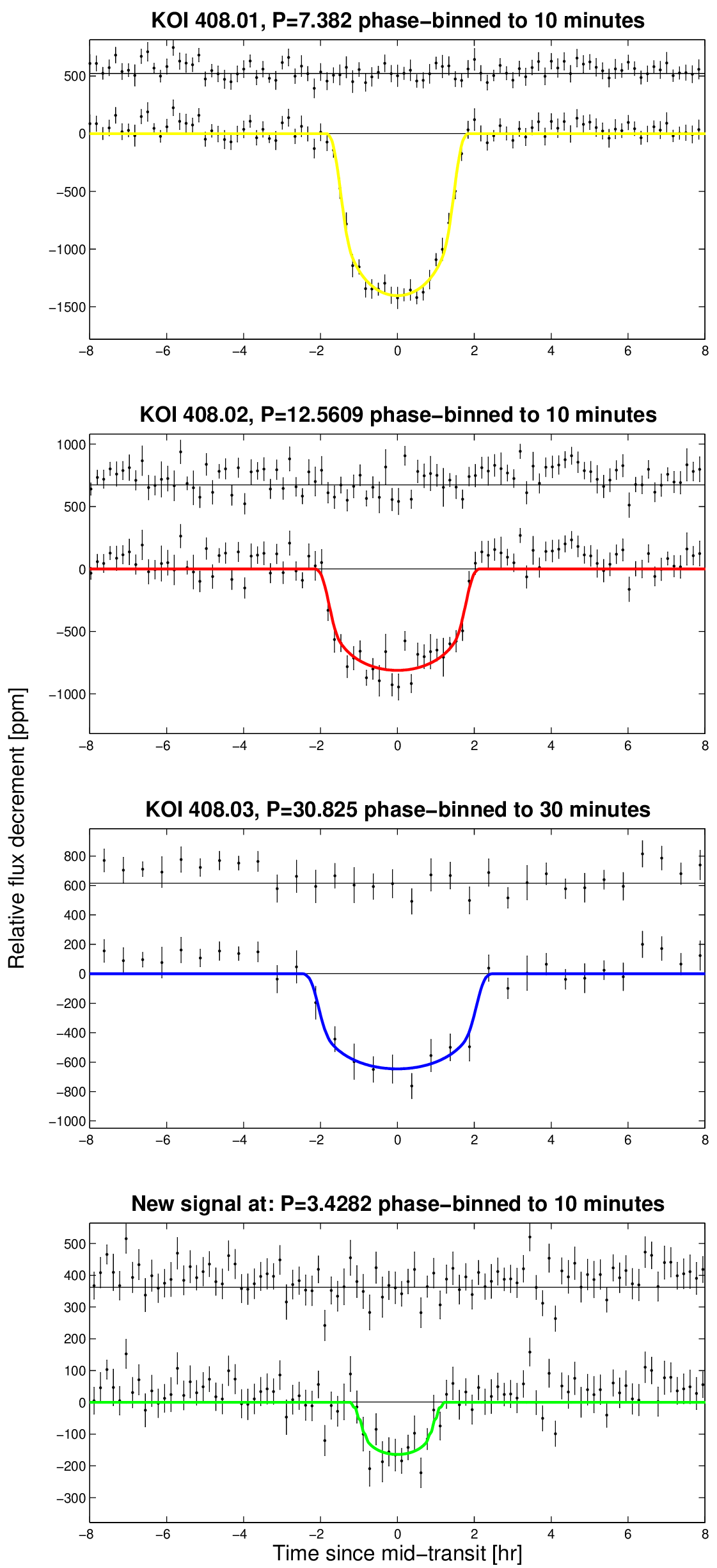}
\caption{Similar to Figure \ref{KOI246fig}.}
\label{KOI408fig}
\end{figure}

\begin{figure}[tbp]\includegraphics[width=0.5\textwidth]{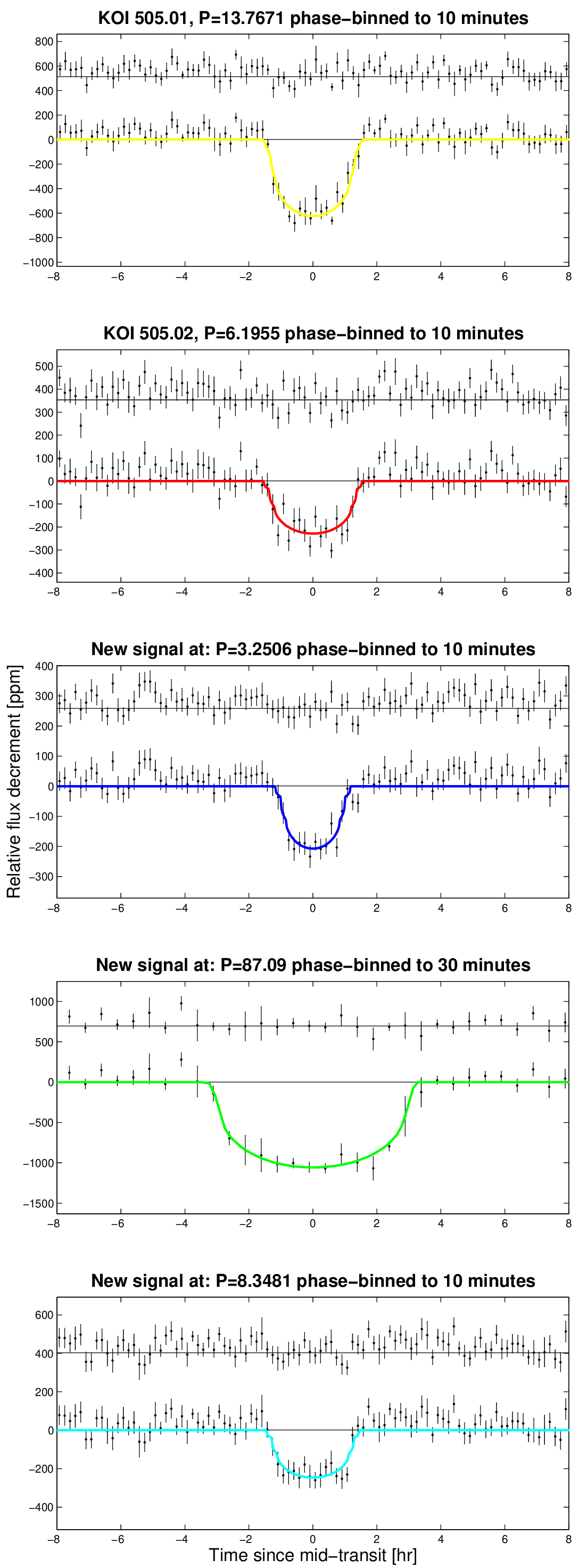}
\caption{Similar to Figure \ref{KOI246fig}.}
\label{KOI505fig}
\end{figure}

\begin{figure}[tbp]\includegraphics[width=0.5\textwidth]{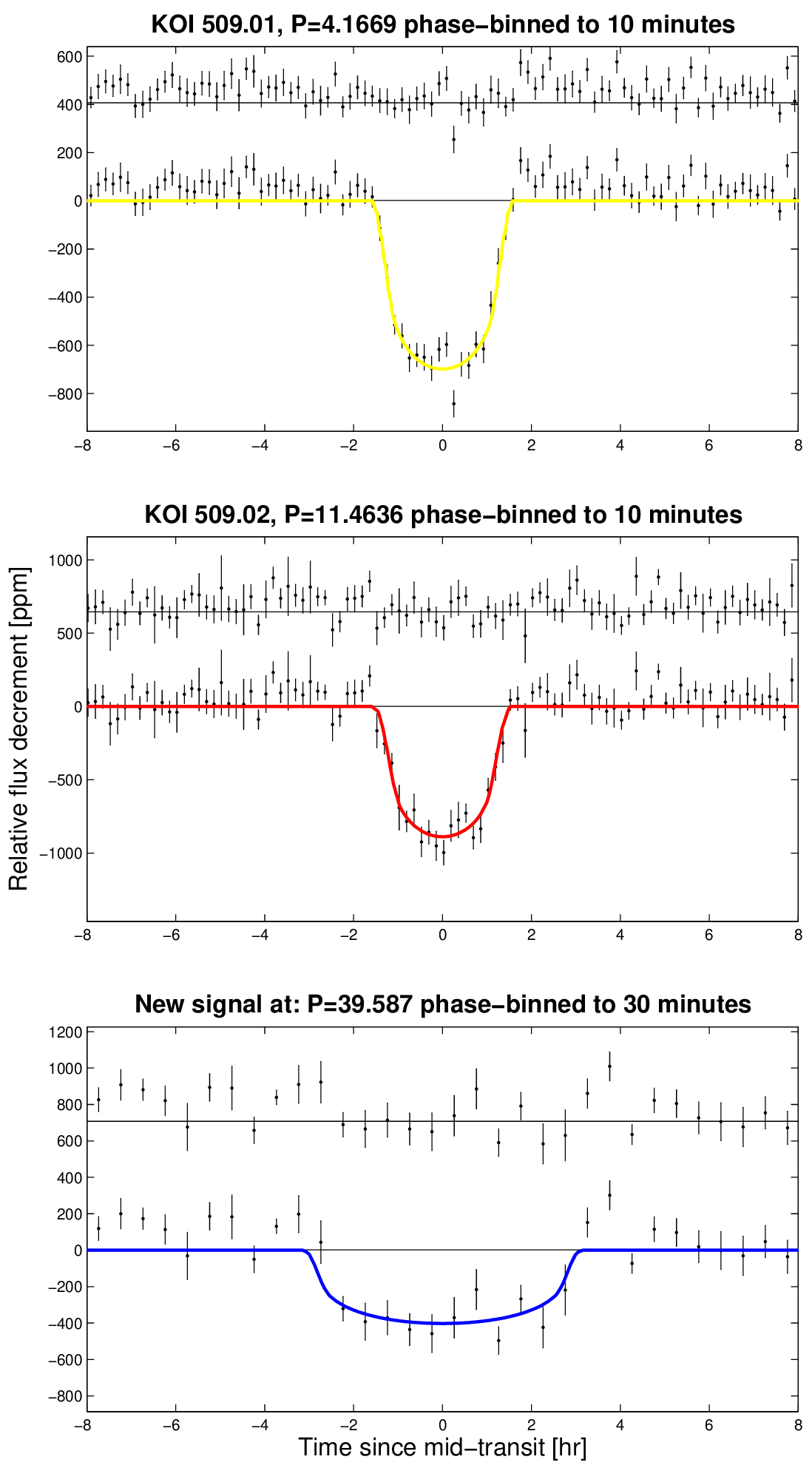}
\caption{Similar to Figure \ref{KOI246fig}.}
\label{KOI509fig}
\end{figure}

\begin{figure}[tbp]\includegraphics[width=0.5\textwidth]{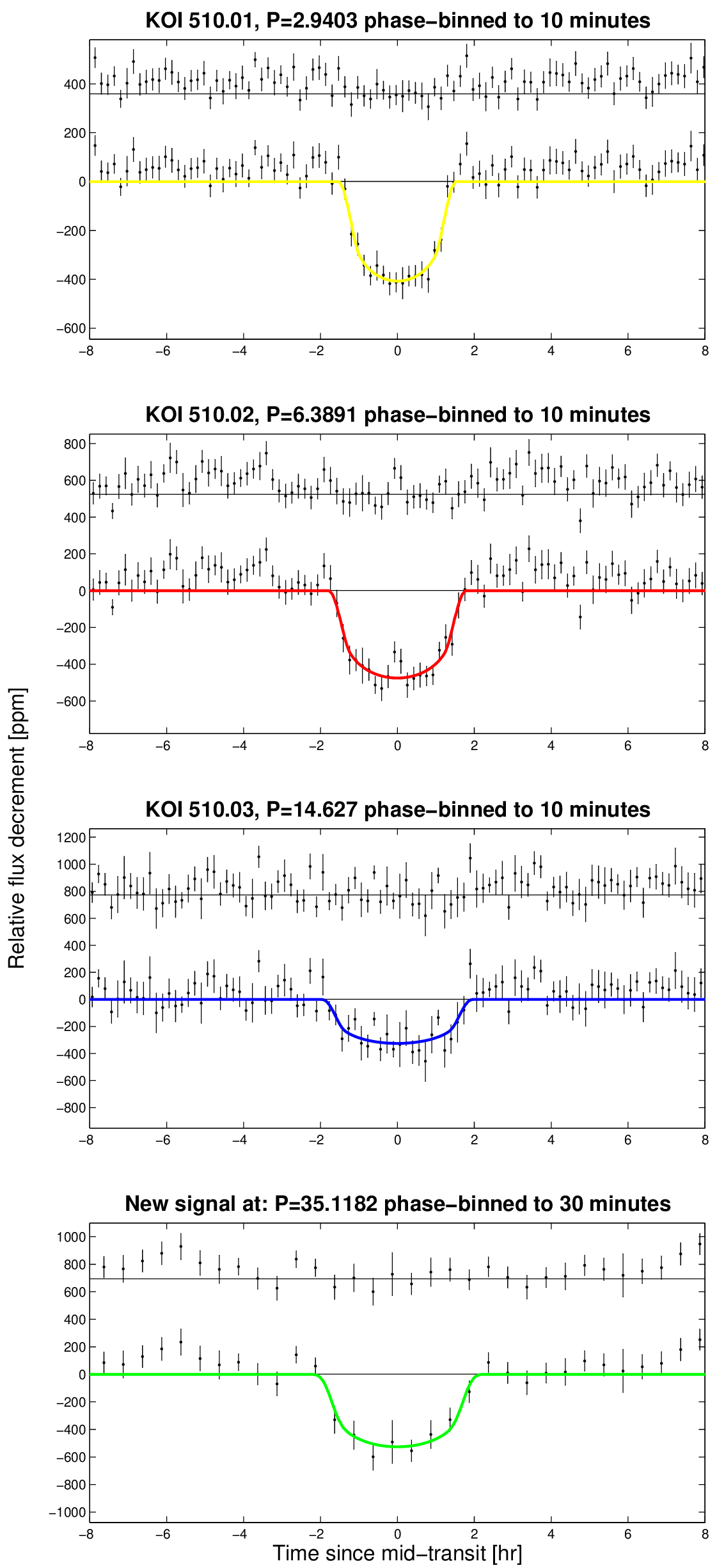}
\caption{Similar to Figure \ref{KOI246fig}.}
\label{KOI510fig}
\end{figure}

\begin{figure}[tbp]\includegraphics[width=0.5\textwidth]{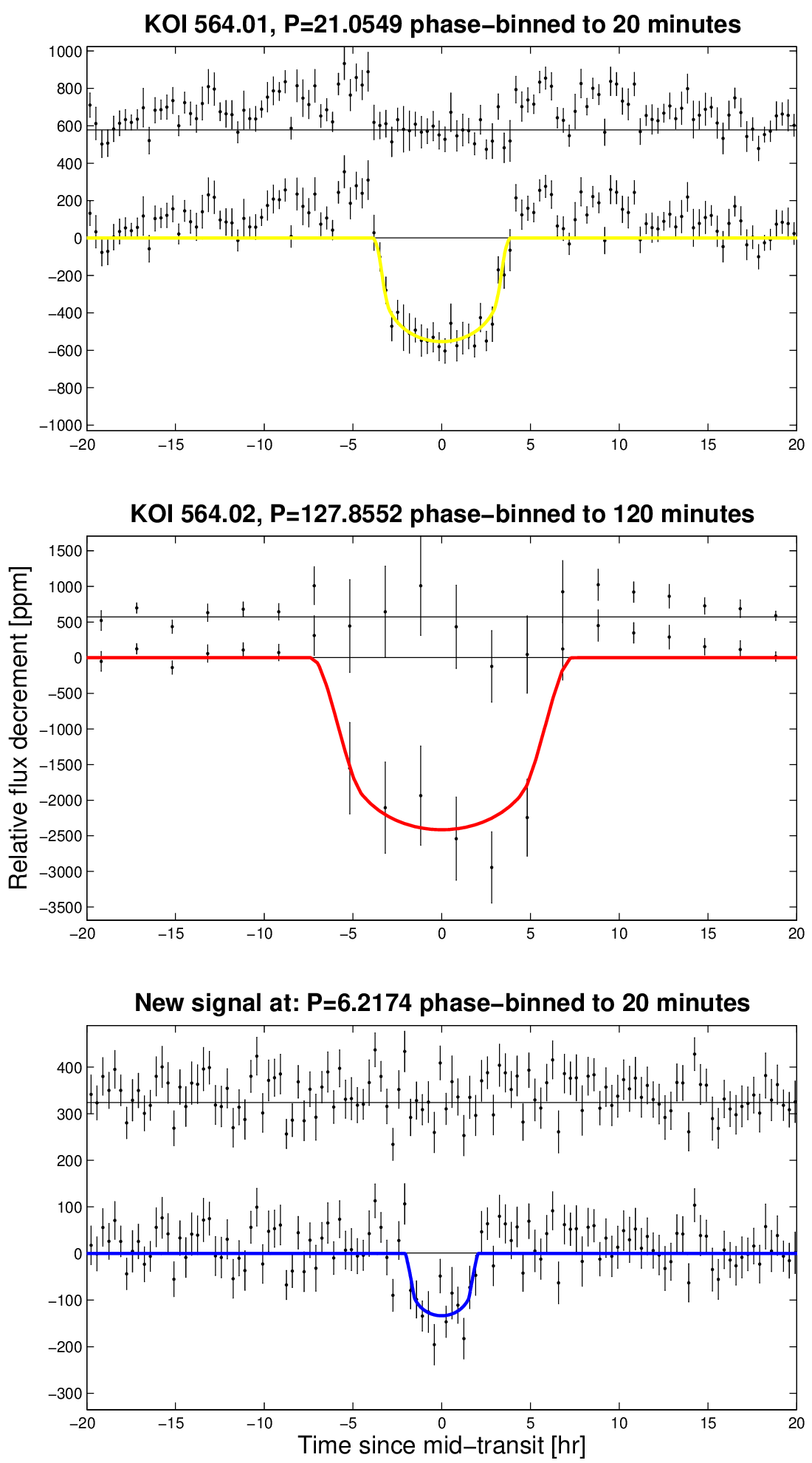}
\caption{Similar to Figure \ref{KOI246fig}. The larger-than-usual errorbars on KOI 564.02 were caused by a particularly unfortunate timing of one of the three events shortly after a discontinuity in the data that affected the filtering.}
\label{KOI564fig}
\end{figure}

\clearpage

\begin{figure}[tbp]\includegraphics[width=0.5\textwidth]{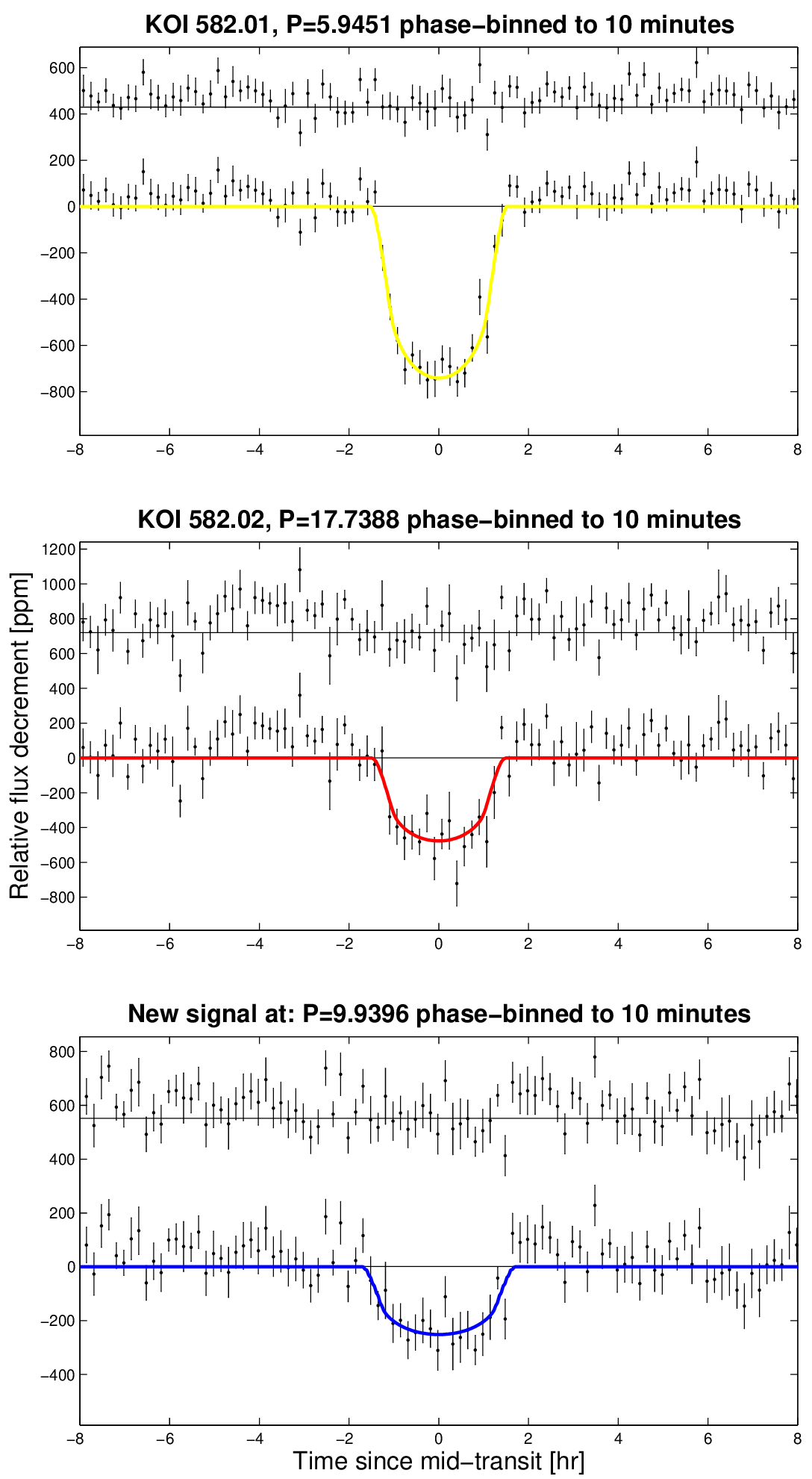}
\caption{Similar to Figure \ref{KOI246fig}.}
\label{KOI582fig}
\end{figure}

\begin{figure}[tbp]\includegraphics[width=0.5\textwidth]{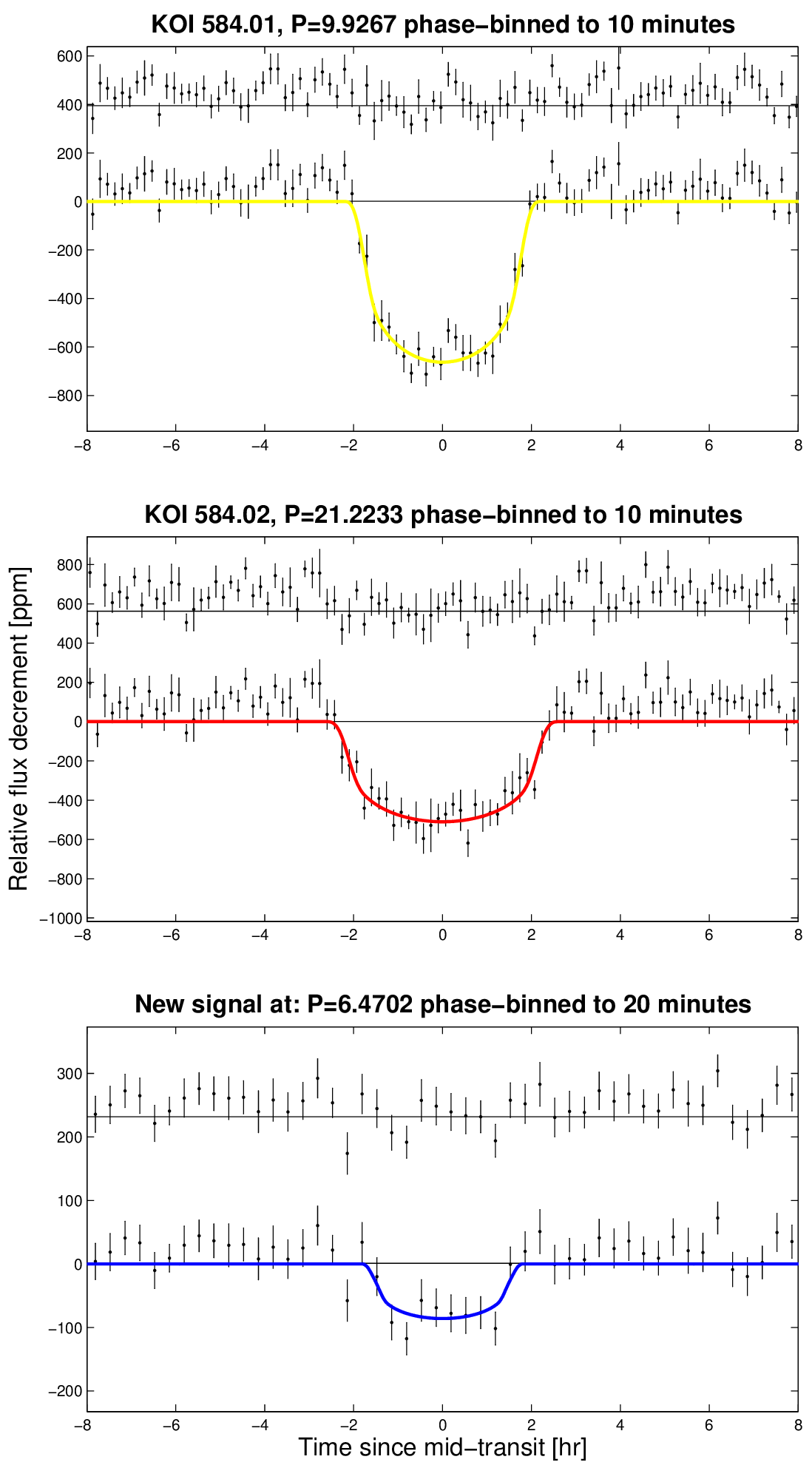}
\caption{Similar to Figure \ref{KOI246fig}. Note an even shallower, and thus much less robust, transit-like signal was detected at phase 0.59 (see text) with respect to the new $6.47$d signal.}
\label{KOI584fig}
\end{figure}

\begin{figure}[tbp]\includegraphics[width=0.5\textwidth]{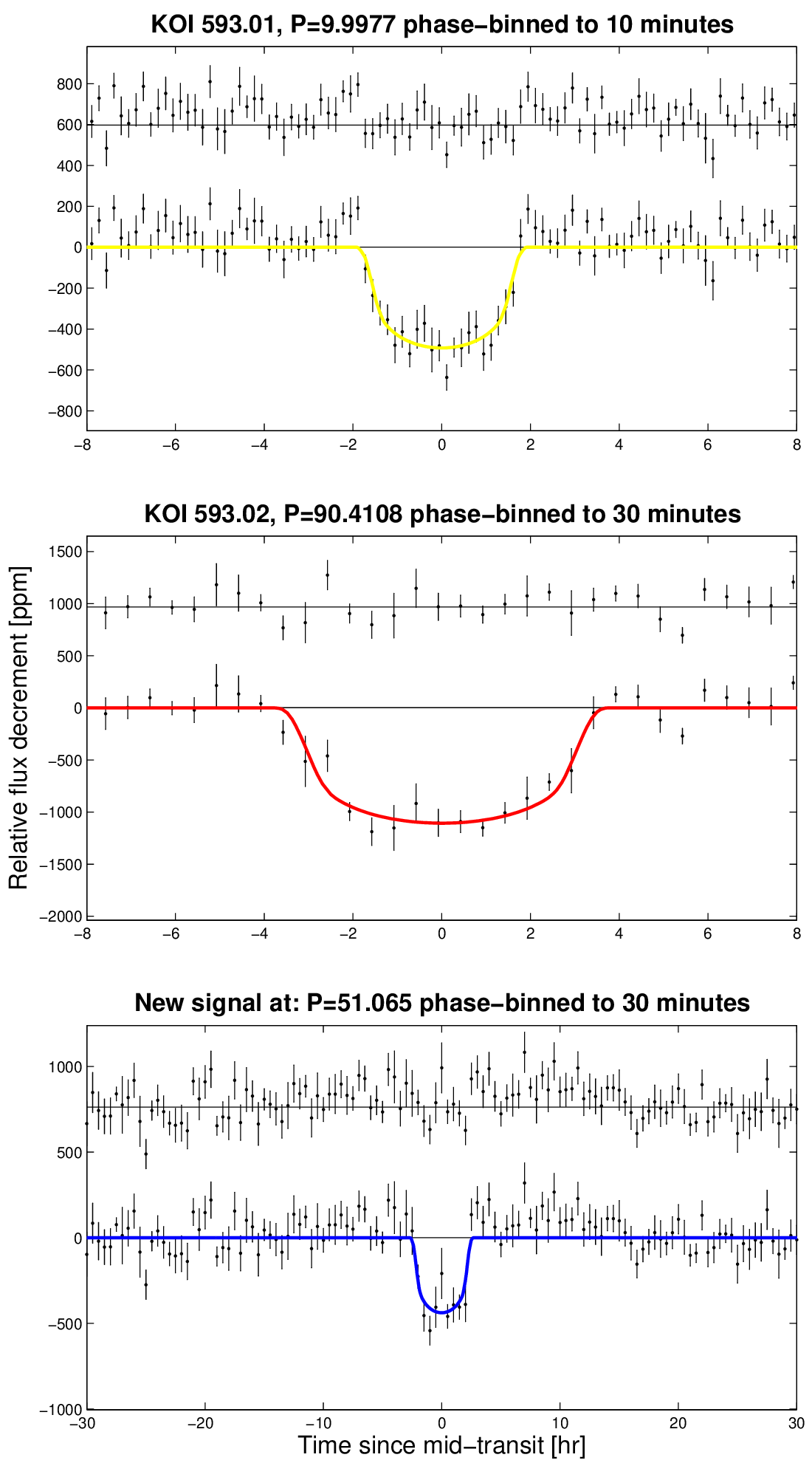}
\caption{Similar to Figure \ref{KOI246fig}.}
\label{KOI593fig}
\end{figure}

\begin{figure}[tbp]\includegraphics[width=0.5\textwidth]{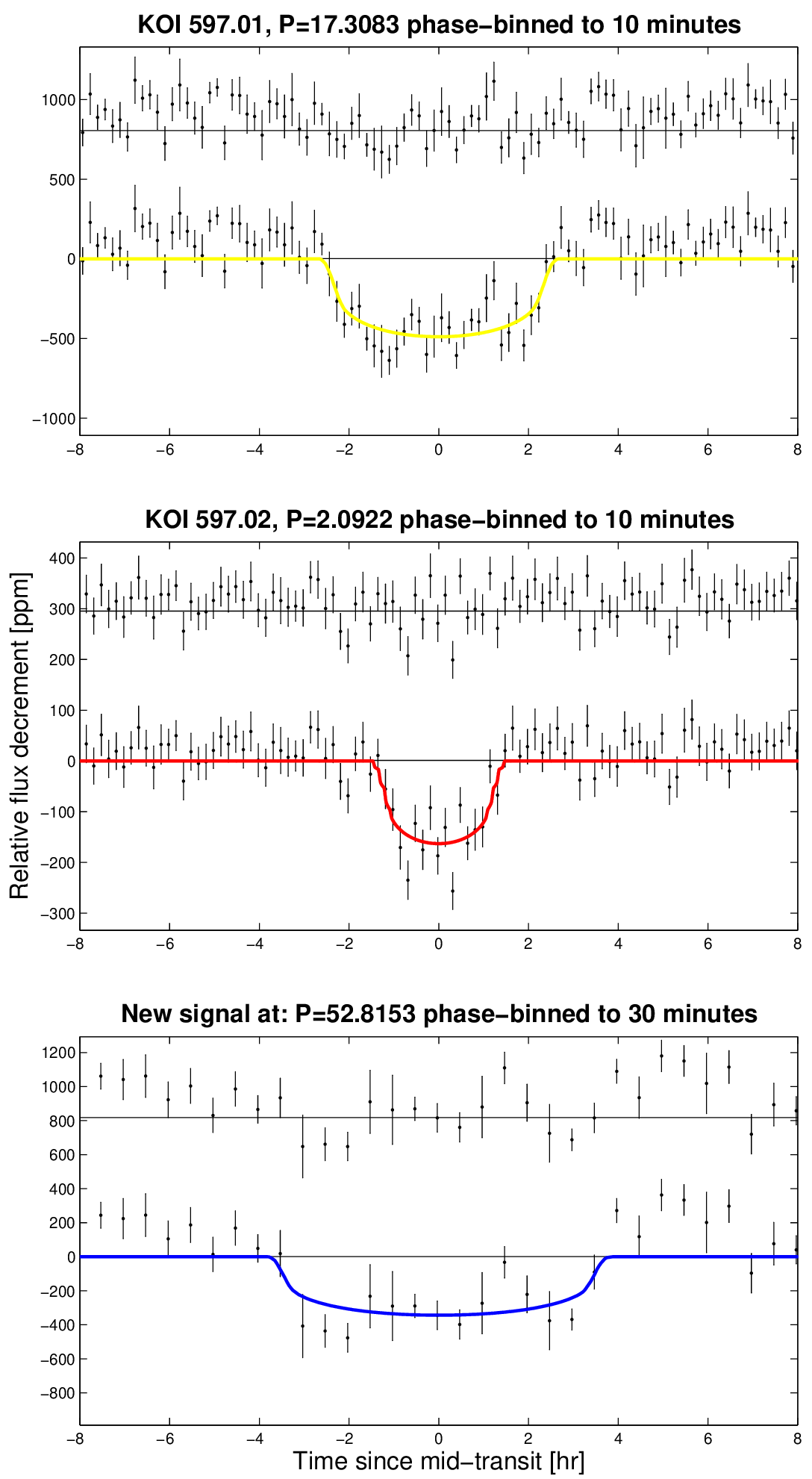}
\caption{Similar to Figure \ref{KOI246fig}.}
\label{KOI597fig}
\end{figure}

\begin{figure}[tbp]\includegraphics[width=0.5\textwidth]{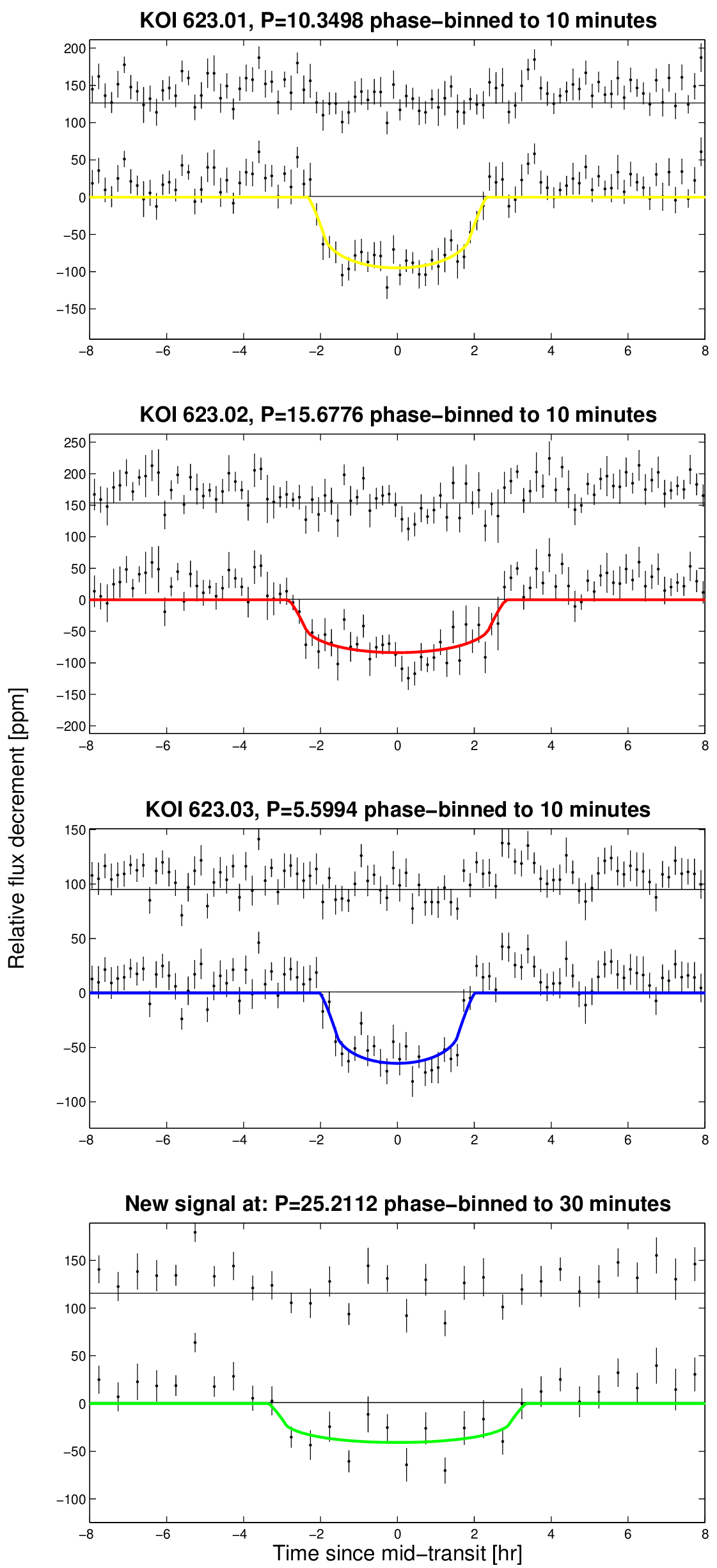}
\caption{Similar to Figure \ref{KOI246fig}.}
\label{KOI623fig}
\end{figure}

\begin{figure}[tbp]\includegraphics[width=0.5\textwidth]{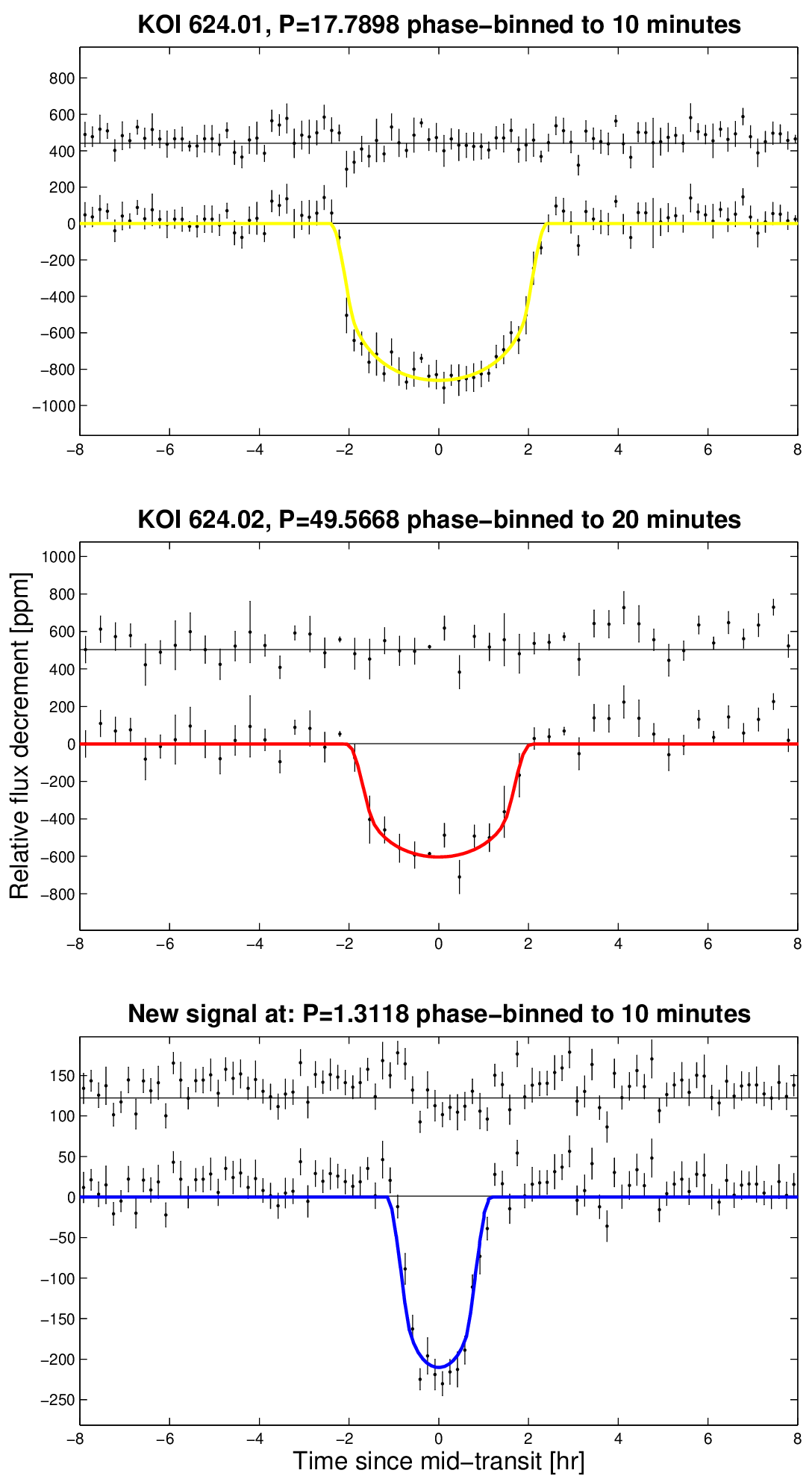}
\caption{Similar to Figure \ref{KOI246fig}.}
\label{KOI624fig}
\end{figure}

\begin{figure}[tbp]\includegraphics[width=0.5\textwidth]{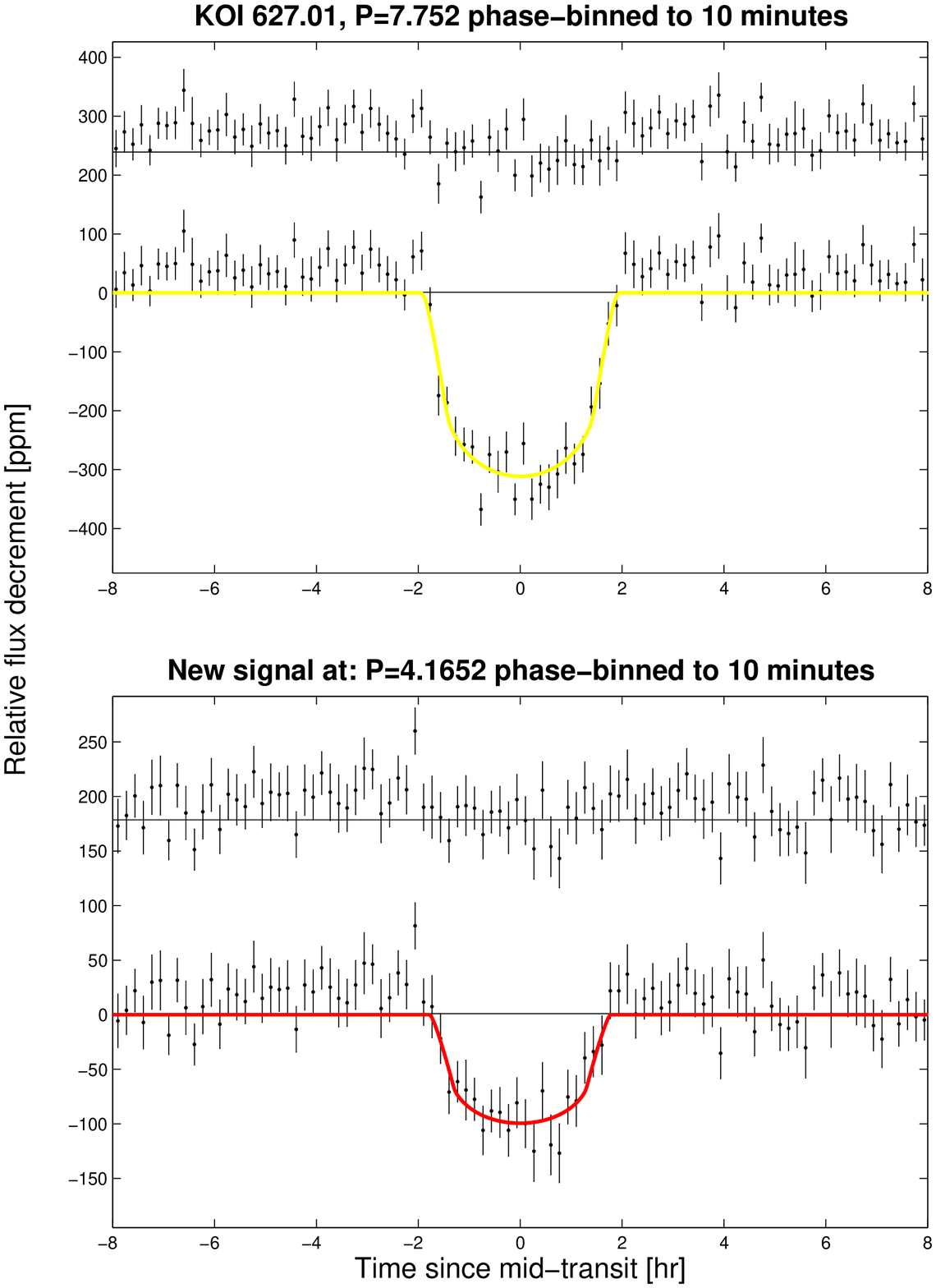}
\caption{Similar to Figure \ref{KOI246fig}.}
\label{KOI627fig}
\end{figure}

\begin{figure}[tbp]\includegraphics[width=0.5\textwidth]{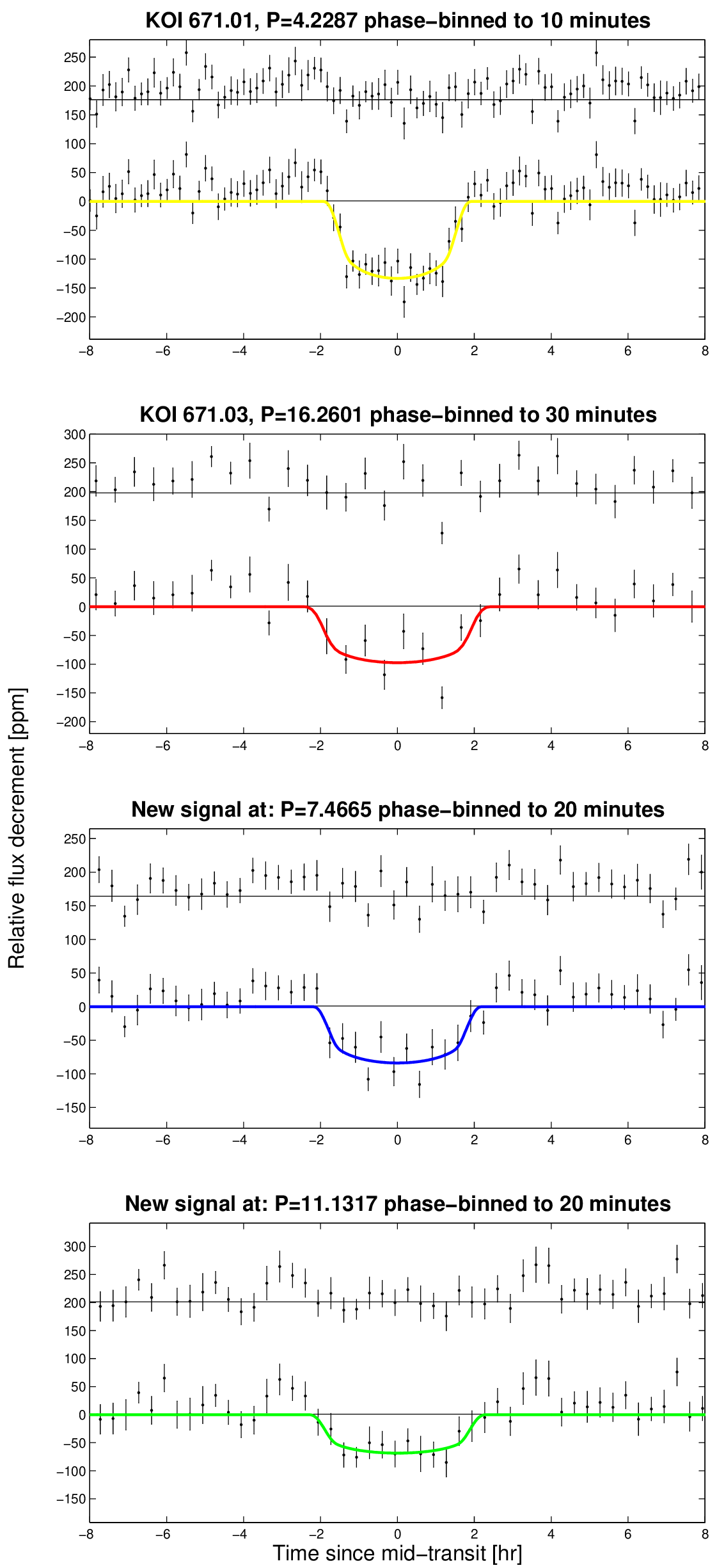}
\caption{Similar to Figure \ref{KOI246fig}.}
\label{KOI671fig}
\end{figure}

\begin{figure}[tbp]\includegraphics[width=0.5\textwidth]{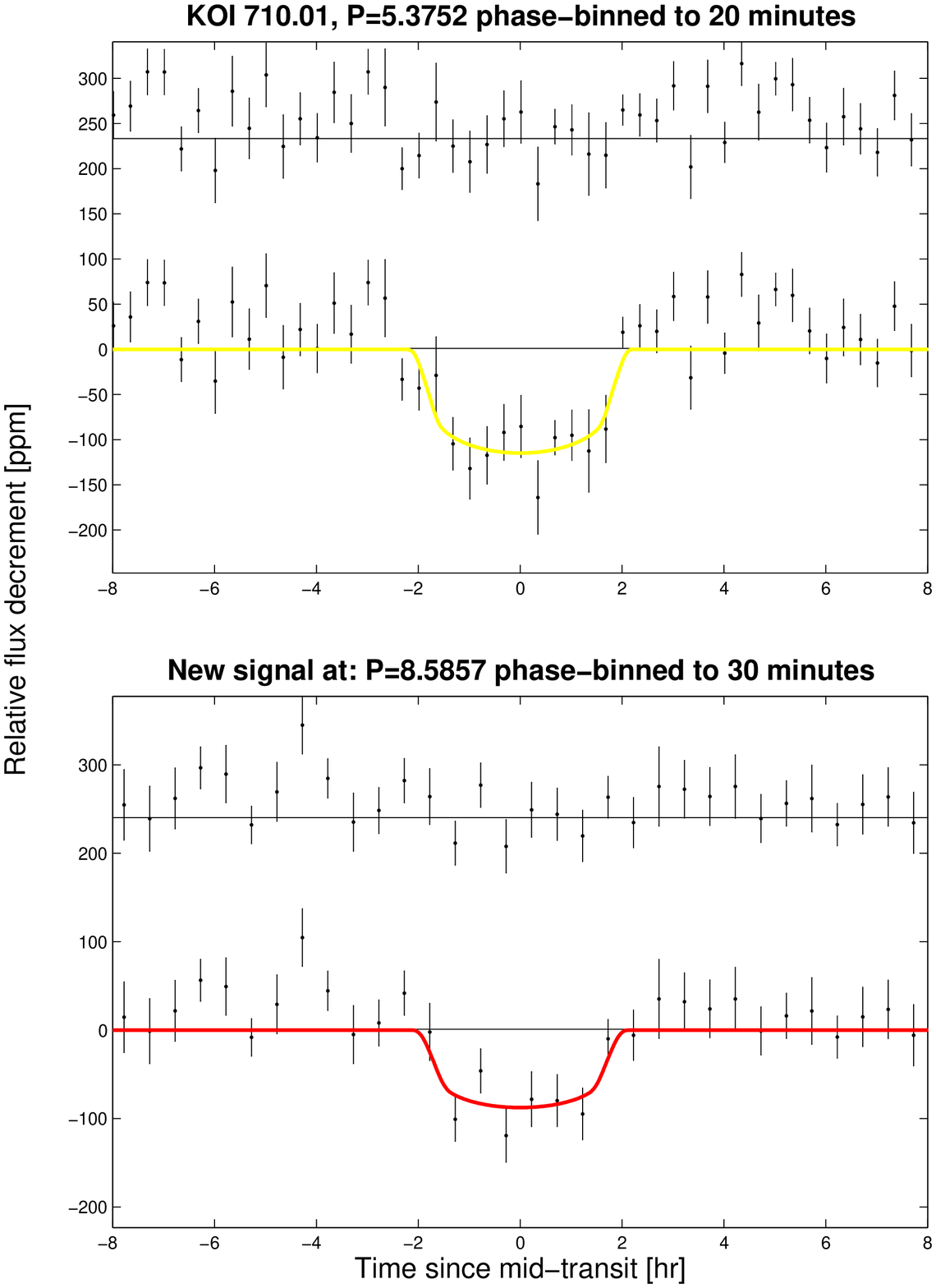}
\caption{Similar to Figure \ref{KOI246fig}.}
\label{KOI710fig}
\end{figure}

\begin{figure}[tbp]\includegraphics[width=0.5\textwidth]{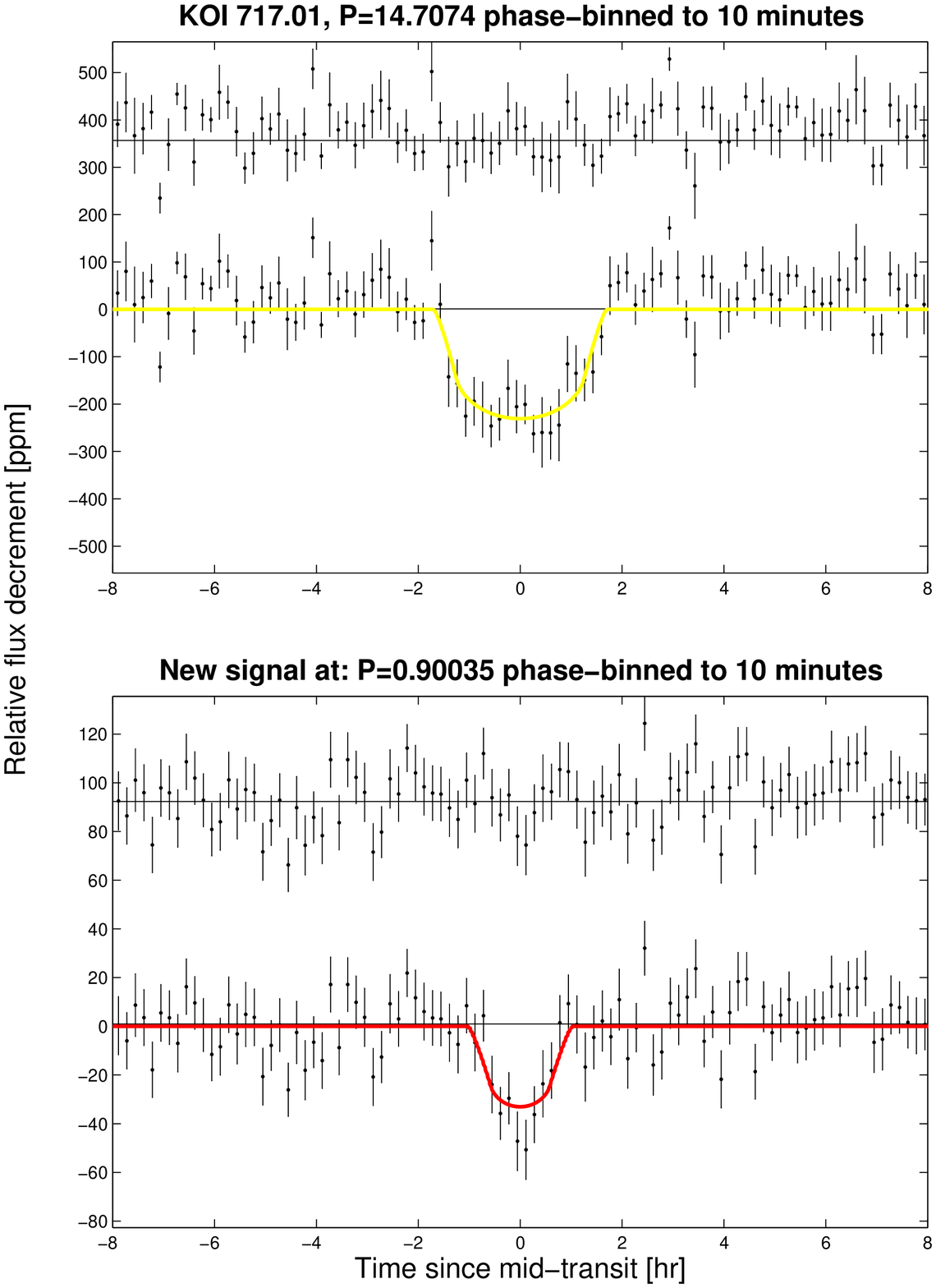}
\caption{Similar to Figure \ref{KOI246fig}.}
\label{KOI717fig}
\end{figure}

\begin{figure}[tbp]\includegraphics[width=0.5\textwidth]{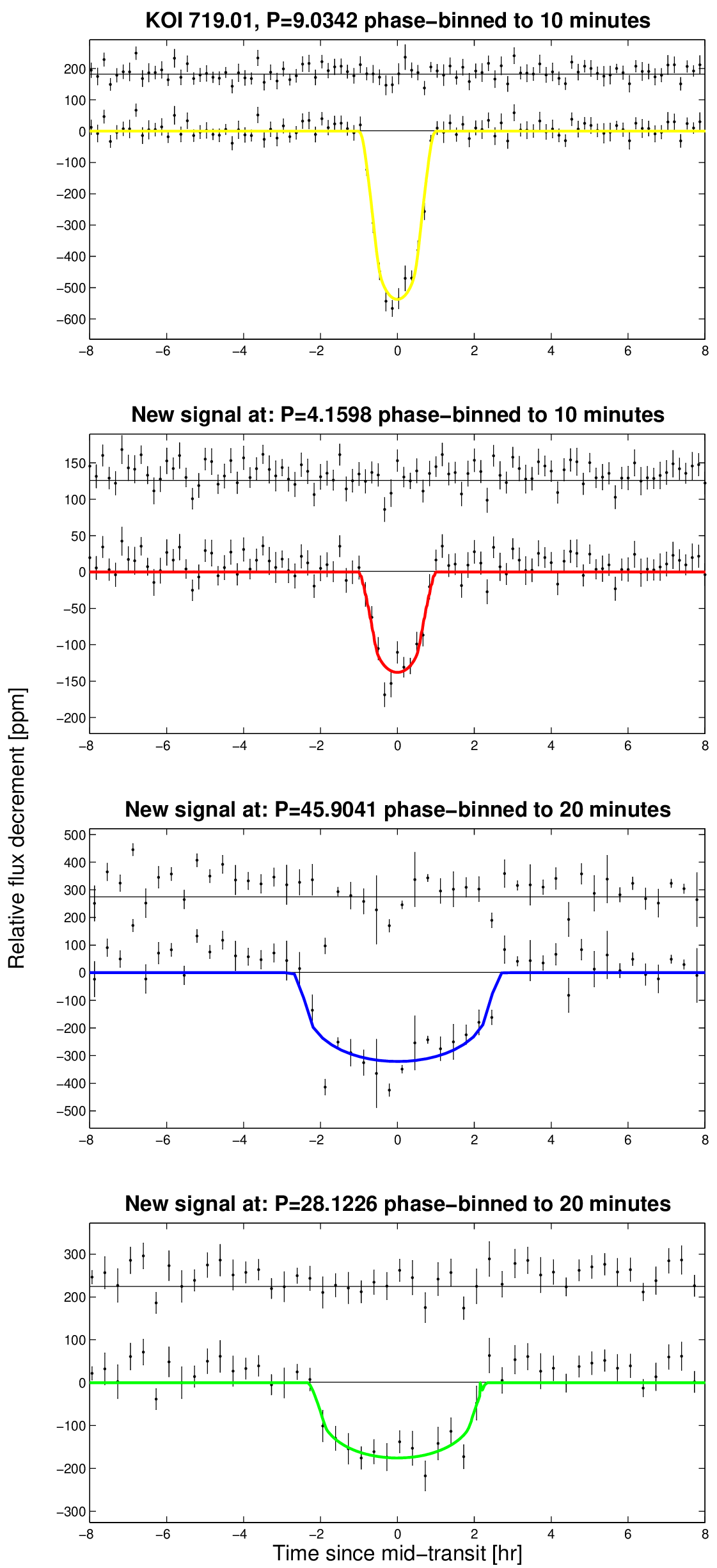}
\caption{Similar to Figure \ref{KOI246fig}.}
\label{KOI719fig}
\end{figure}

\begin{figure}[tbp]\includegraphics[width=0.5\textwidth]{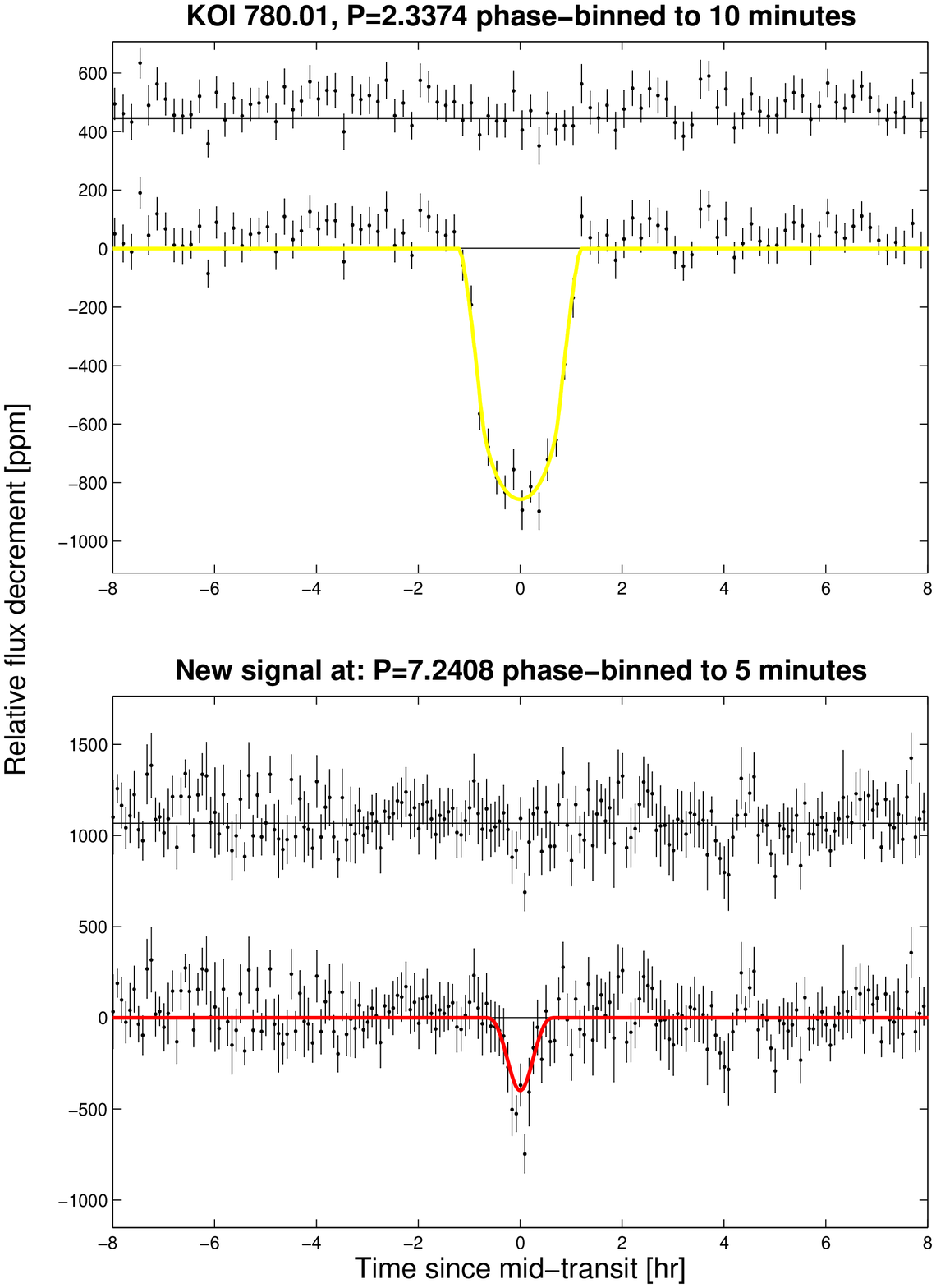}
\caption{Similar to Figure \ref{KOI246fig}.}
\label{KOI780fig}
\end{figure}

\begin{figure}[tbp]\includegraphics[width=0.5\textwidth]{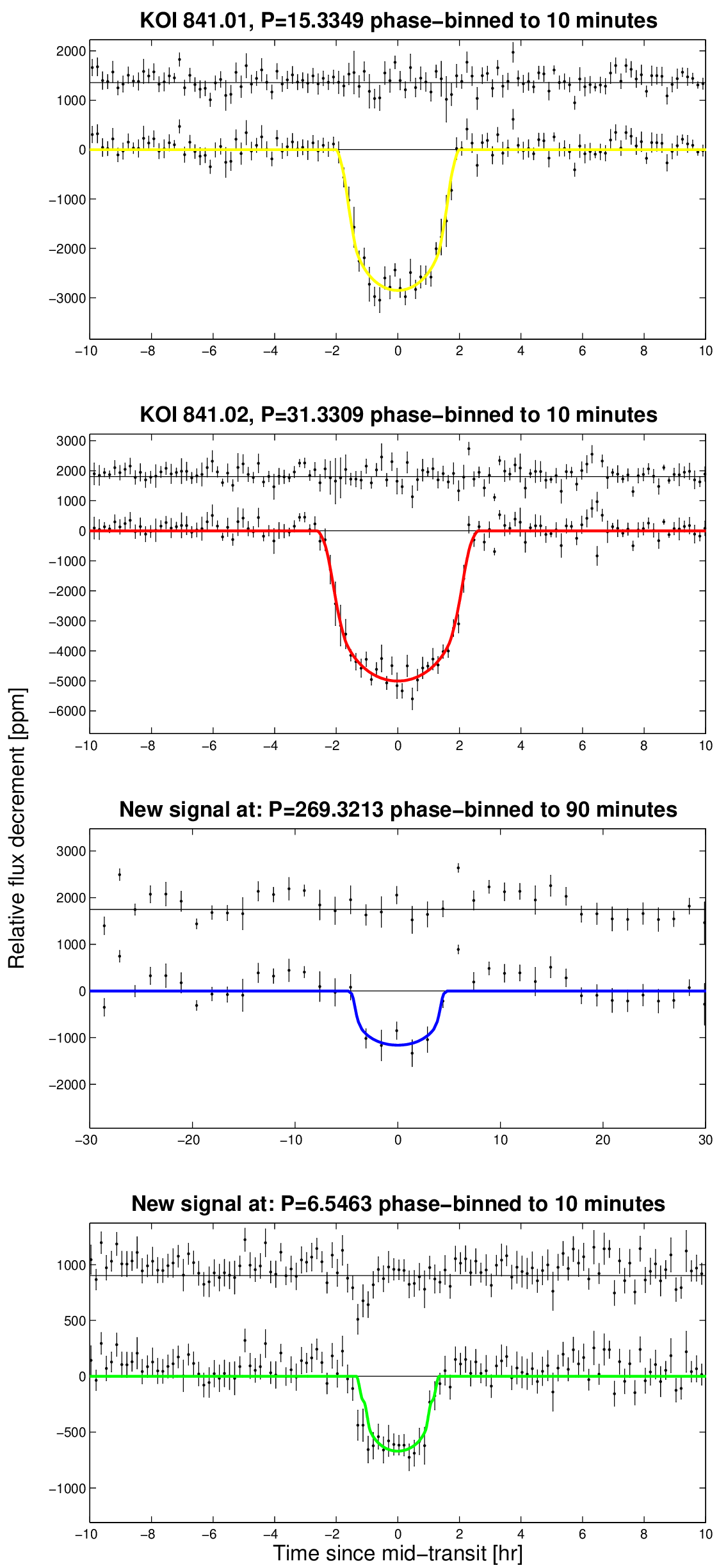}
\caption{Similar to Figure \ref{KOI246fig}.}
\label{KOI841fig}
\end{figure}

\begin{figure}[tbp]\includegraphics[width=0.5\textwidth]{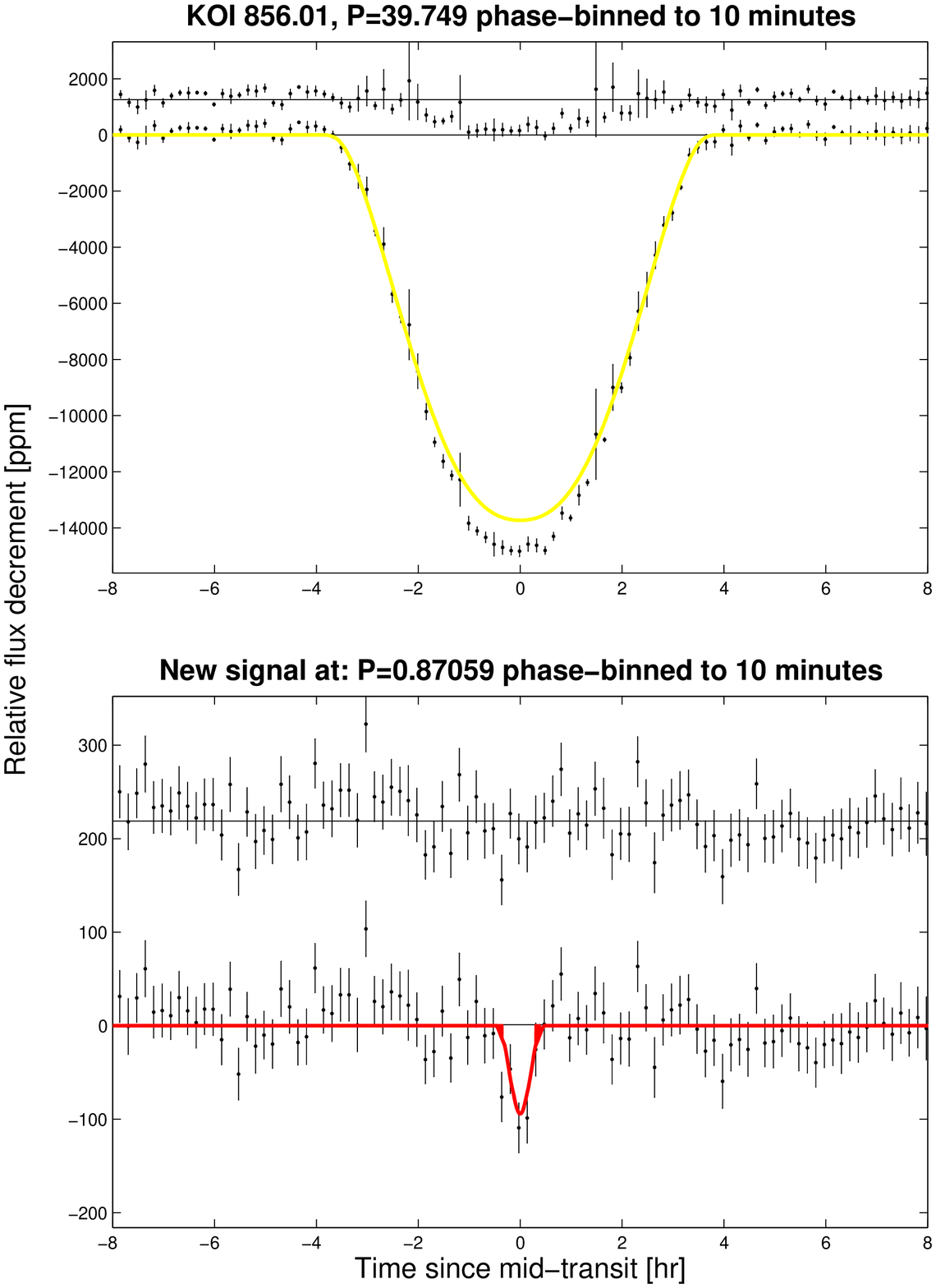}
\caption{Similar to Figure \ref{KOI246fig}. The observed offset between the data and model is caused by three outlier data points beyond the presented view.}
\label{KOI1060fig}
\end{figure}

\begin{figure}[tbp]\includegraphics[width=0.5\textwidth]{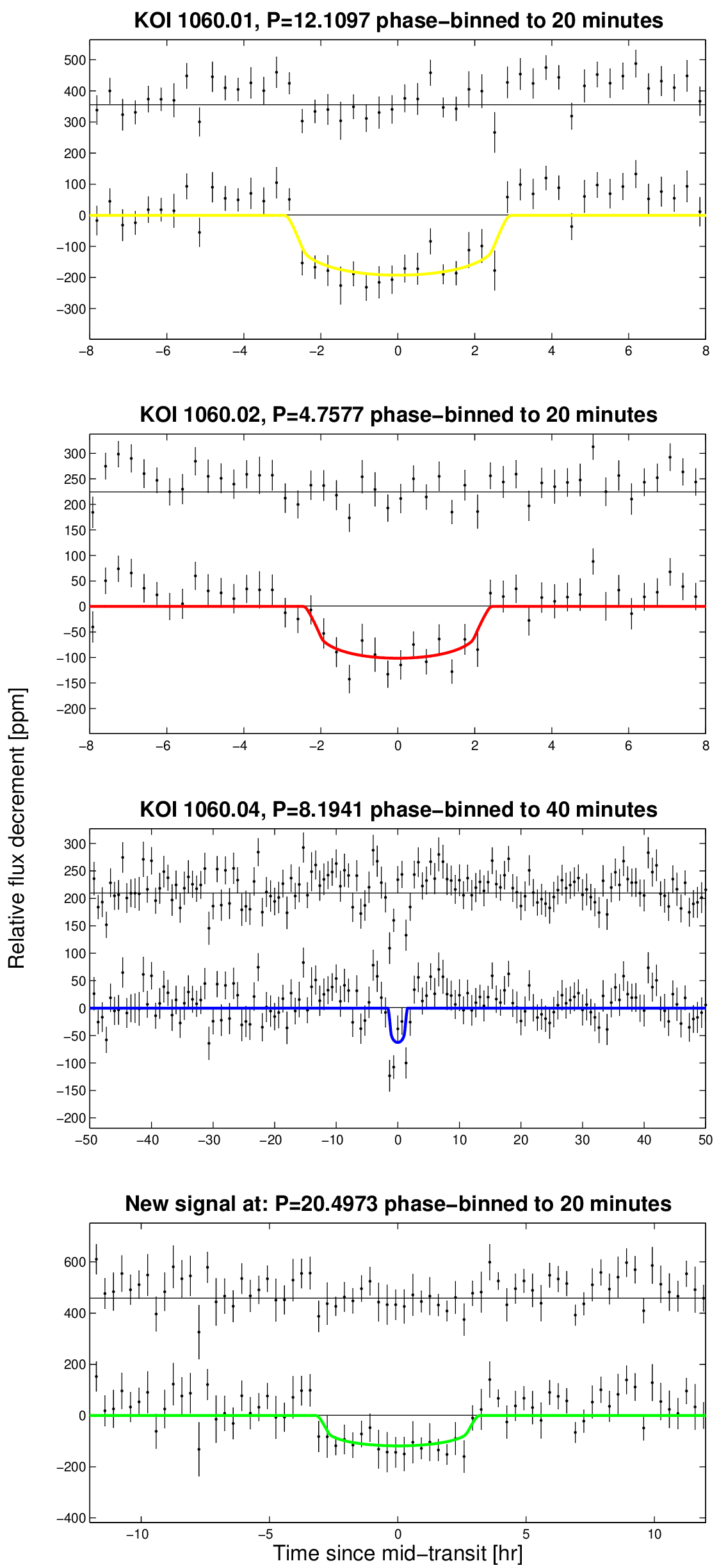}
\caption{Similar to Figure \ref{KOI246fig}. We note the long window for KOI 1060.04, spanning about half the orbital period was chosen to highlight the significance of this shallow signal.}
\label{KOI1060fig}
\end{figure}

\begin{figure}[tbp]\includegraphics[width=0.5\textwidth]{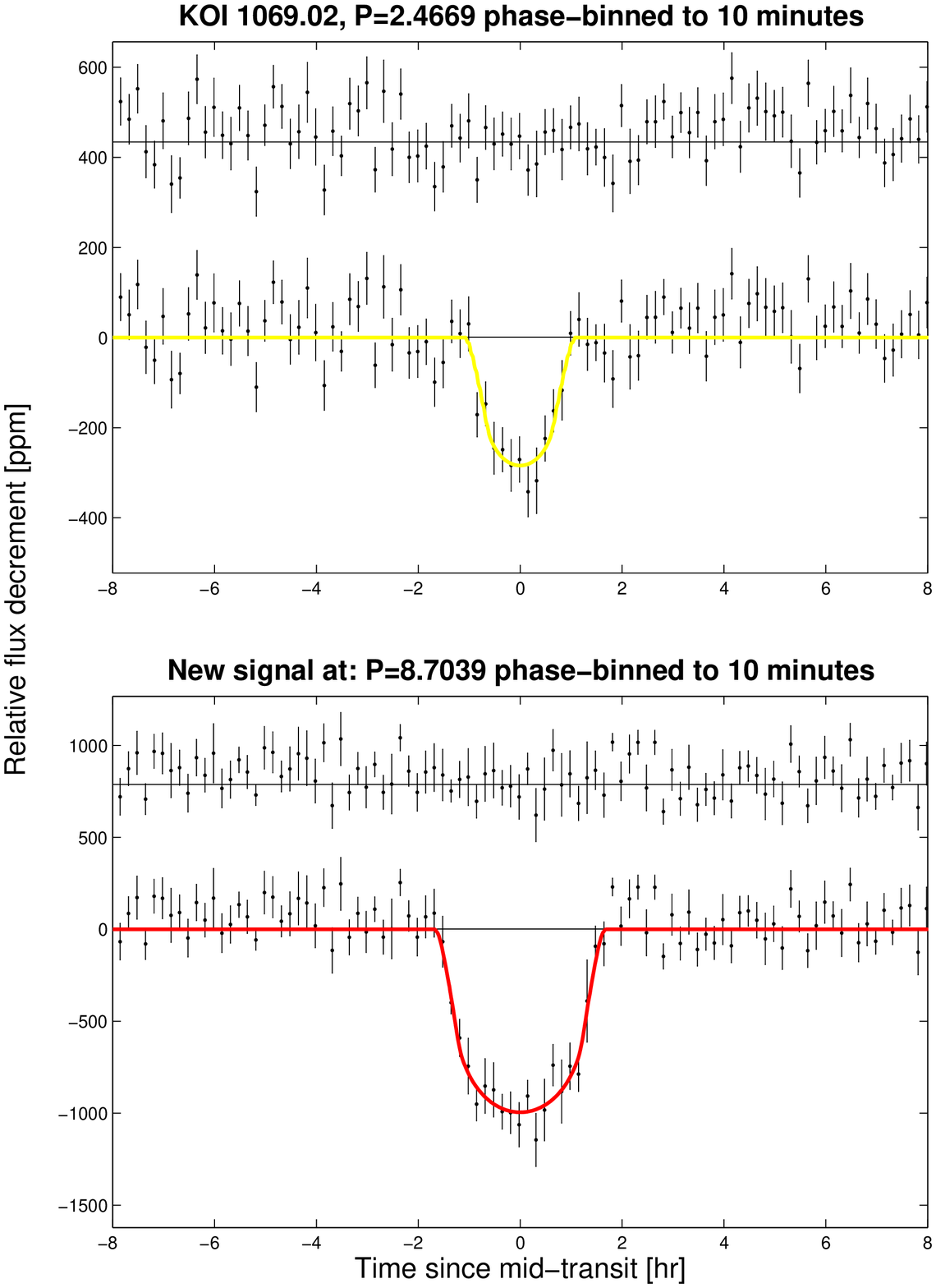}
\caption{Similar to Figure \ref{KOI246fig}.}
\label{KOI1069fig}
\end{figure}

\begin{figure}[tbp]\includegraphics[width=0.5\textwidth]{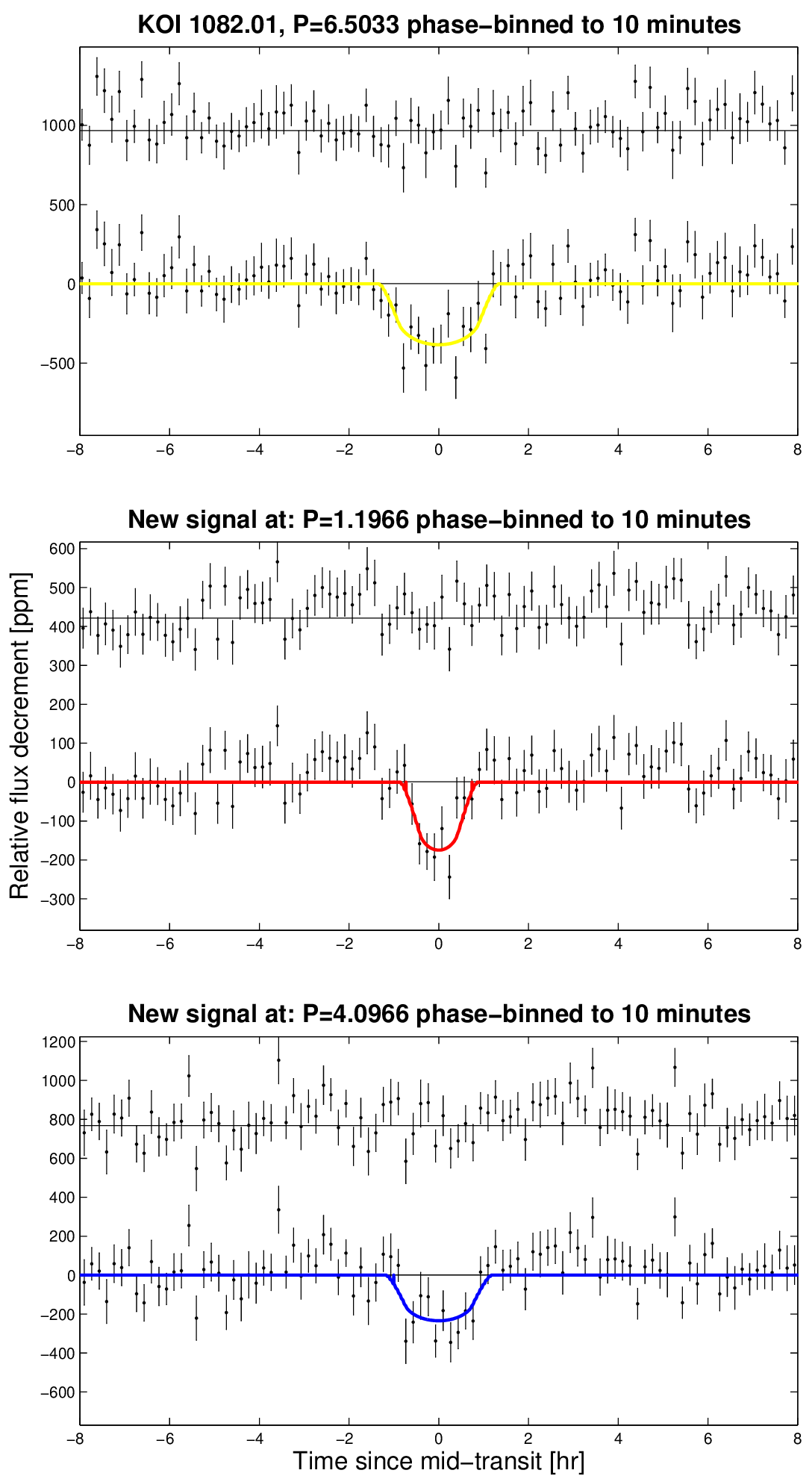}
\caption{Similar to Figure \ref{KOI246fig}.}
\label{KOI1082fig}
\end{figure}

\begin{figure}[tbp]\includegraphics[width=0.5\textwidth]{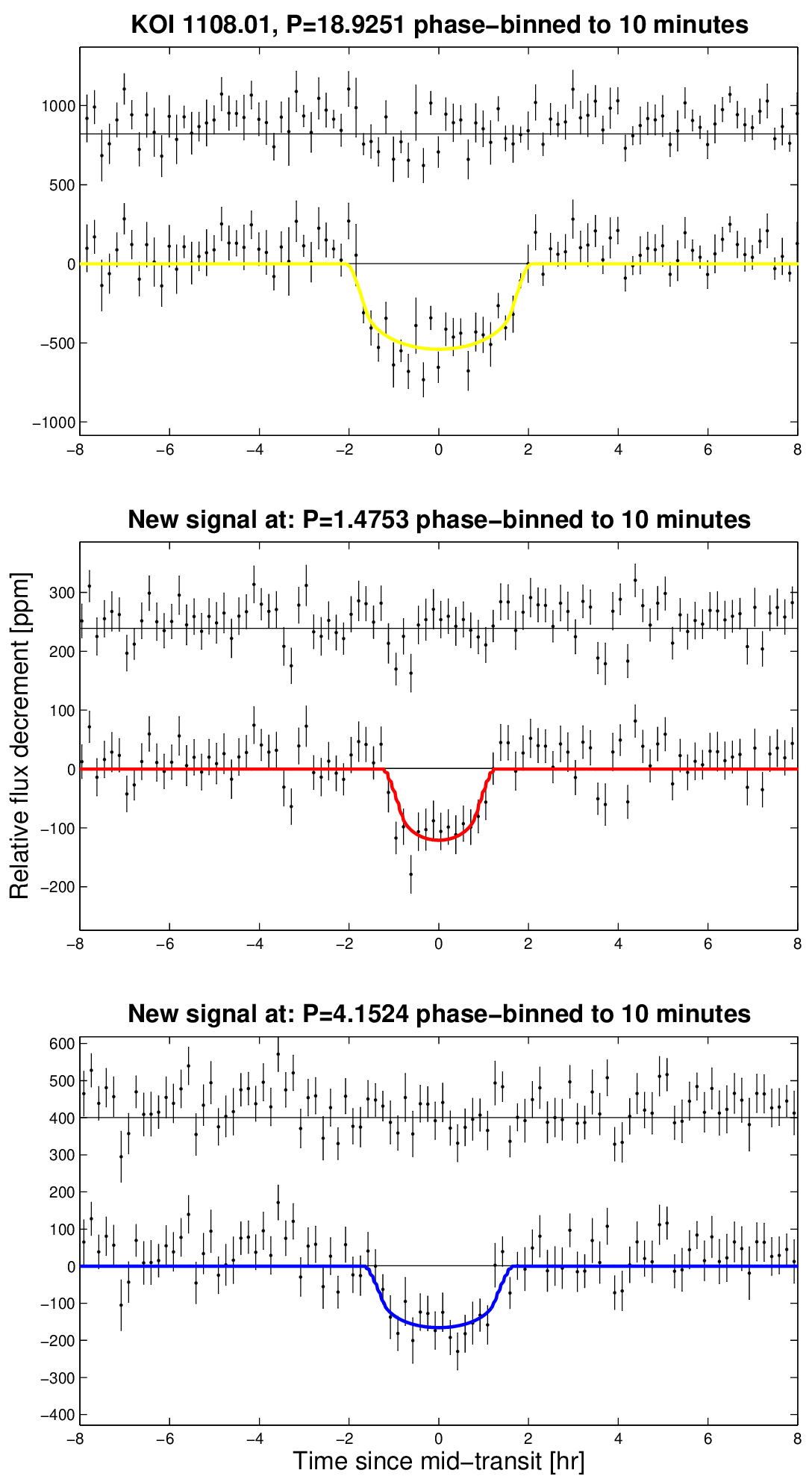}
\caption{Similar to Figure \ref{KOI246fig}.}
\label{KOI1108fig}
\end{figure}

\clearpage

\begin{figure}[tbp]\includegraphics[width=0.5\textwidth]{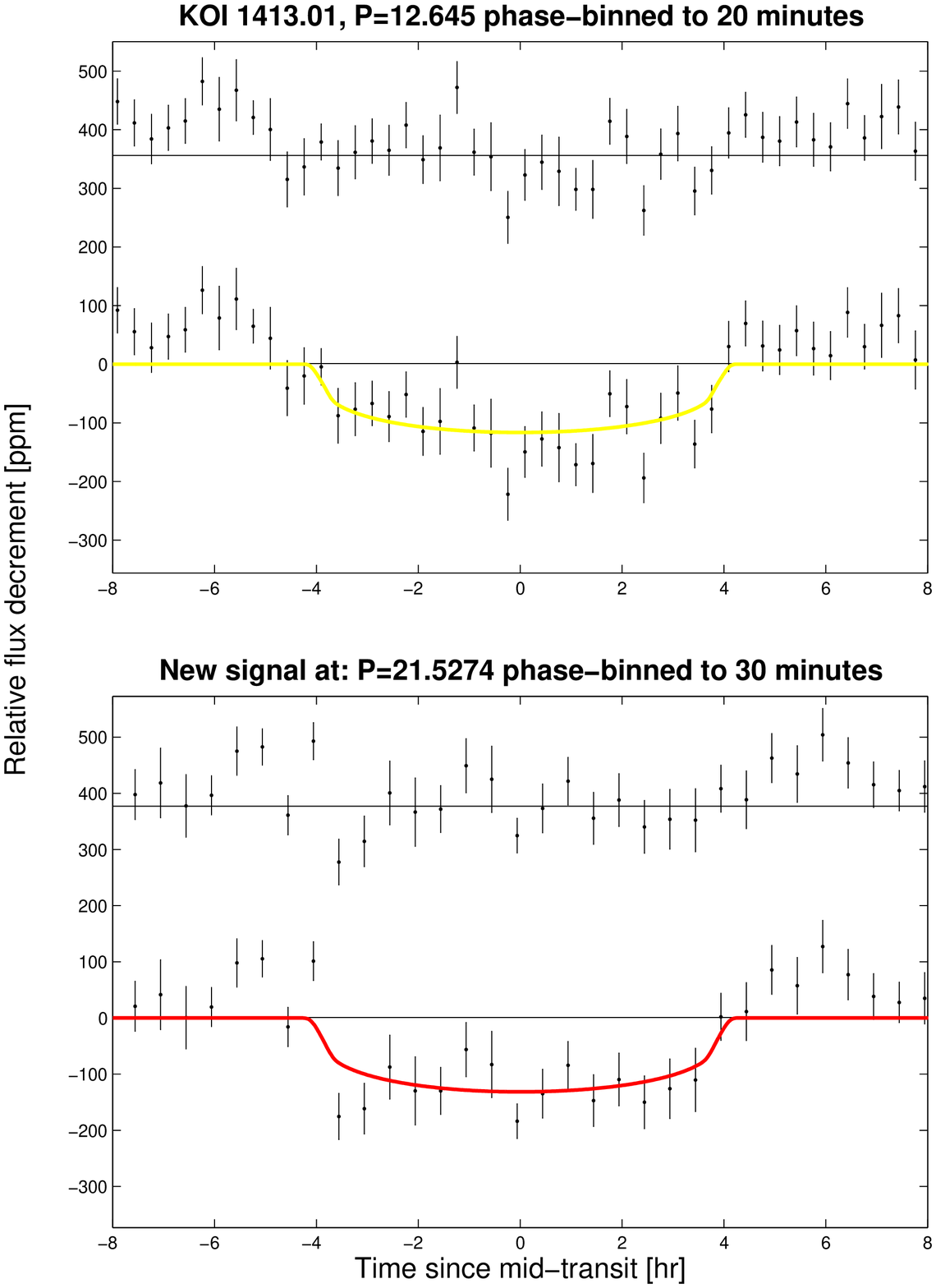}
\caption{Similar to Figure \ref{KOI246fig}.}
\label{KOI1413fig}
\end{figure}

\begin{figure}[tbp]\includegraphics[width=0.5\textwidth]{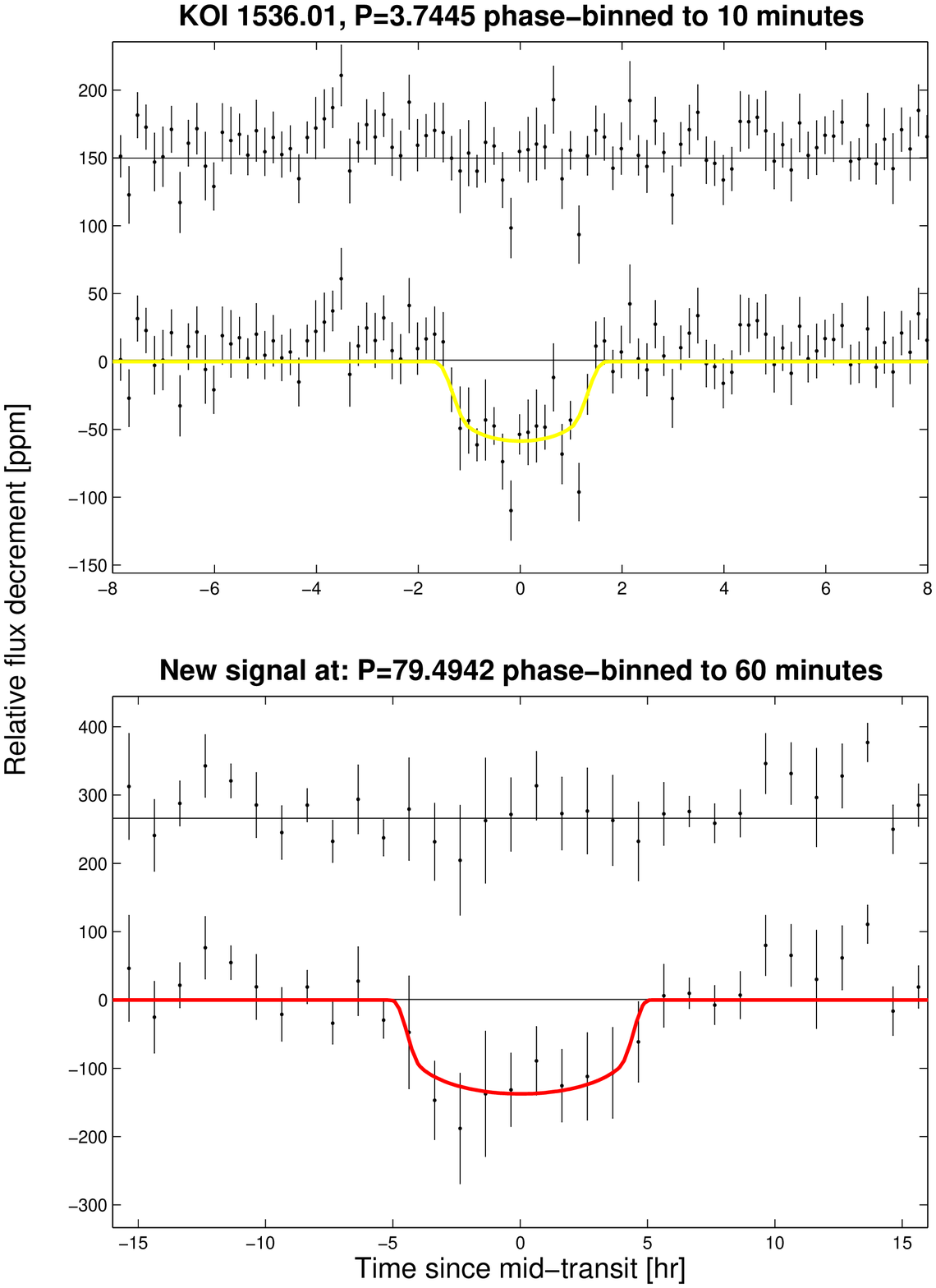}
\caption{Similar to Figure \ref{KOI246fig}.}
\label{KOI1536fig}
\end{figure}

\begin{figure}[tbp]\includegraphics[width=0.5\textwidth]{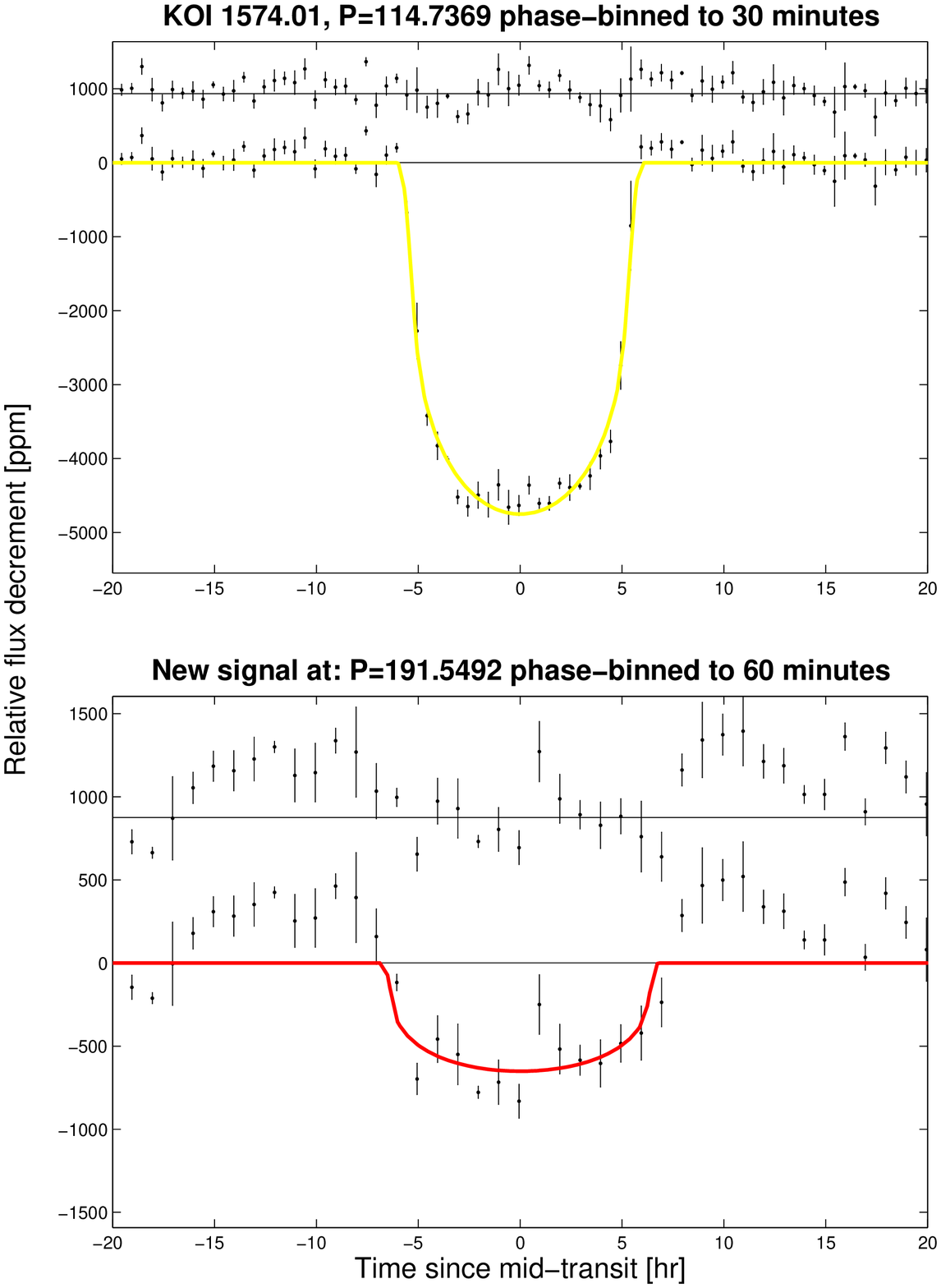}
\caption{Similar to Figure \ref{KOI246fig}.}
\label{KOI1574fig}
\end{figure}

\begin{figure}[tbp]\includegraphics[width=0.5\textwidth]{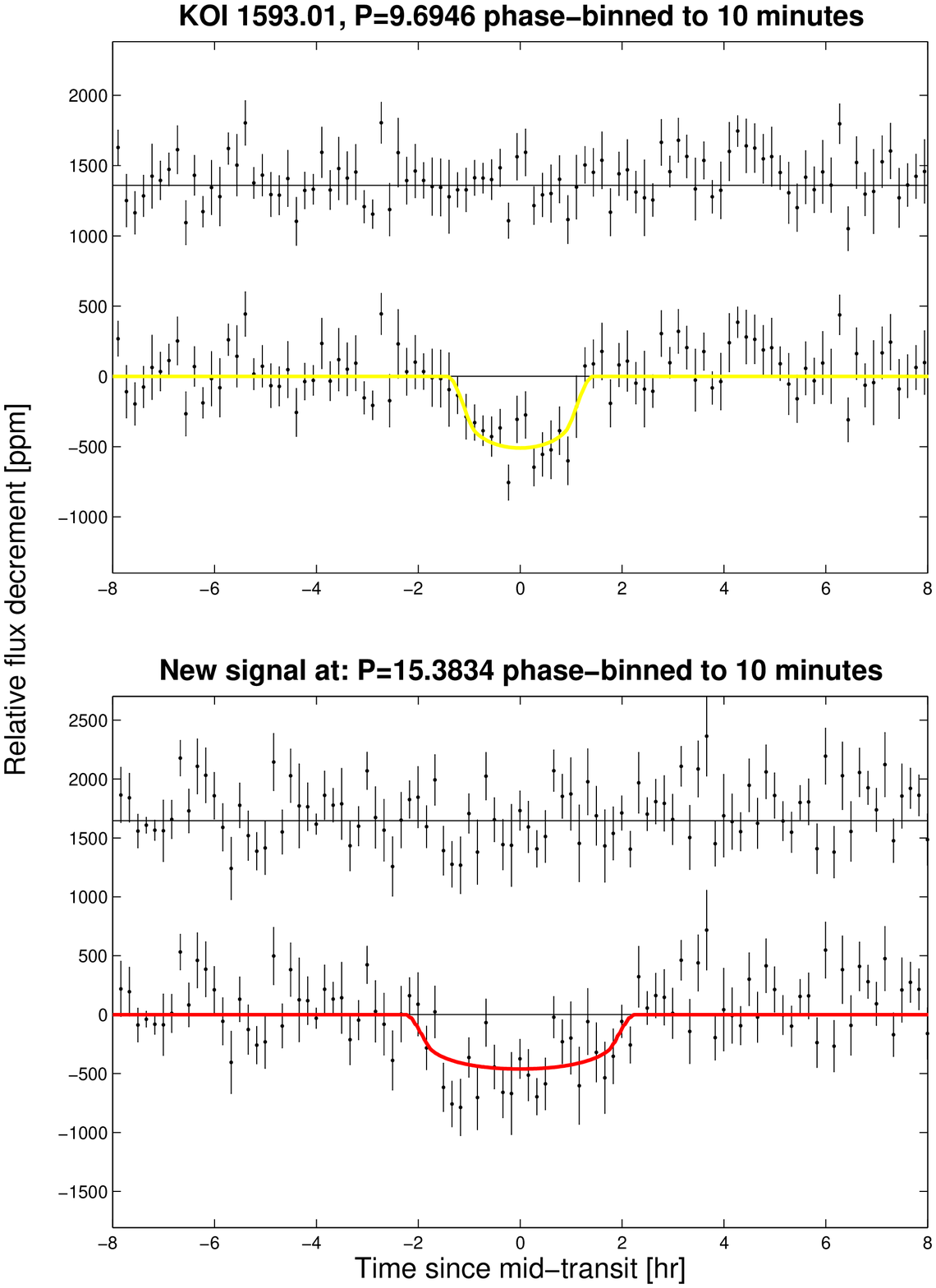}
\caption{Similar to Figure \ref{KOI246fig}.}
\label{KOI1593fig}
\end{figure}

\begin{figure}[tbp]\includegraphics[width=0.5\textwidth]{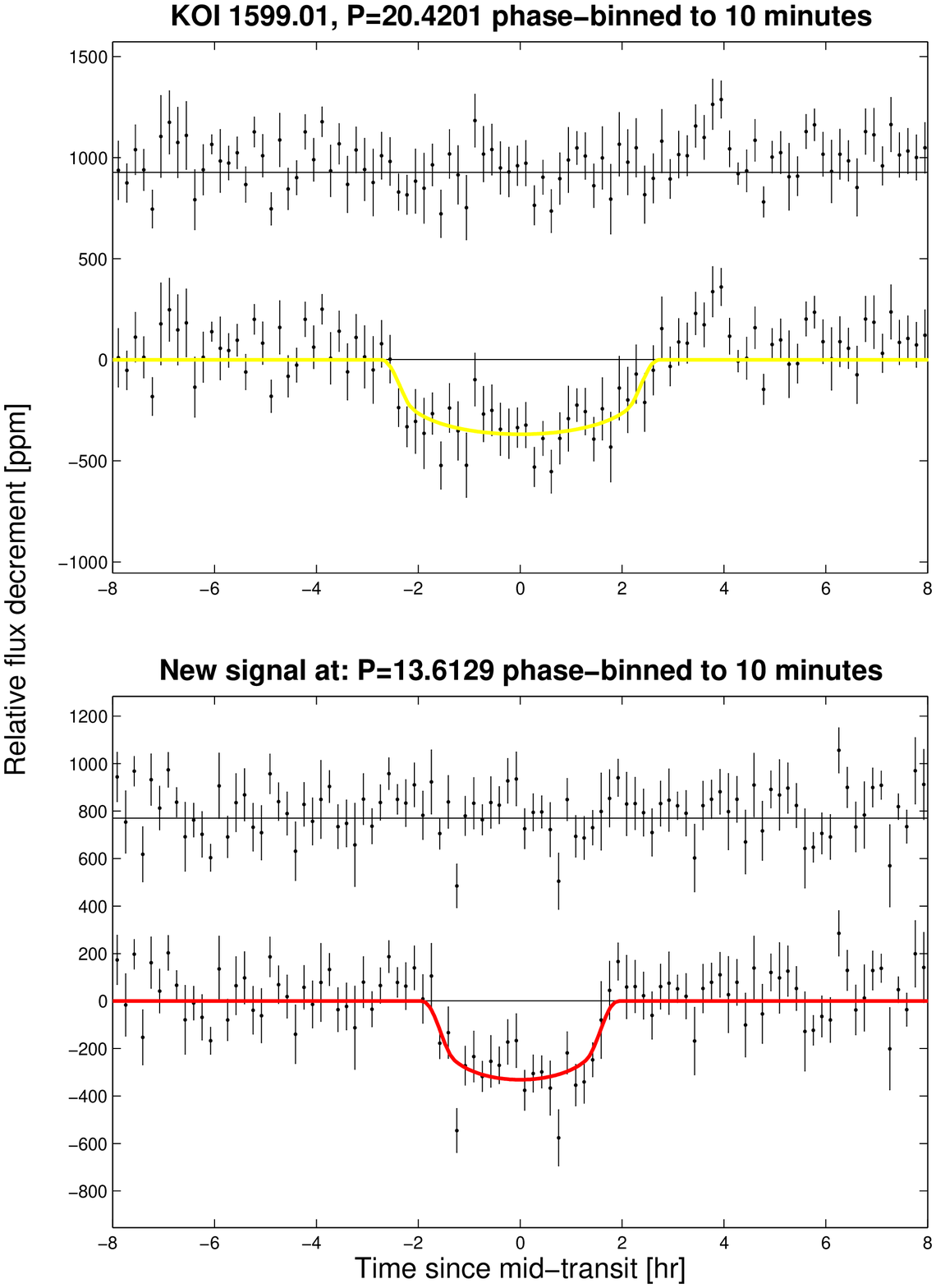}
\caption{Similar to Figure \ref{KOI246fig}.}
\label{KOI1599fig}
\end{figure}

\begin{figure}[tbp]\includegraphics[width=0.5\textwidth]{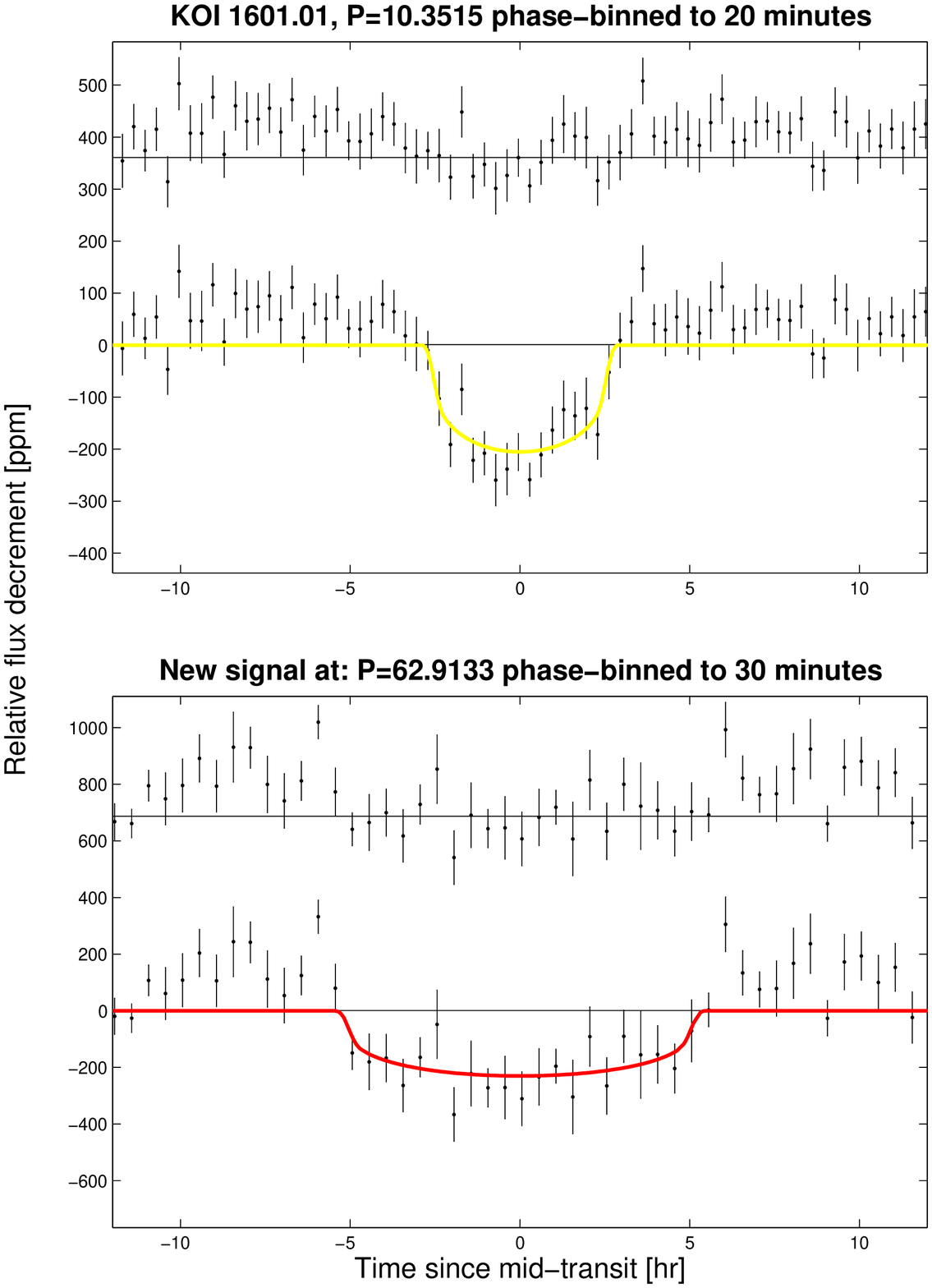}
\caption{Similar to Figure \ref{KOI246fig}.}
\label{KOI1601fig}
\end{figure}

\begin{figure}[tbp]\includegraphics[width=0.5\textwidth]{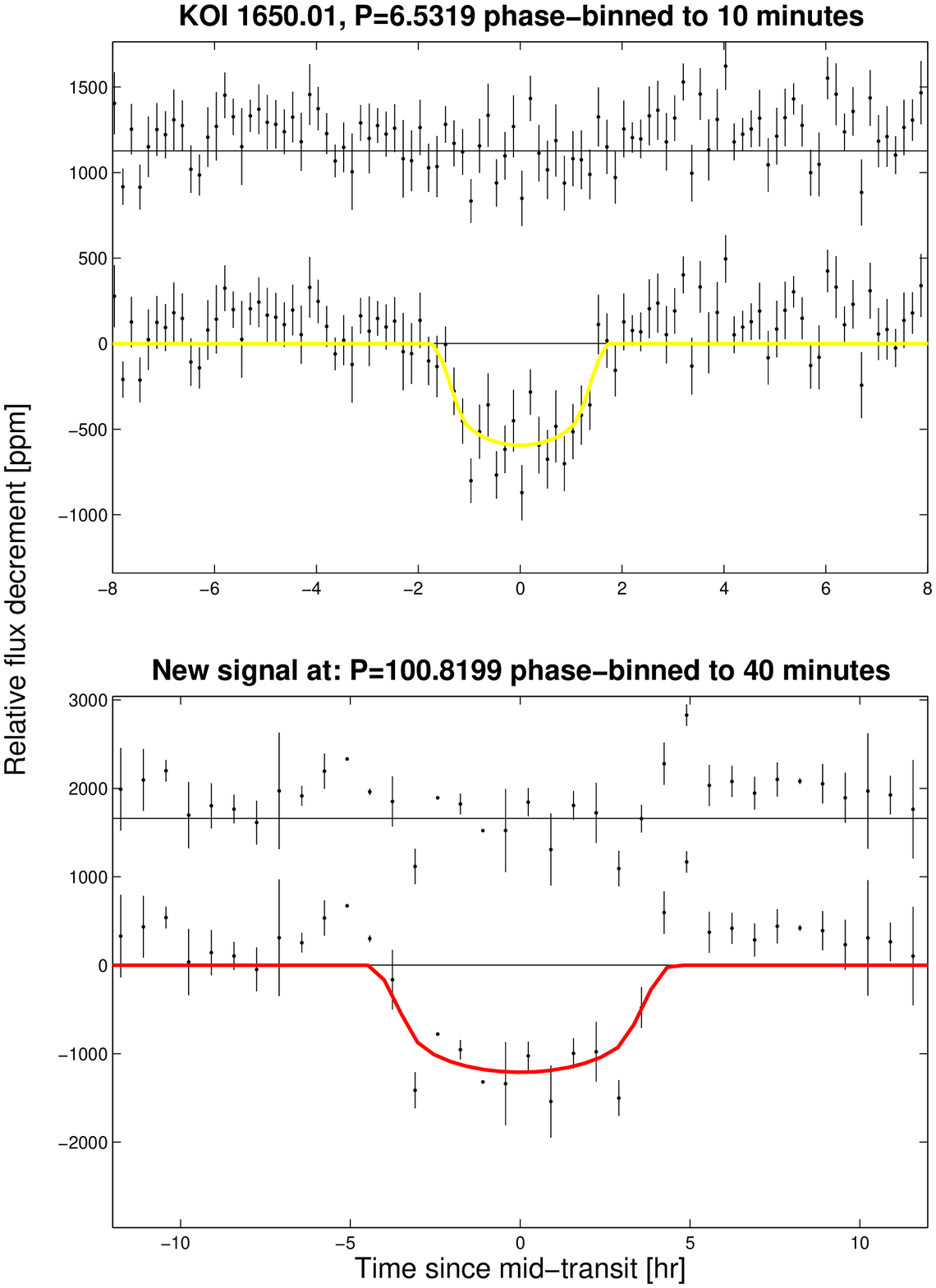}
\caption{Similar to Figure \ref{KOI246fig}.}
\label{KOI1650fig}
\end{figure}

\begin{figure}[tbp]\includegraphics[width=0.5\textwidth]{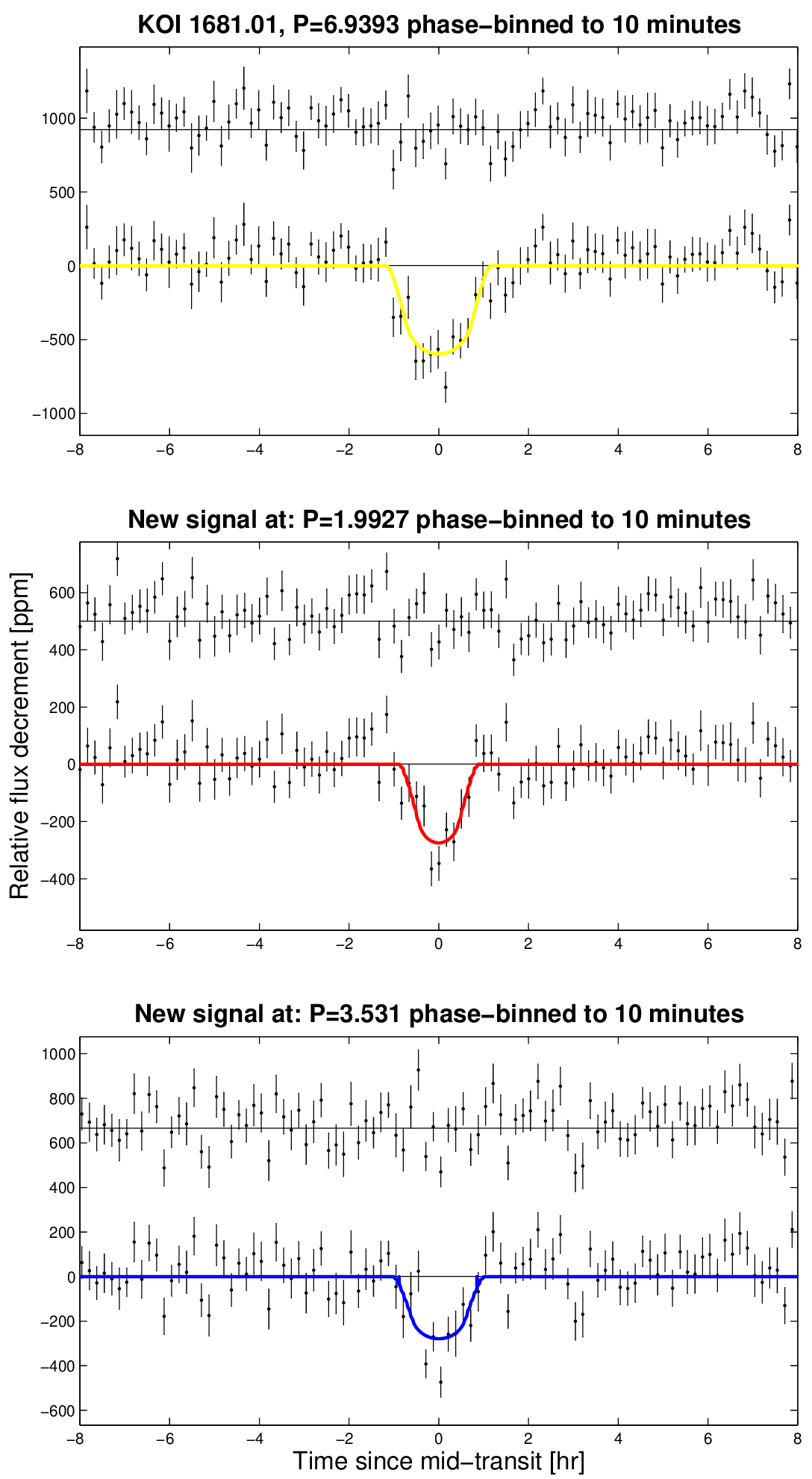}
\caption{Similar to Figure \ref{KOI246fig}.}
\label{KOI1681fig}
\end{figure}

\begin{figure}[tbp]\includegraphics[width=0.5\textwidth]{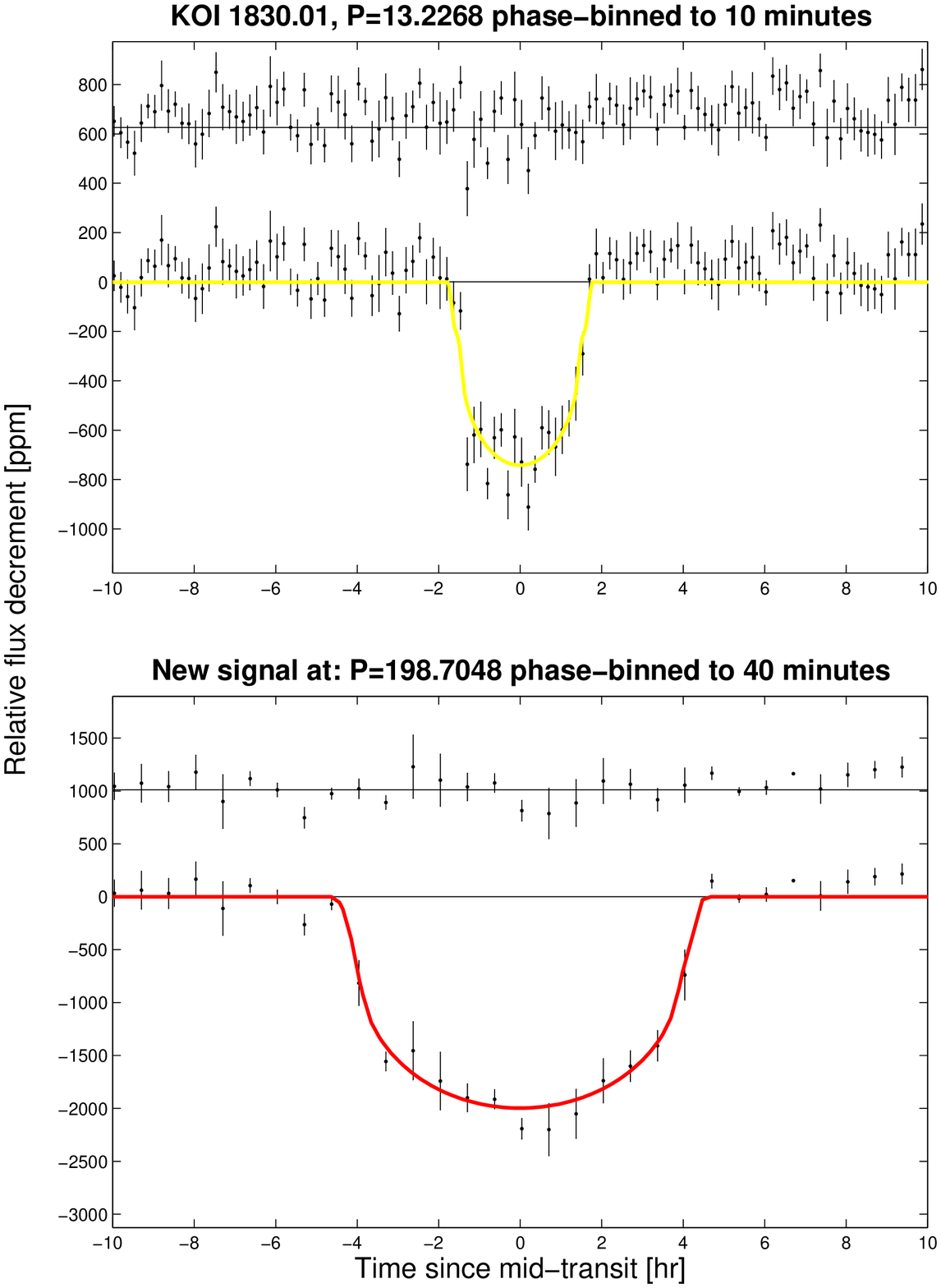}
\caption{Similar to Figure \ref{KOI246fig}.}
\label{KOI1830fig}
\end{figure}

\begin{figure}[tbp]\includegraphics[width=0.5\textwidth]{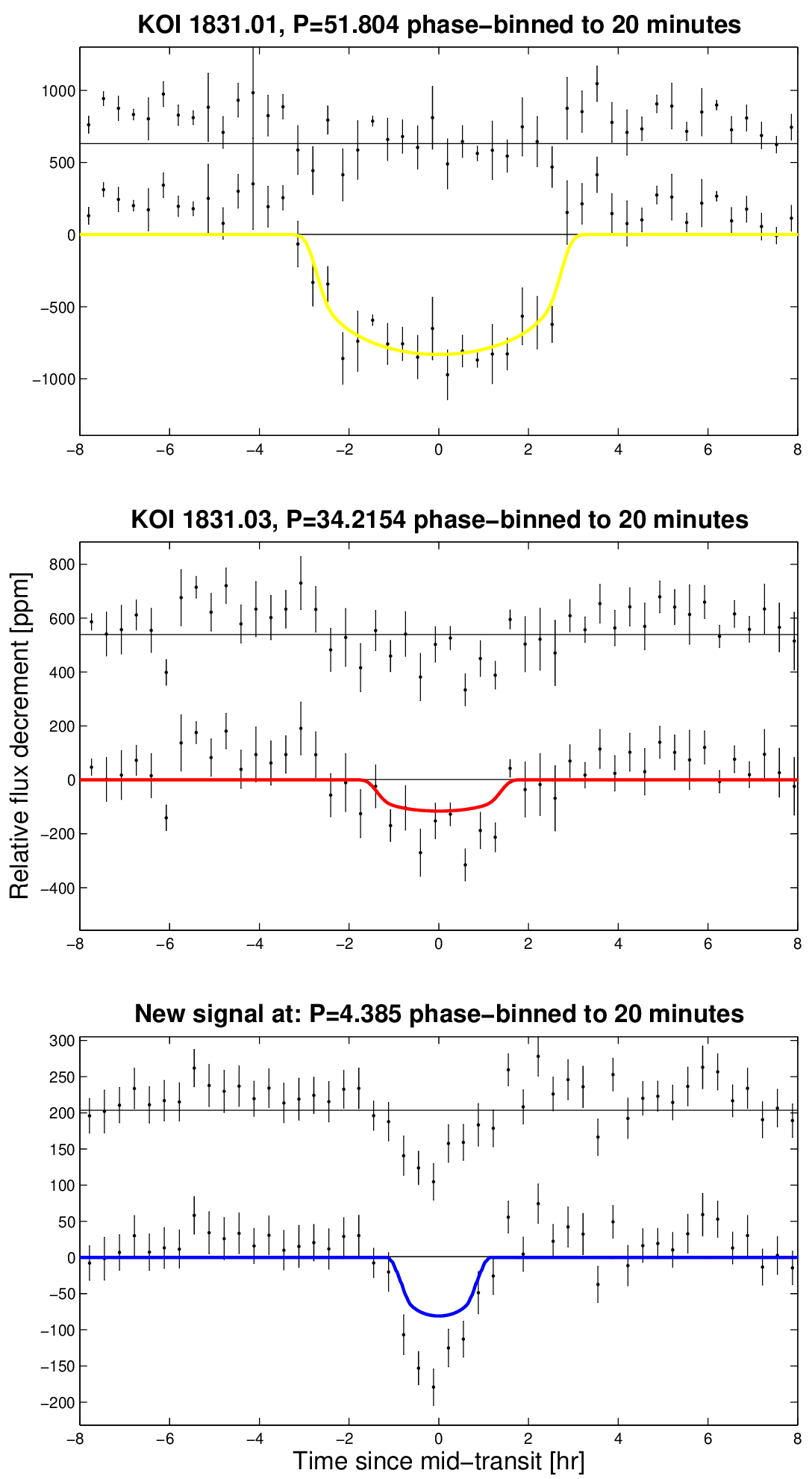}
\caption{Similar to Figure \ref{KOI246fig}.}
\label{KOI1831fig}
\end{figure}

\begin{figure}[tbp]\includegraphics[width=0.5\textwidth]{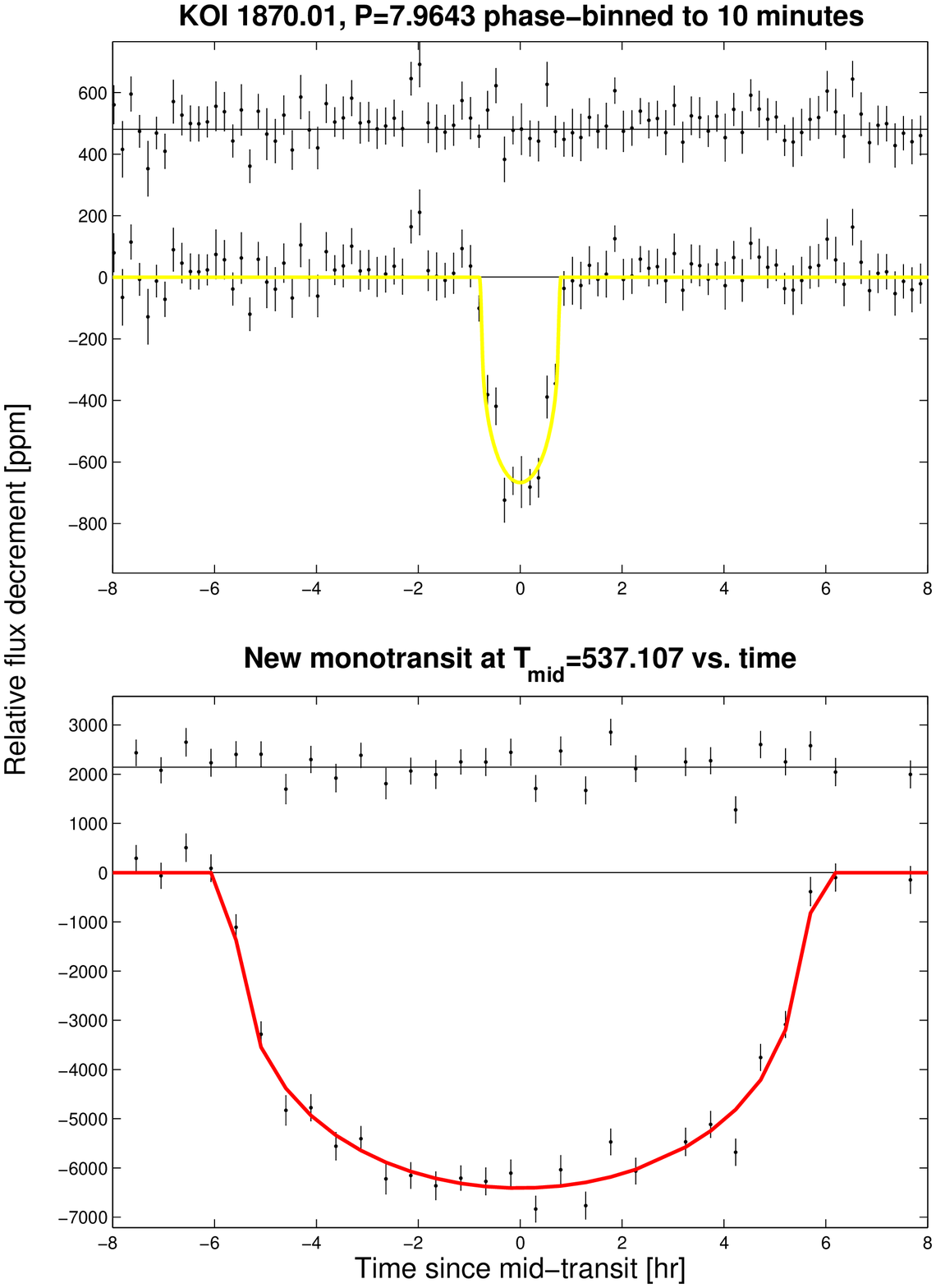}
\caption{Similar to Figure \ref{KOI246fig}.}
\label{KOI1870fig}
\end{figure}

\begin{figure}[tbp]\includegraphics[width=0.5\textwidth]{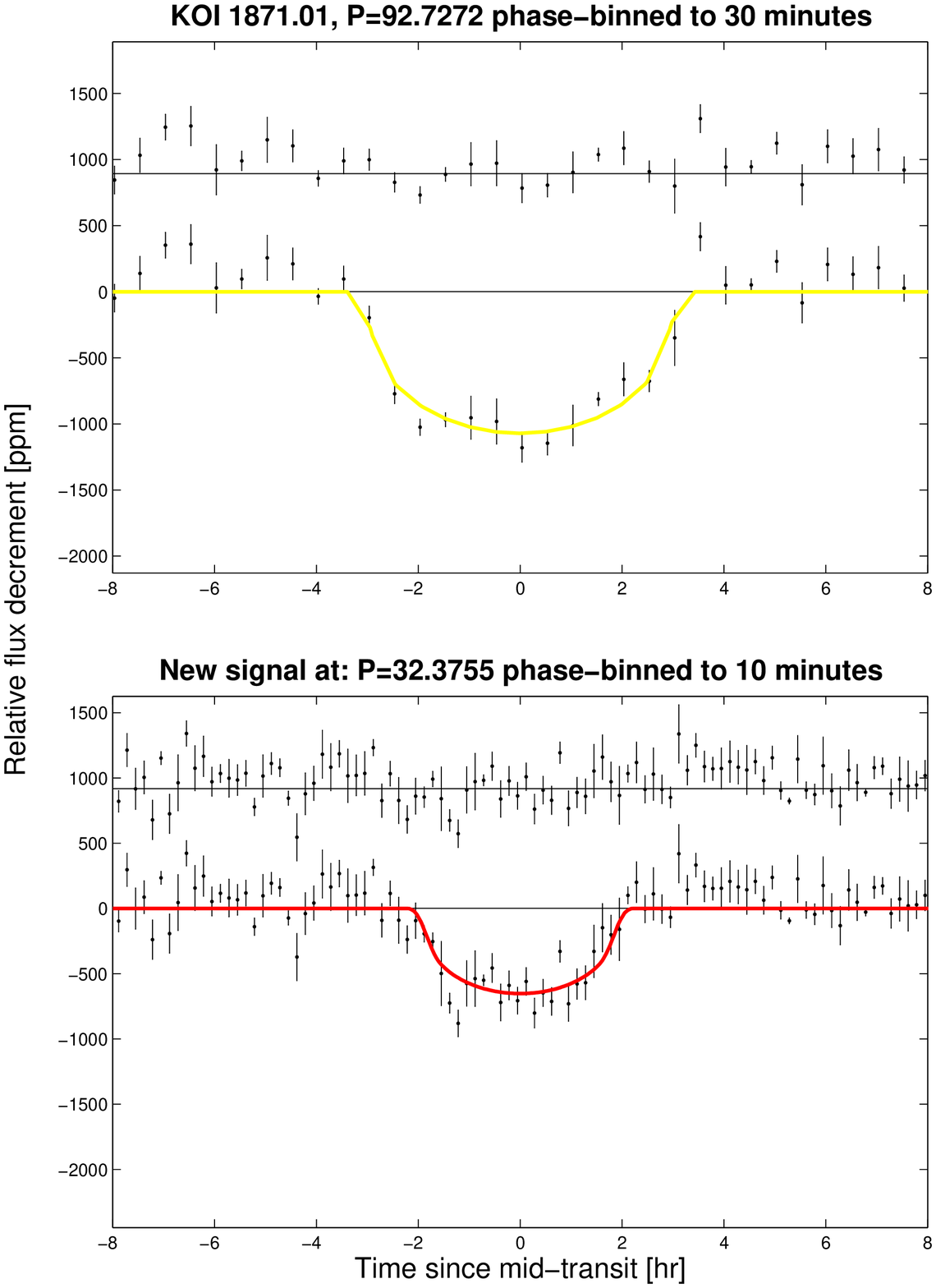}
\caption{Similar to Figure \ref{KOI246fig}.}
\label{KOI1871fig}
\end{figure}

\begin{figure}[tbp]\includegraphics[width=0.5\textwidth]{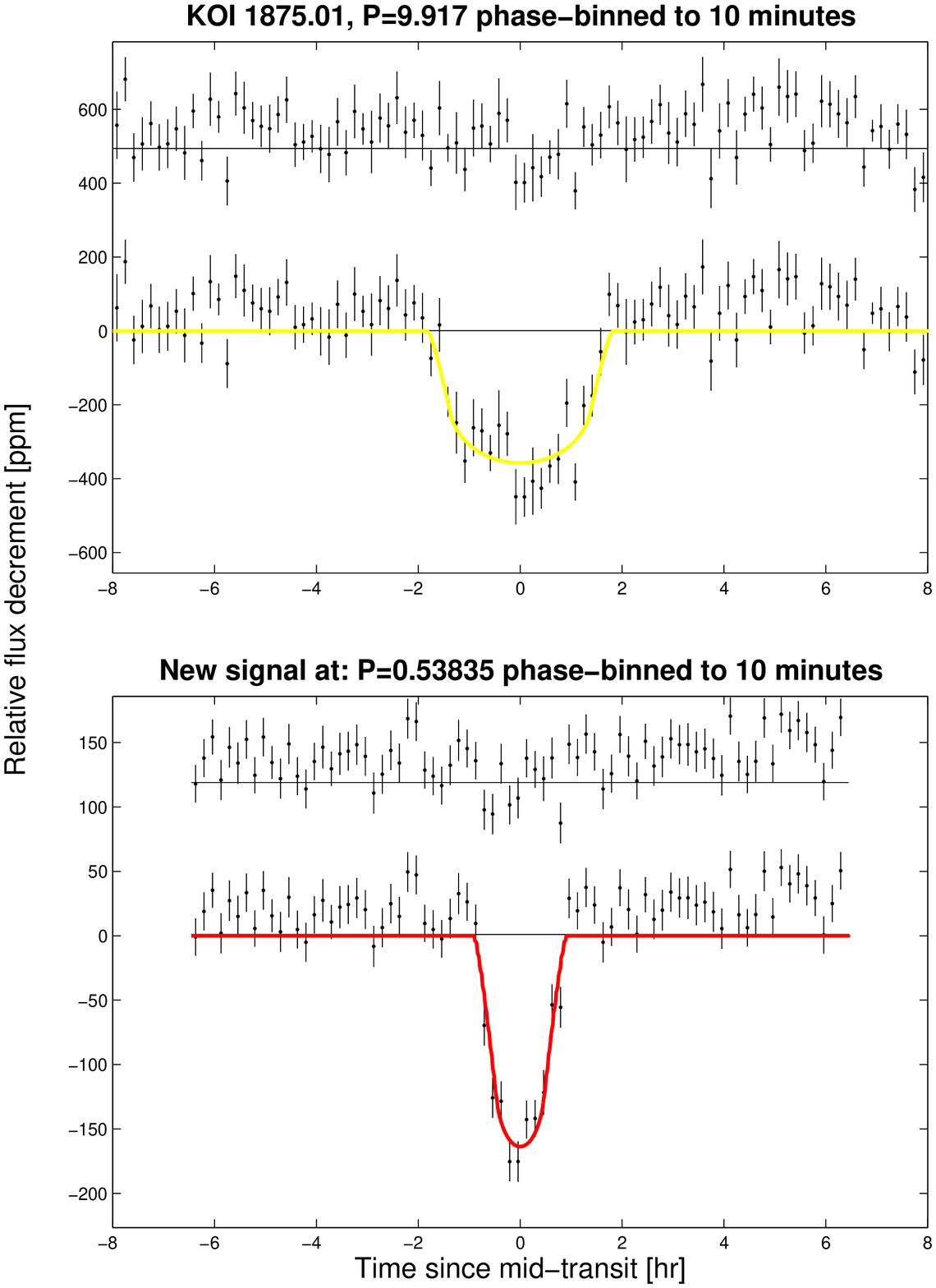}
\caption{Similar to Figure \ref{KOI246fig}.}
\label{KOI1875fig}
\end{figure}

\begin{figure}[tbp]\includegraphics[width=0.5\textwidth]{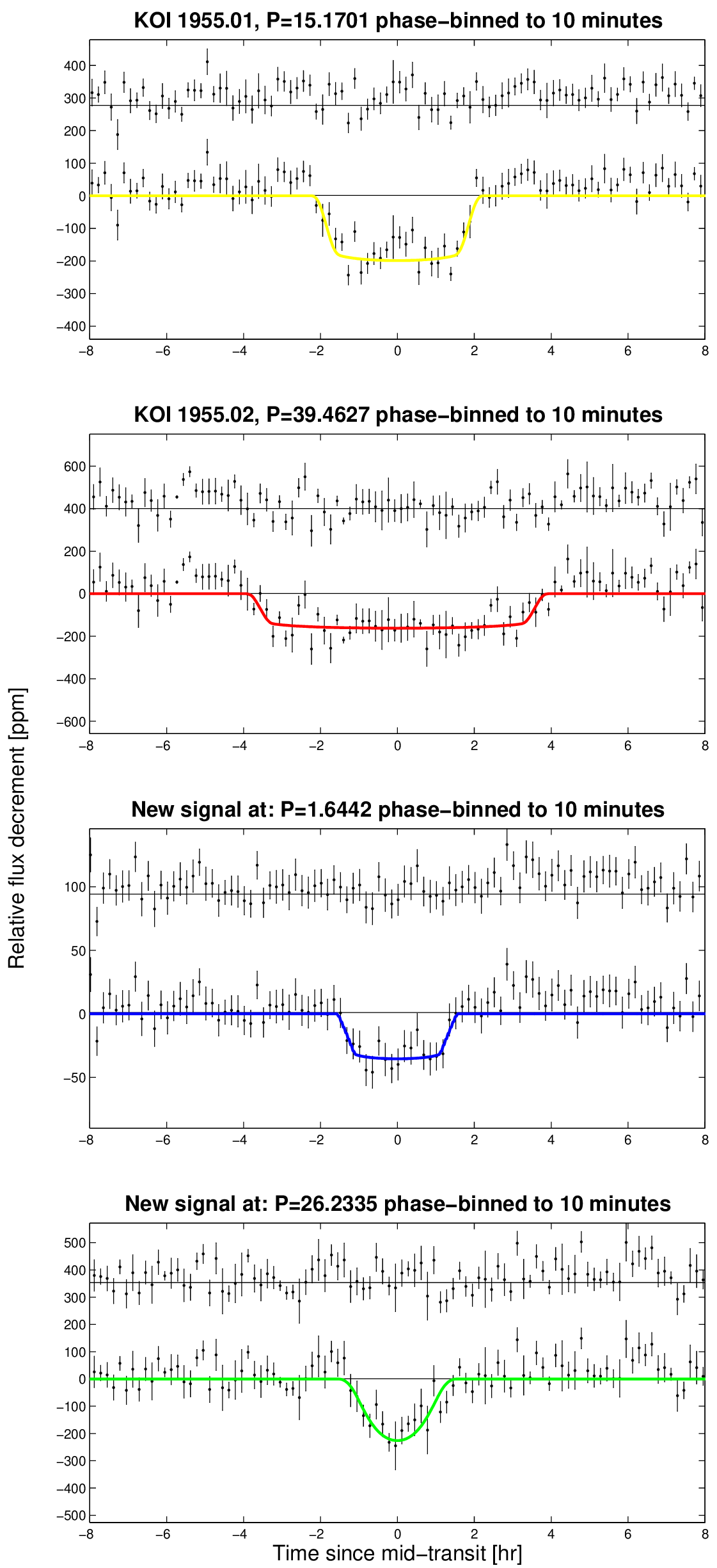}
\caption{Similar to Figure \ref{KOI246fig}.}
\label{KOI1955fig}
\end{figure}

\begin{figure}[tbp]\includegraphics[width=0.5\textwidth]{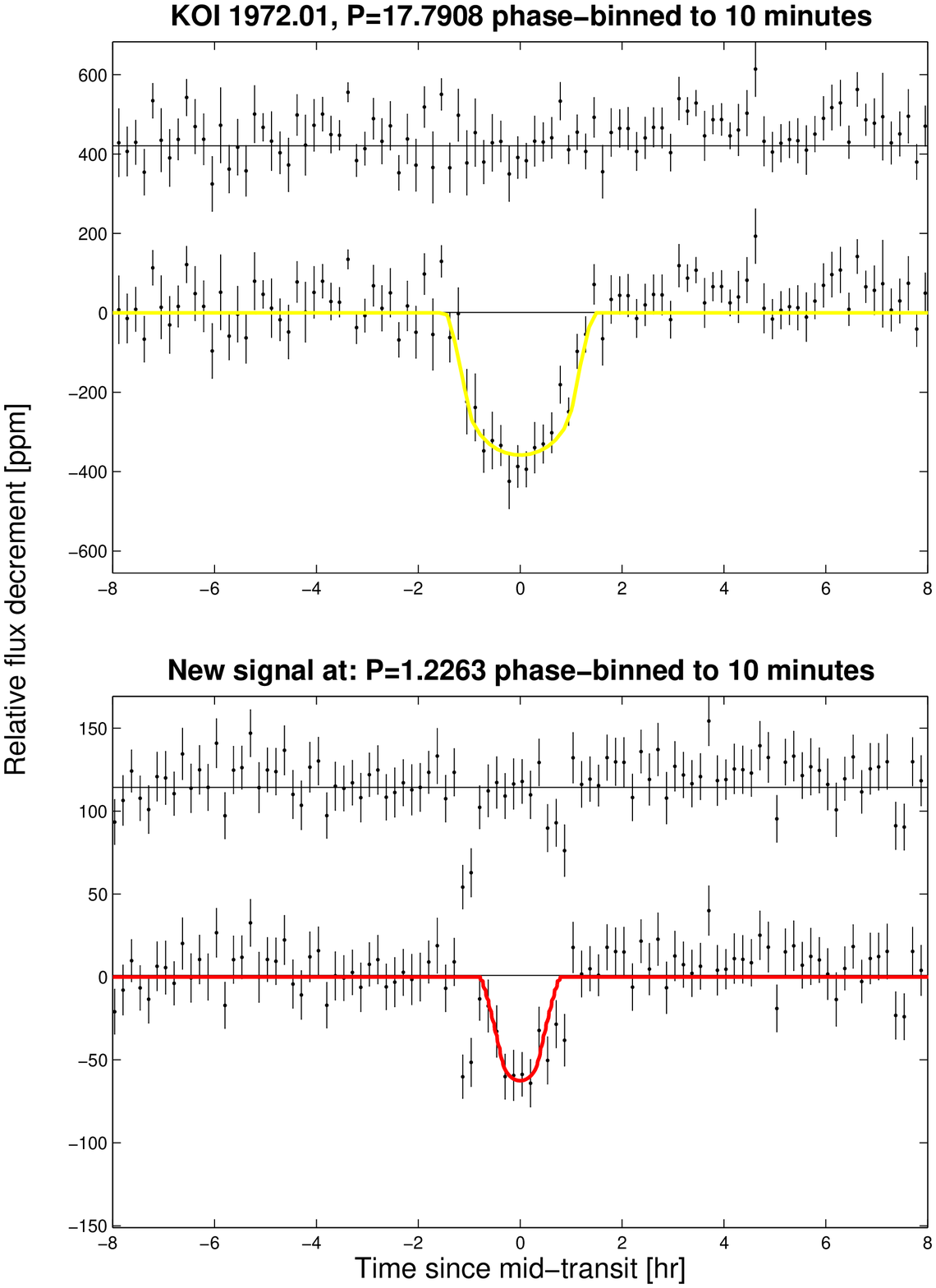}
\caption{Similar to Figure \ref{KOI246fig}.}
\label{KOI1972fig}
\end{figure}

\begin{figure}[tbp]\includegraphics[width=0.5\textwidth]{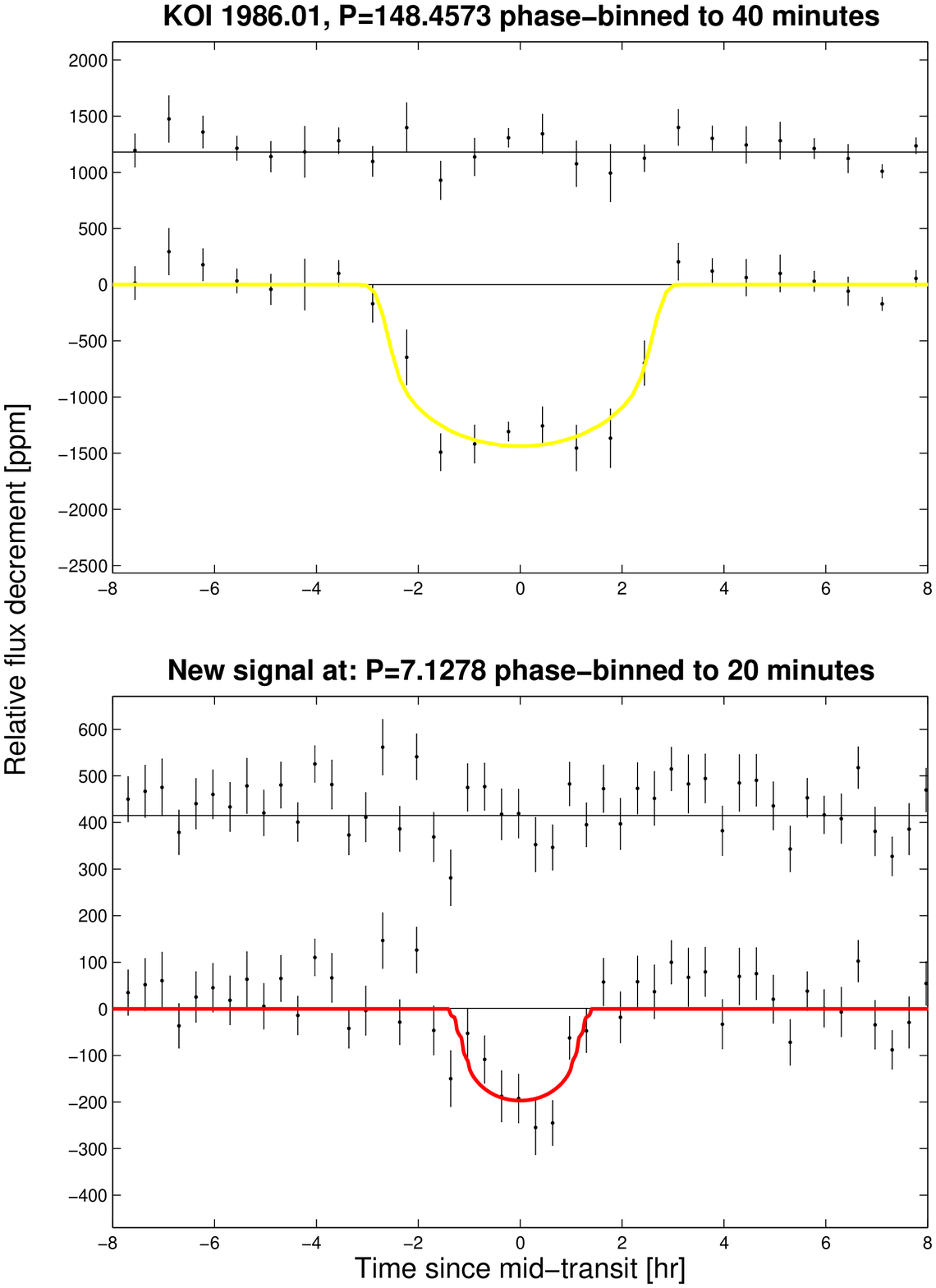}
\caption{Similar to Figure \ref{KOI246fig}.}
\label{KOI1986fig}
\end{figure}

\begin{figure}[tbp]\includegraphics[width=0.5\textwidth]{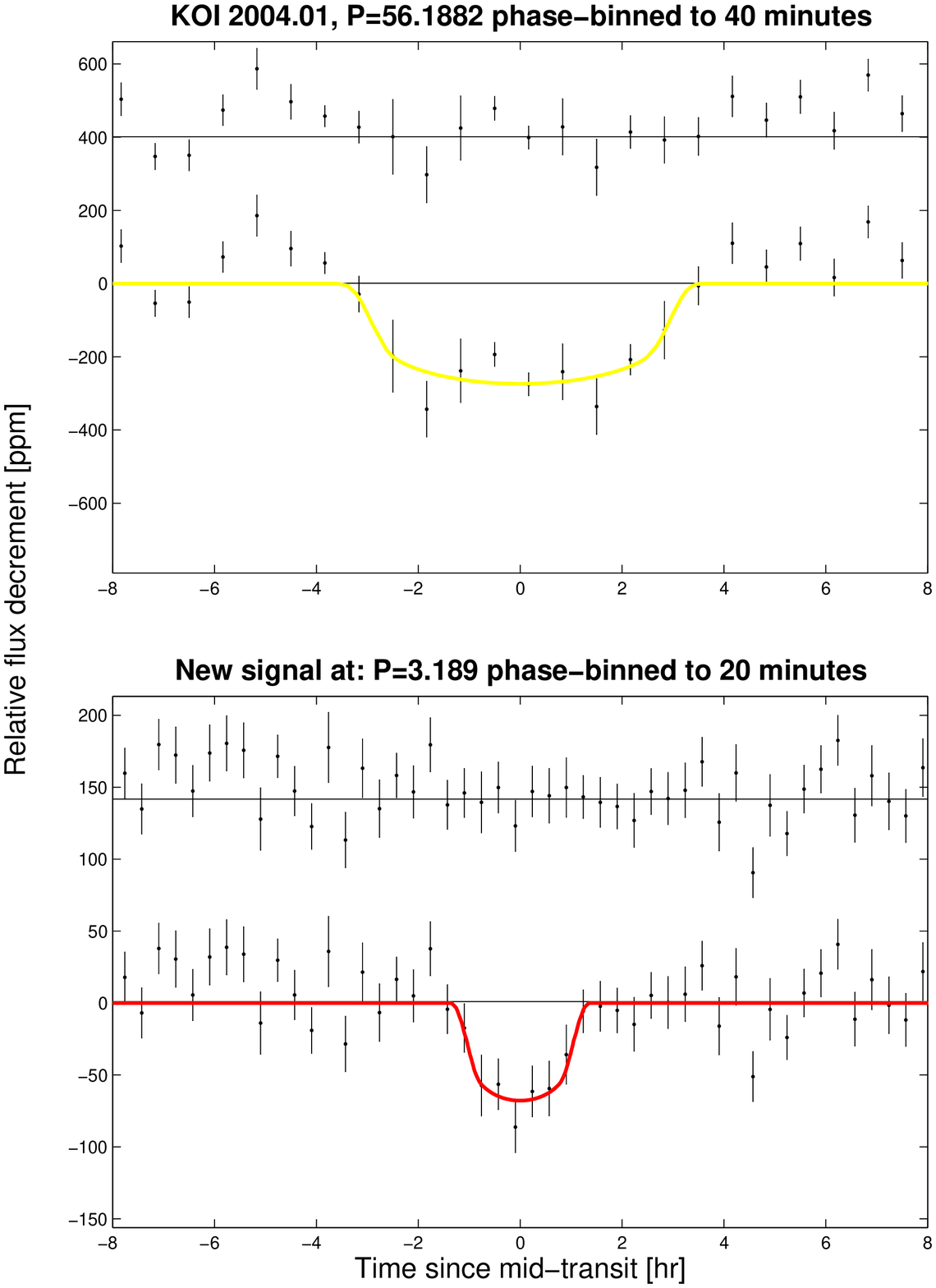}
\caption{Similar to Figure \ref{KOI246fig}.}
\label{KOI2004fig}
\end{figure}

\begin{figure}[tbp]\includegraphics[width=0.5\textwidth]{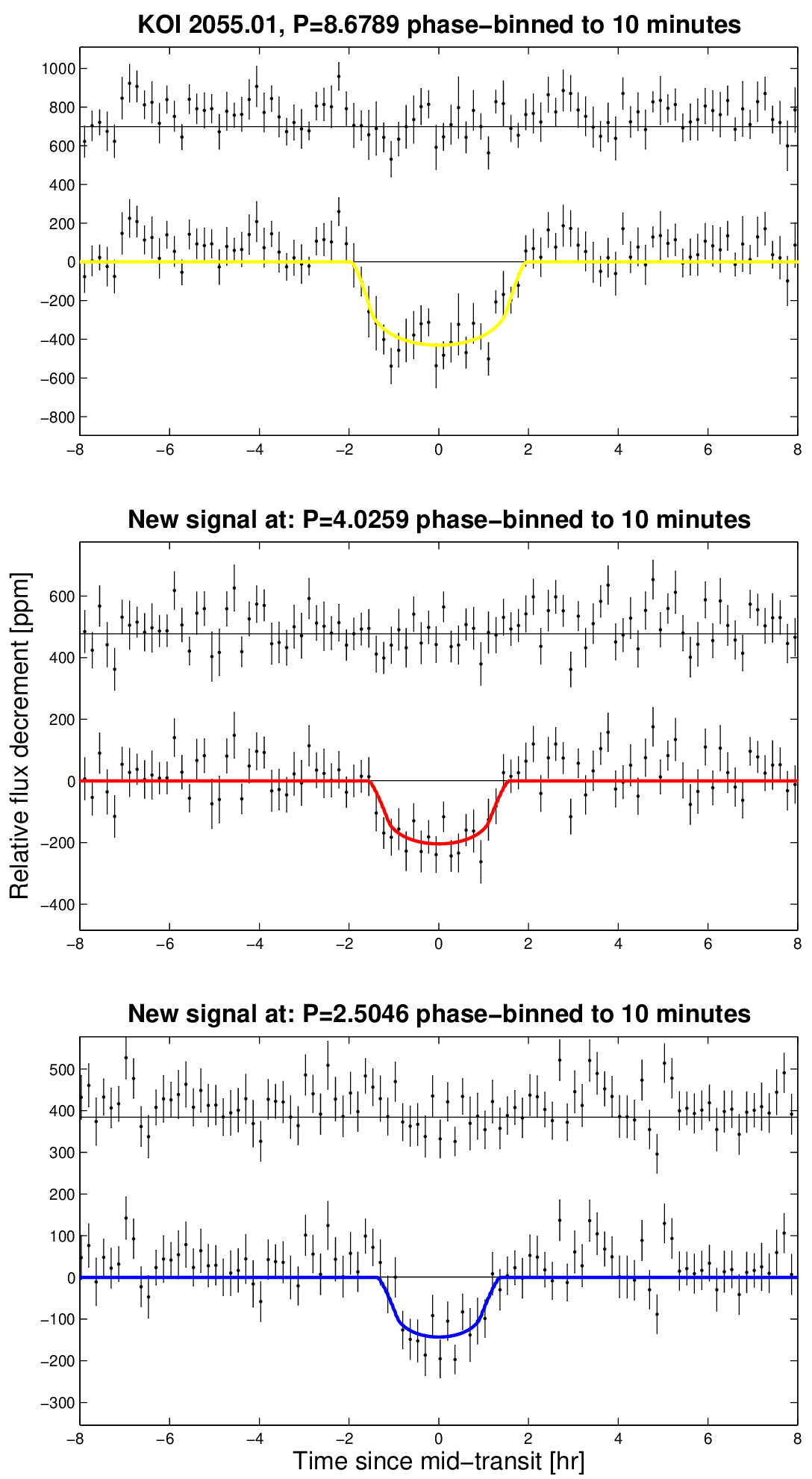}
\caption{Similar to Figure \ref{KOI246fig}.}
\label{KOI2055fig}
\end{figure}

\clearpage

\begin{figure}[tbp]\includegraphics[width=0.5\textwidth]{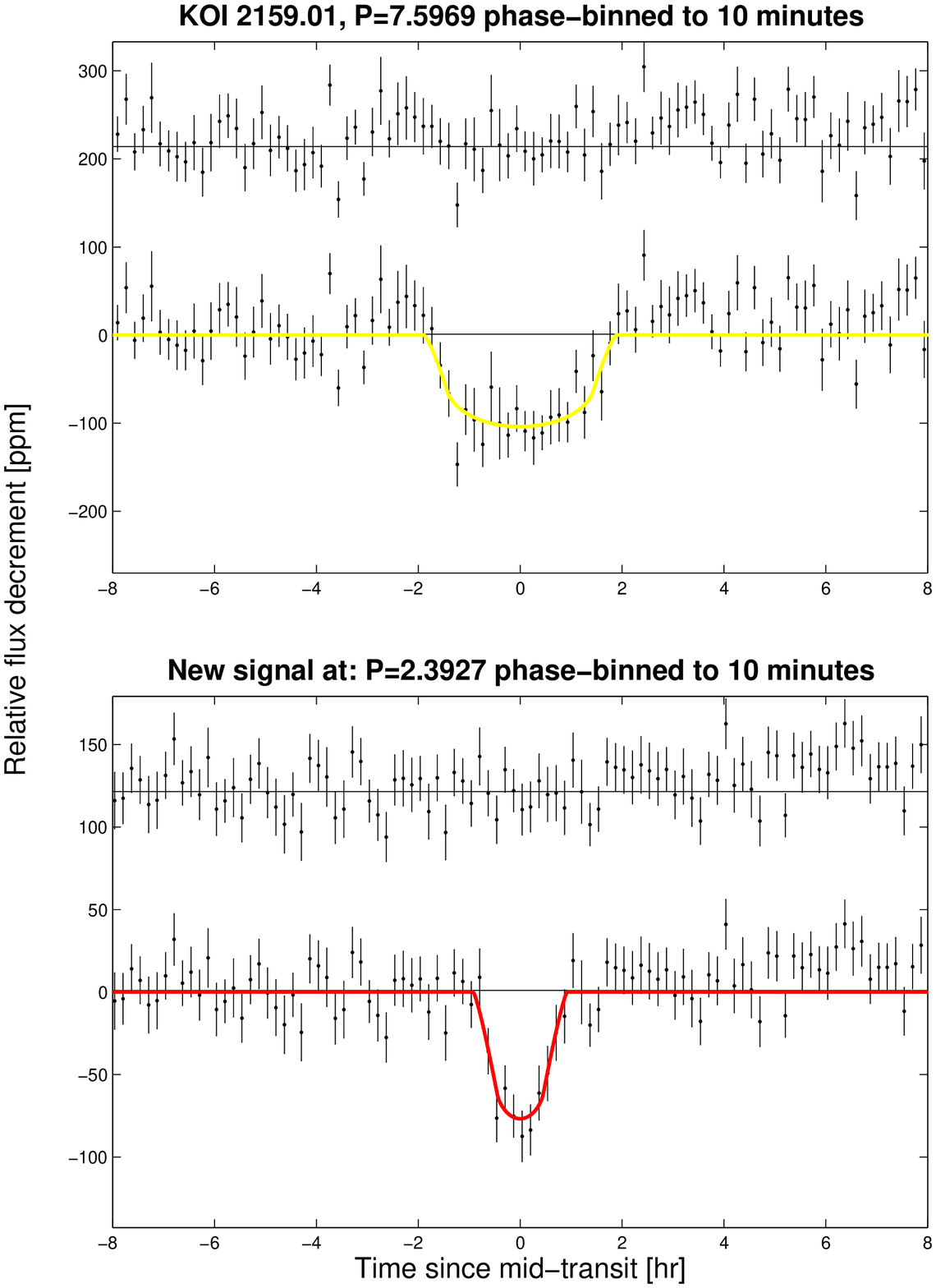}
\caption{Similar to Figure \ref{KOI246fig}.}
\label{KOI2159fig}
\end{figure}

\begin{figure}[tbp]\includegraphics[width=0.5\textwidth]{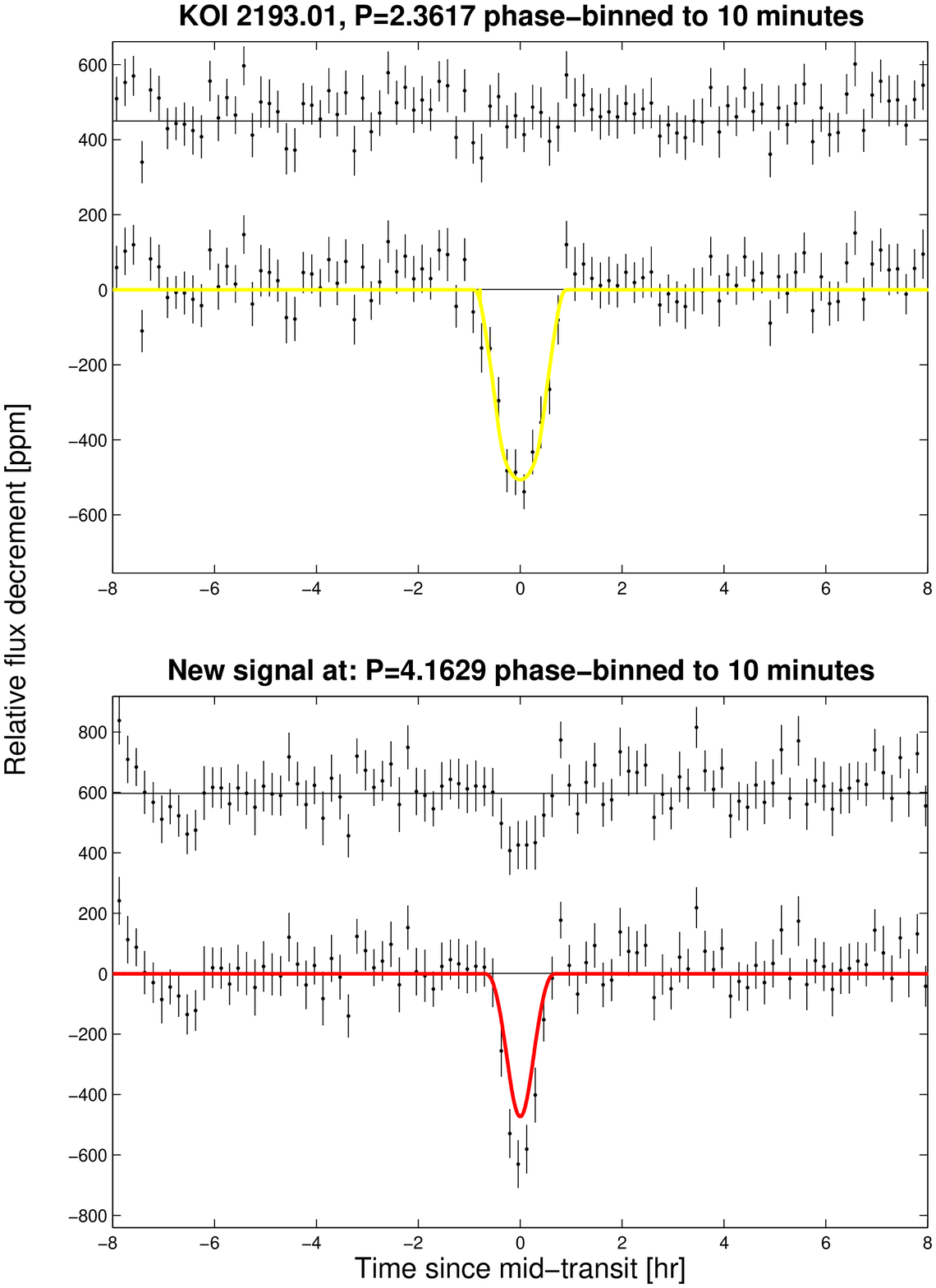}
\caption{Similar to Figure \ref{KOI246fig}.}
\label{KOI2193fig}
\end{figure}

\begin{figure}[tbp]\includegraphics[width=0.5\textwidth]{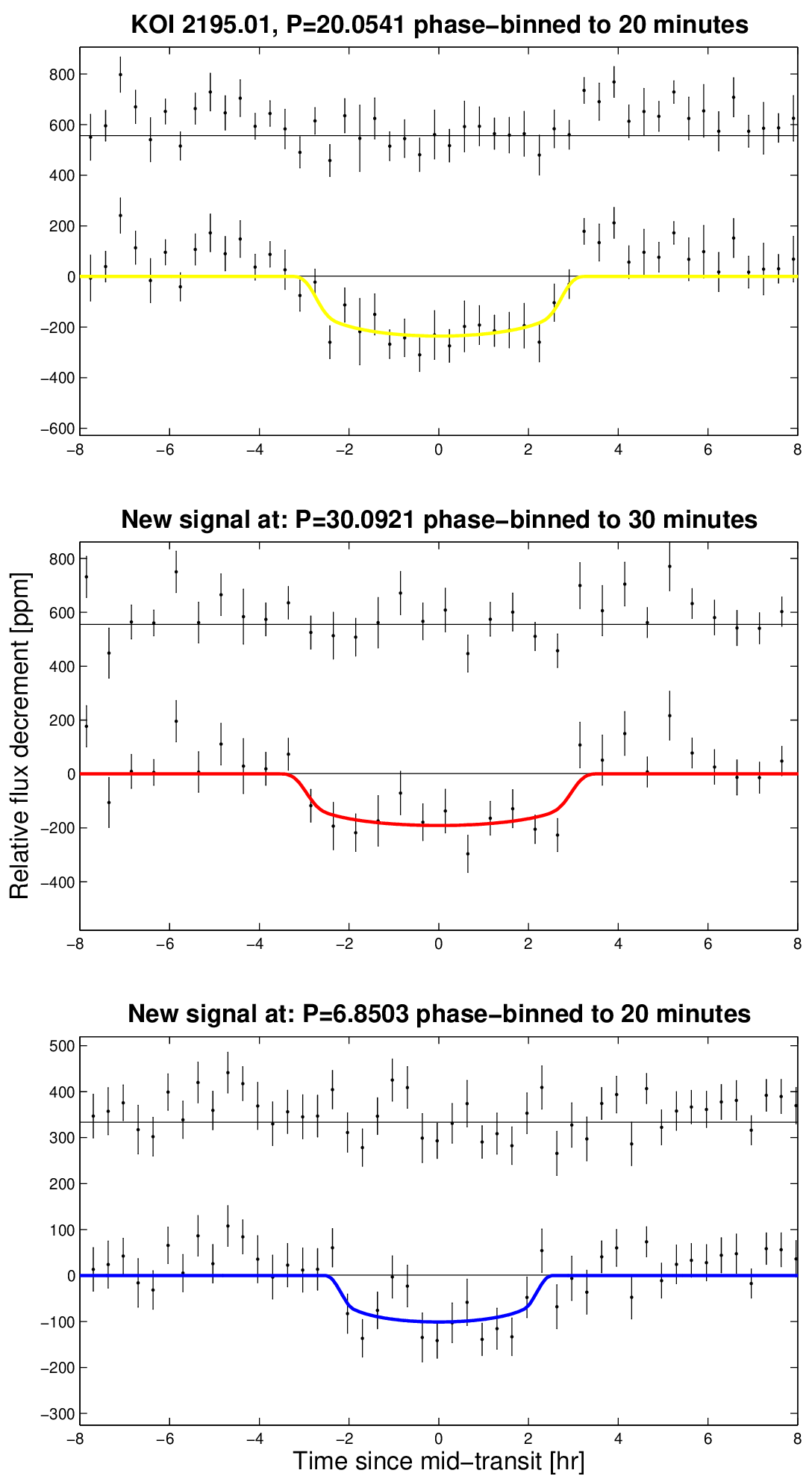}
\caption{Similar to Figure \ref{KOI246fig}.}
\label{KOI2195fig}
\end{figure}

\begin{figure}[tbp]\includegraphics[width=0.5\textwidth]{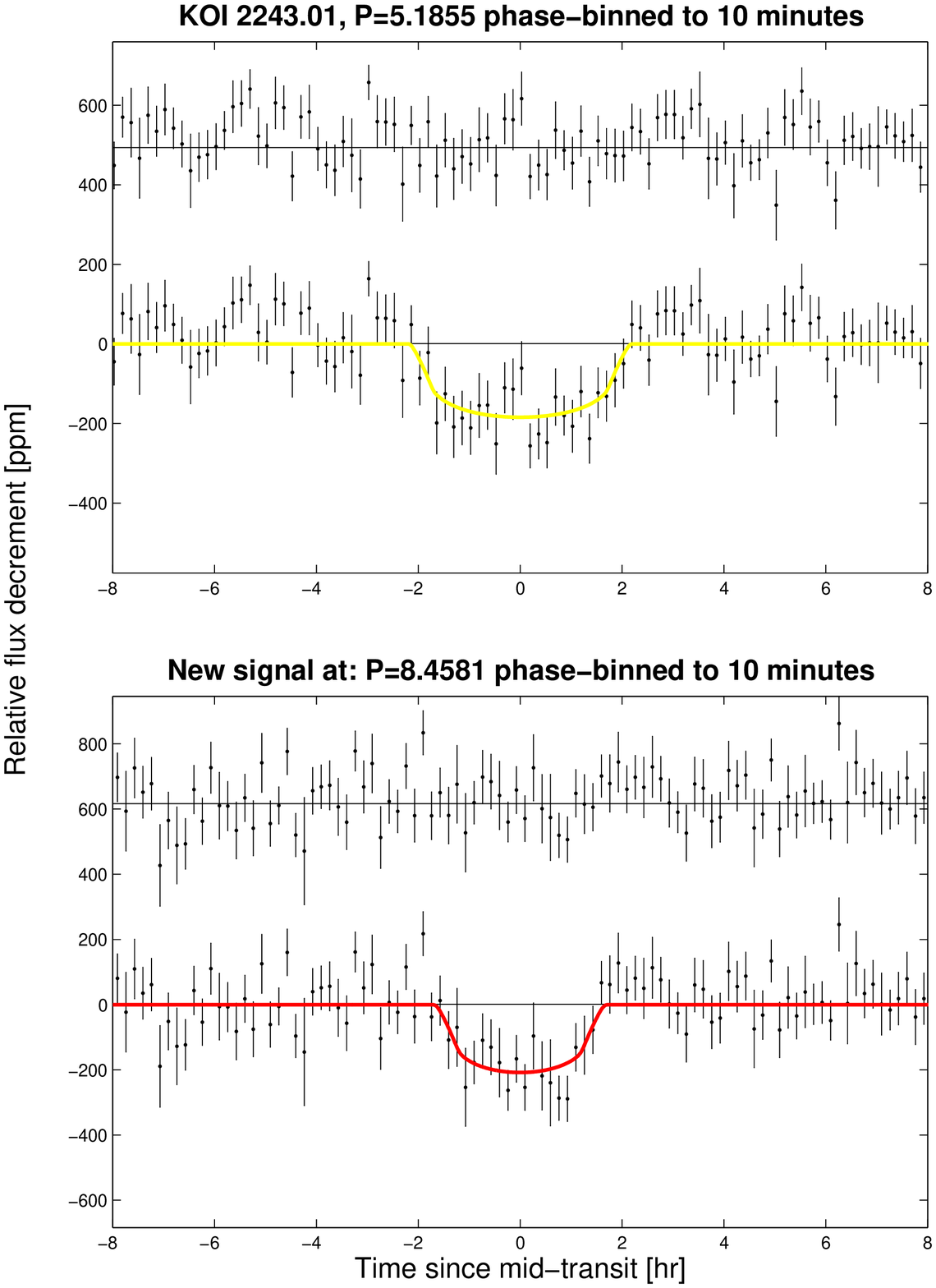}
\caption{Similar to Figure \ref{KOI246fig}.}
\label{KOI2243fig}
\end{figure}

\begin{figure}[tbp]\includegraphics[width=0.5\textwidth]{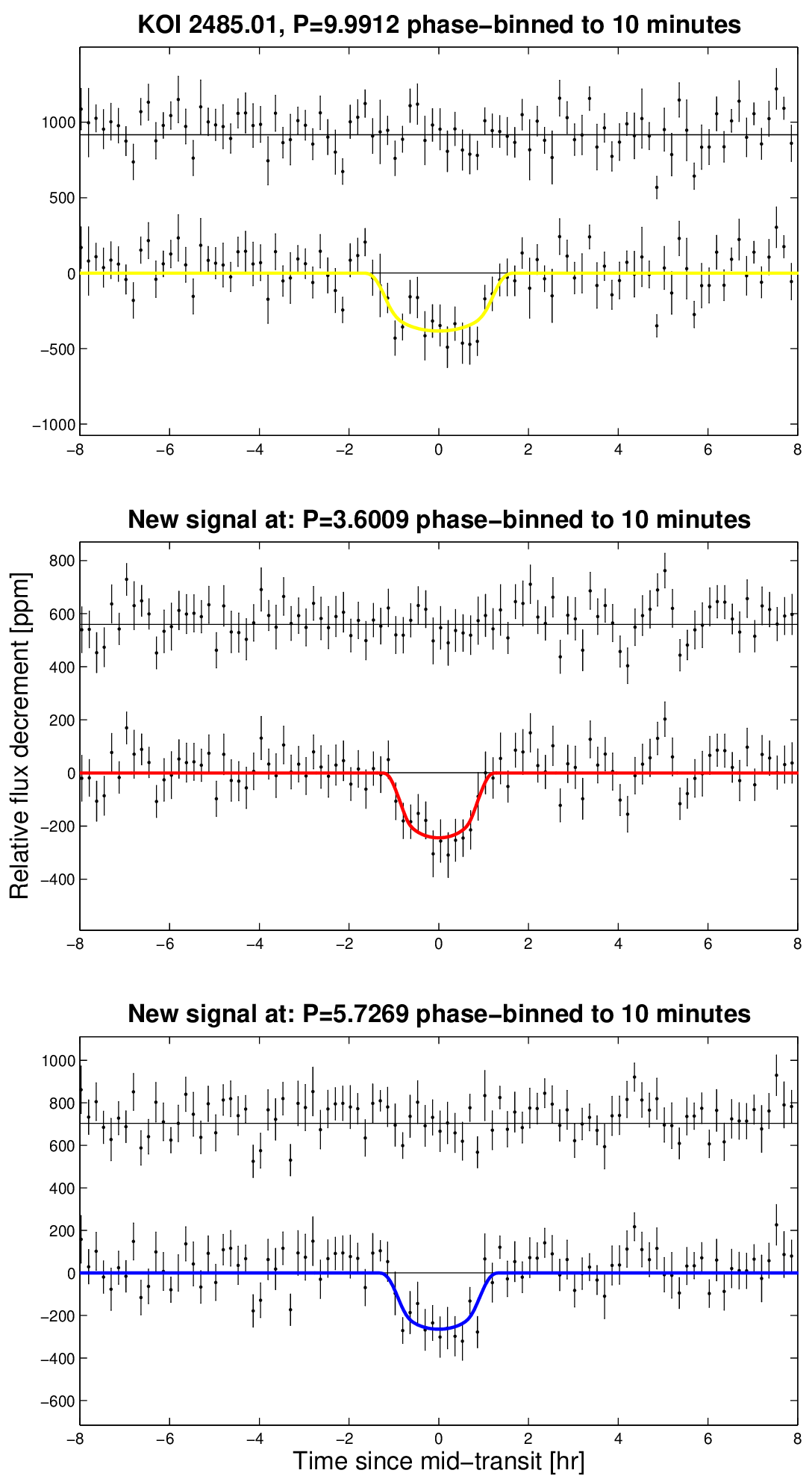}
\caption{Similar to Figure \ref{KOI246fig}.}
\label{KOI2485fig}
\end{figure}

\begin{figure}[tbp]\includegraphics[width=0.5\textwidth]{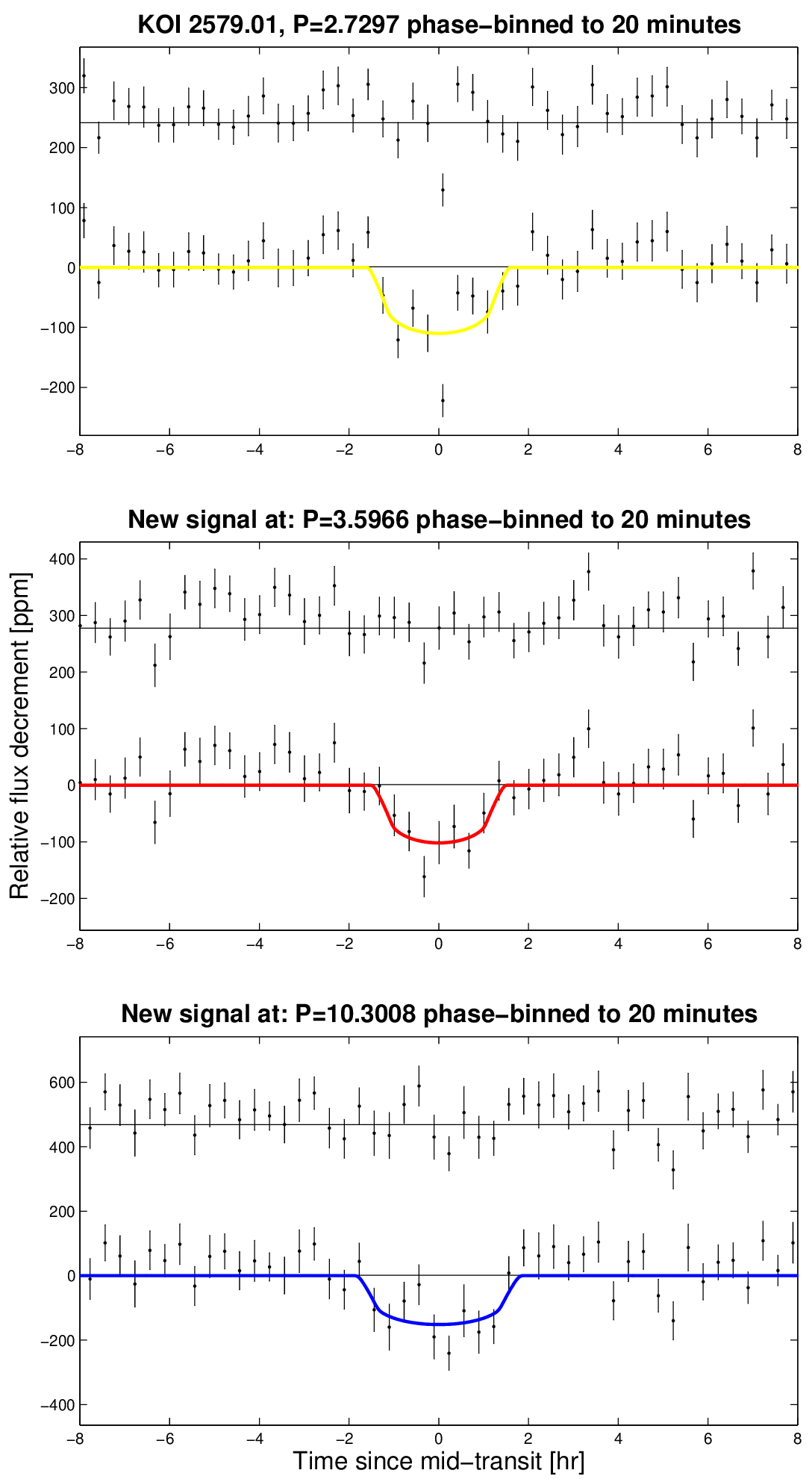}
\caption{Similar to Figure \ref{KOI246fig}.}
\label{KOI2579fig}
\end{figure}

\begin{figure}[tbp]\includegraphics[width=0.5\textwidth]{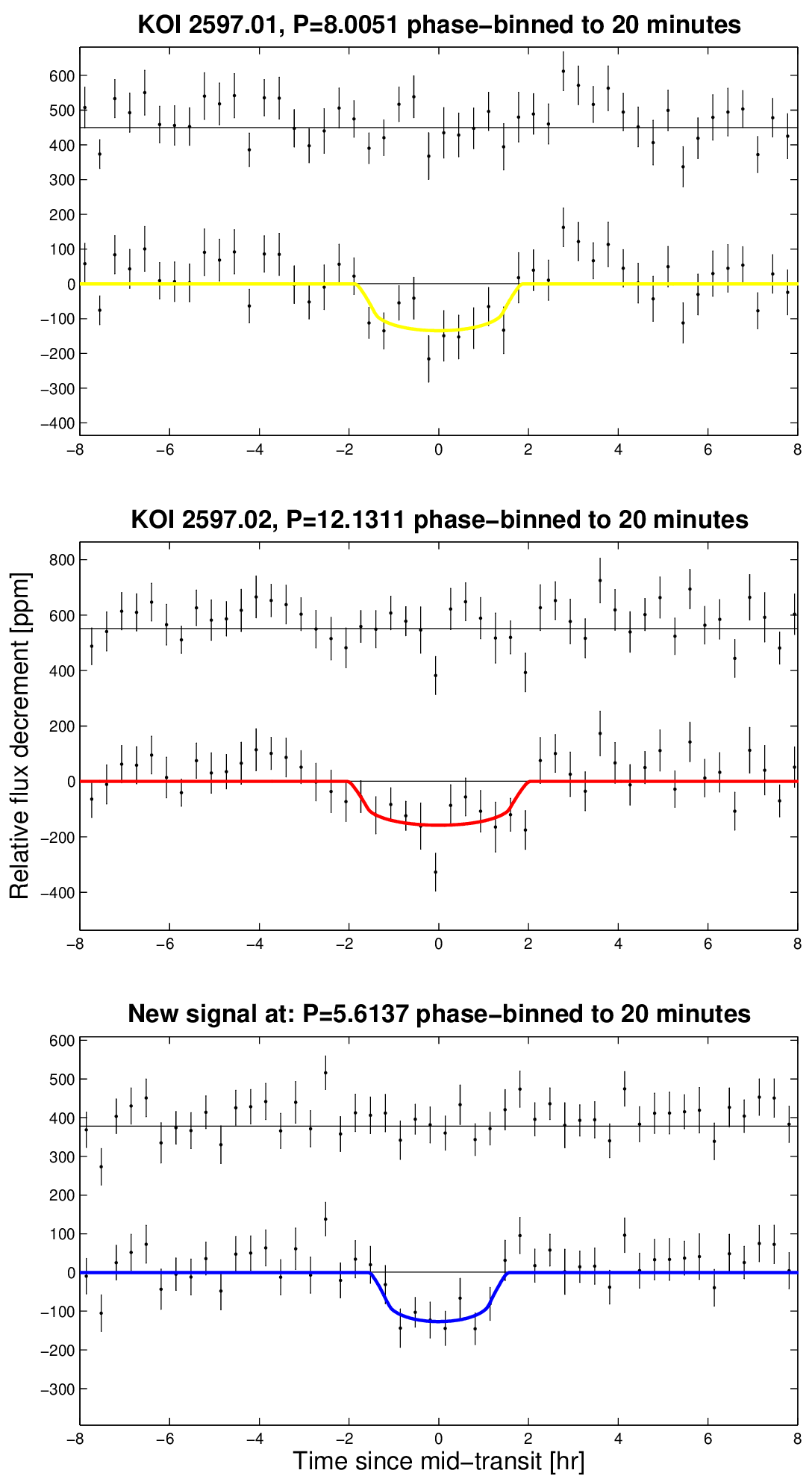}
\caption{Similar to Figure \ref{KOI246fig}.}
\label{KOI2597fig}
\end{figure}

\clearpage
\section{Appendix B}
\label{appB}
This appendix contains figures that have strong similarity to others that appear in the main text that are related to the identification of KOIs as EBs or otherwise to the discussion of KOIs with no new detections.

\begin{figure}[tbp]\includegraphics[width=0.5\textwidth]{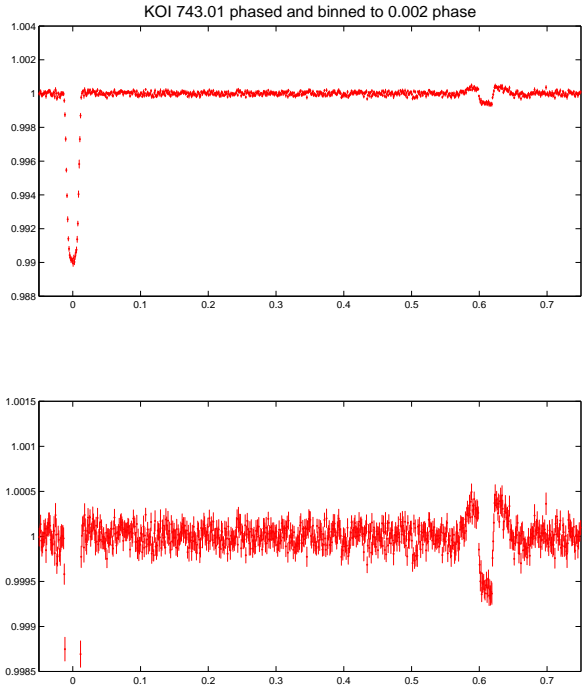}
\caption{Phase-folded and binned LC of KOI 743.01. Top: showing the 743.01 signal at phase 0.  Bottom: expanded scale of the same data, showing a secondary eclipse near phase 0.61.}
\label{KOI743secondary}
\end{figure}

\begin{figure}[tbp]\includegraphics[width=0.5\textwidth]{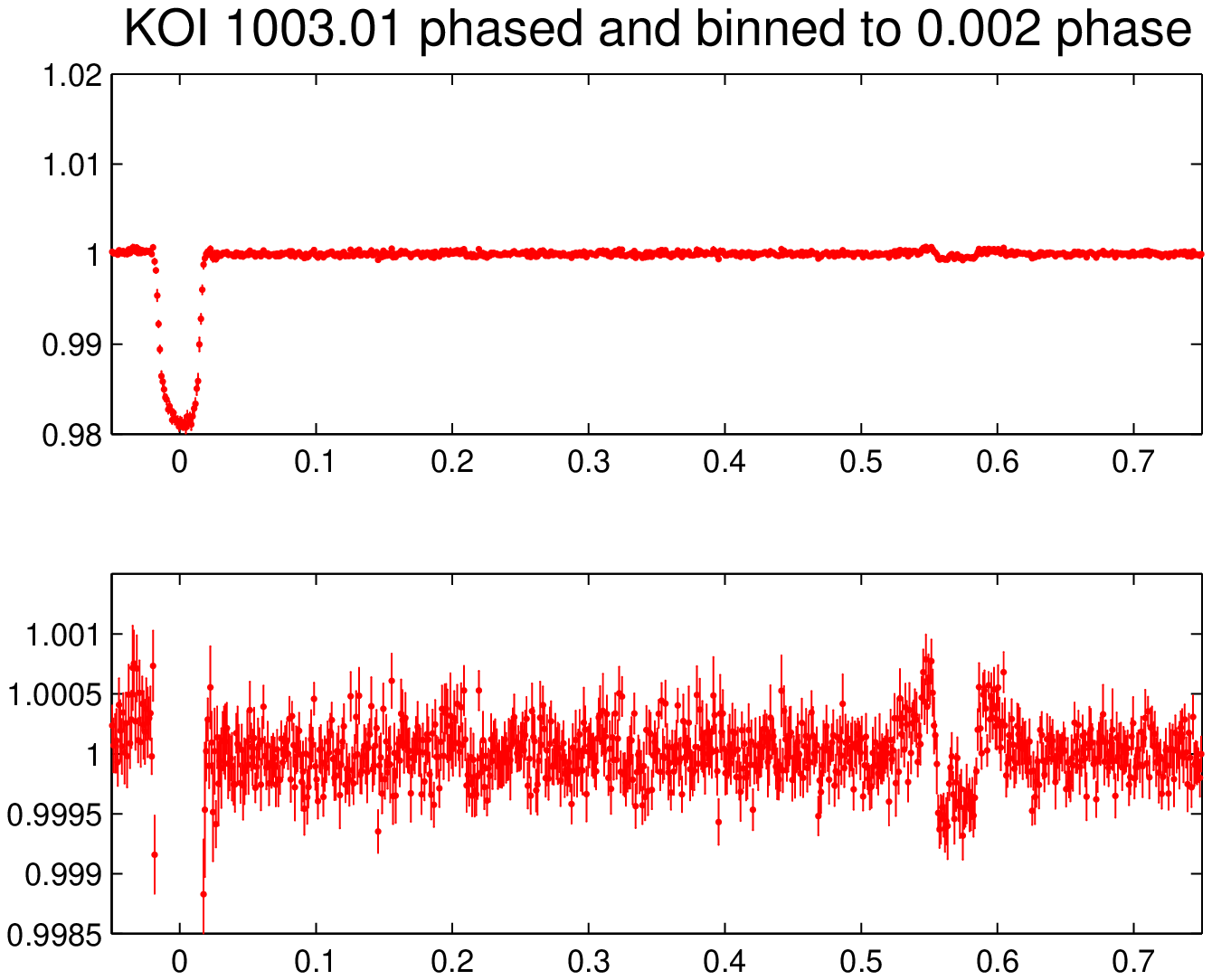}
\caption{Phase-folded and binned LC of KOI 1003.01. Top: showing the 1003.01 signal at phase 0.  Bottom: expanded scale of the same data, showing a secondary eclipse near phase 0.57.}
\label{KOI1003secondary}
\end{figure}

\begin{figure}[tbp]\includegraphics[width=0.5\textwidth]{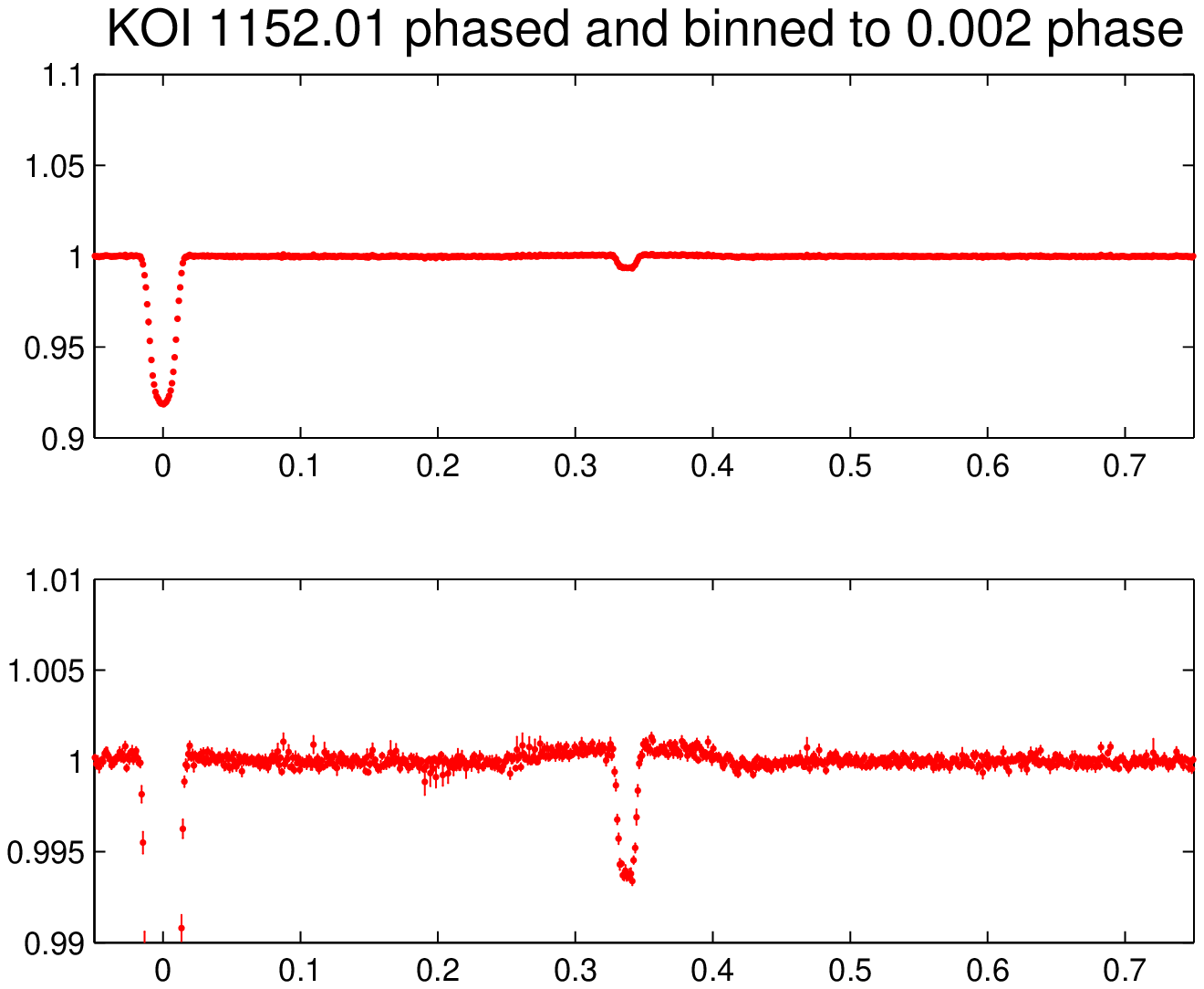}
\caption{Phase-folded and binned LC of KOI 1152.01. Top: showing the 1152.01 signal at phase 0.  Bottom: expanded scale of the same data, showing a secondary eclipse near phase 0.338.}
\label{KOI1152secondary}
\end{figure}

\begin{figure}[tbp]\includegraphics[width=0.5\textwidth]{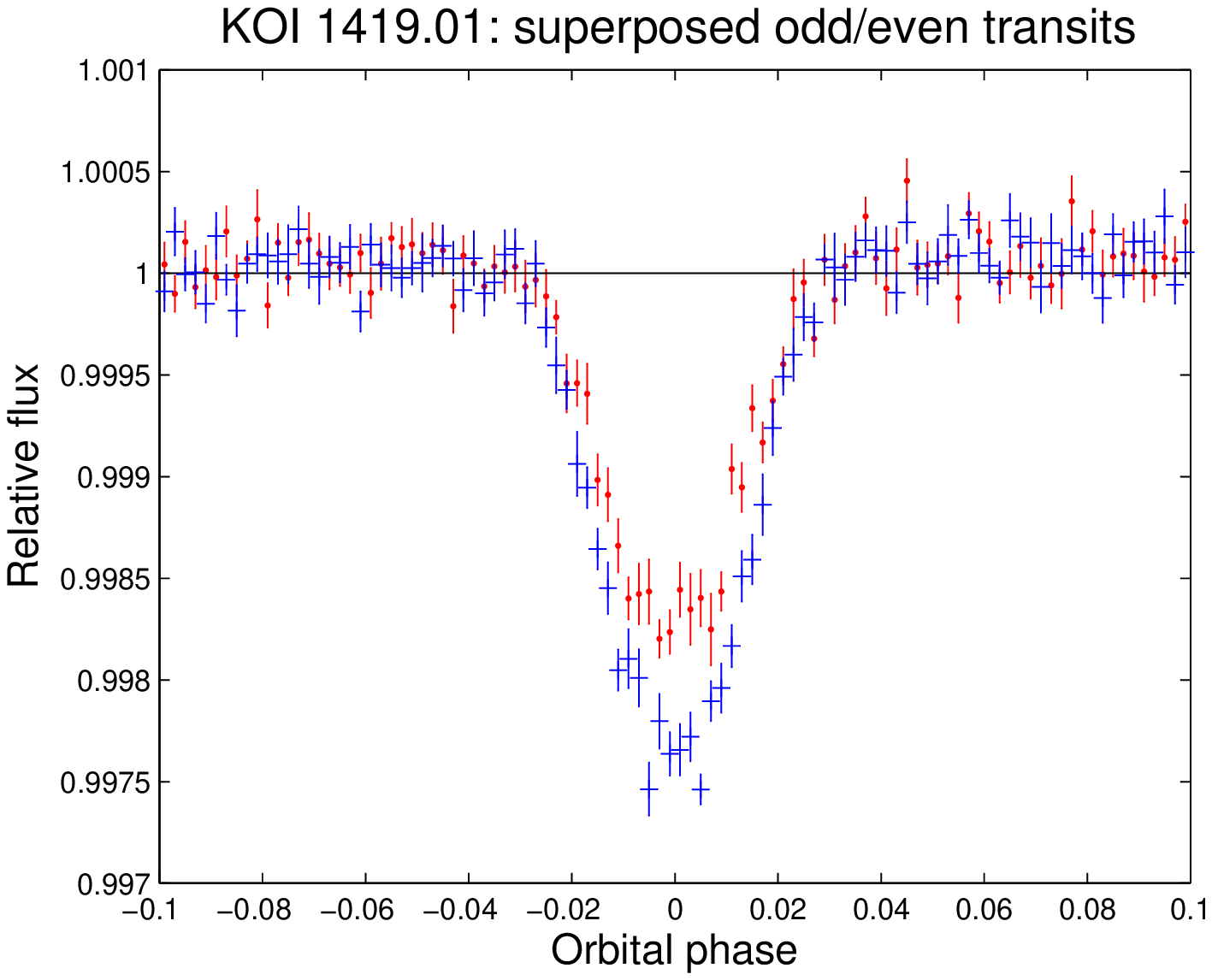}
\caption{Similar to Figure \ref{KOI225oddeven}.}
\label{KOI1419oddeven}
\end{figure}

\begin{figure}[tbp]\includegraphics[width=0.5\textwidth]{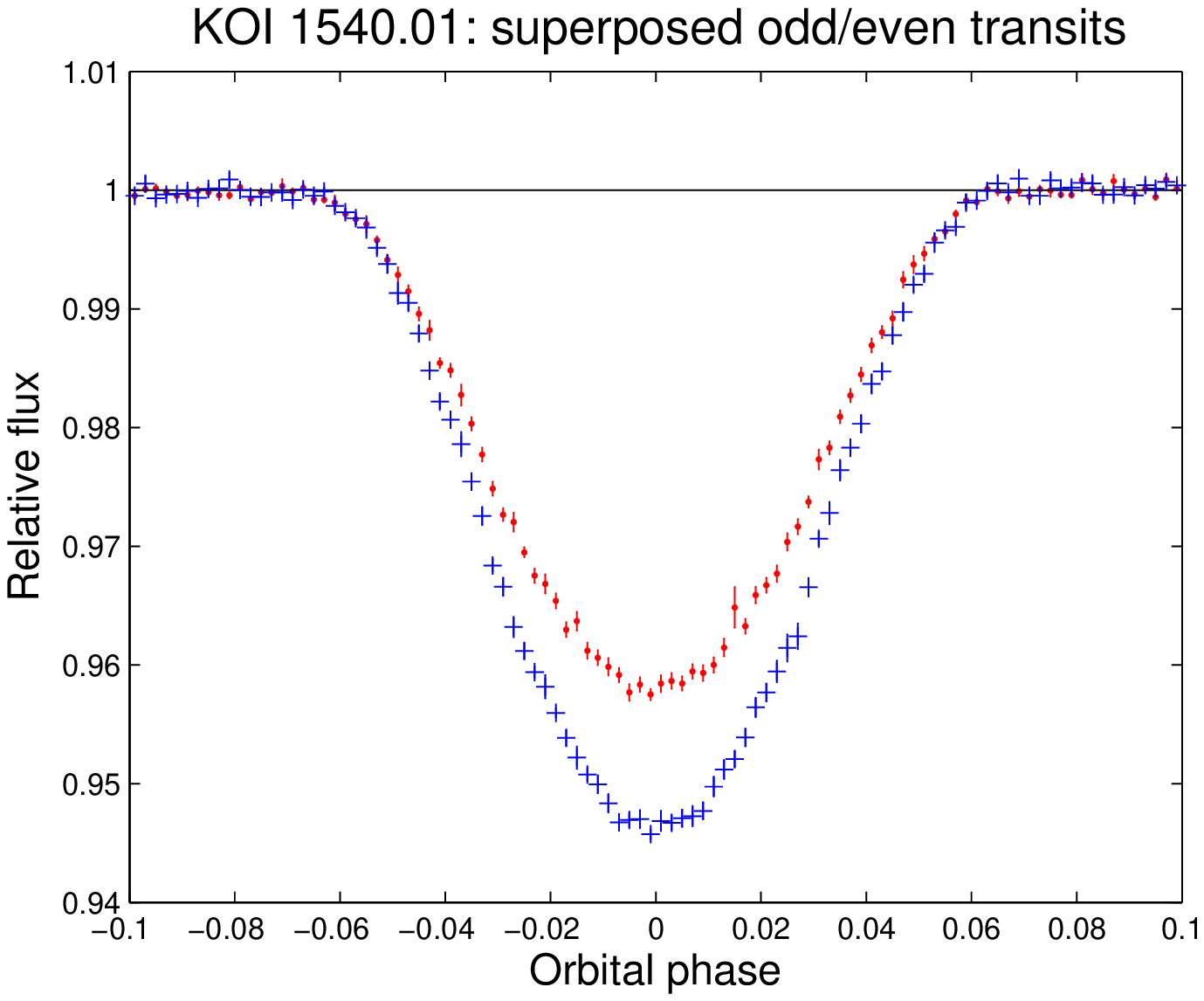}
\caption{Similar to Figure \ref{KOI225oddeven}}
\label{KOI1540oddeven}
\end{figure}

\begin{figure}[tbp]\includegraphics[width=0.5\textwidth]{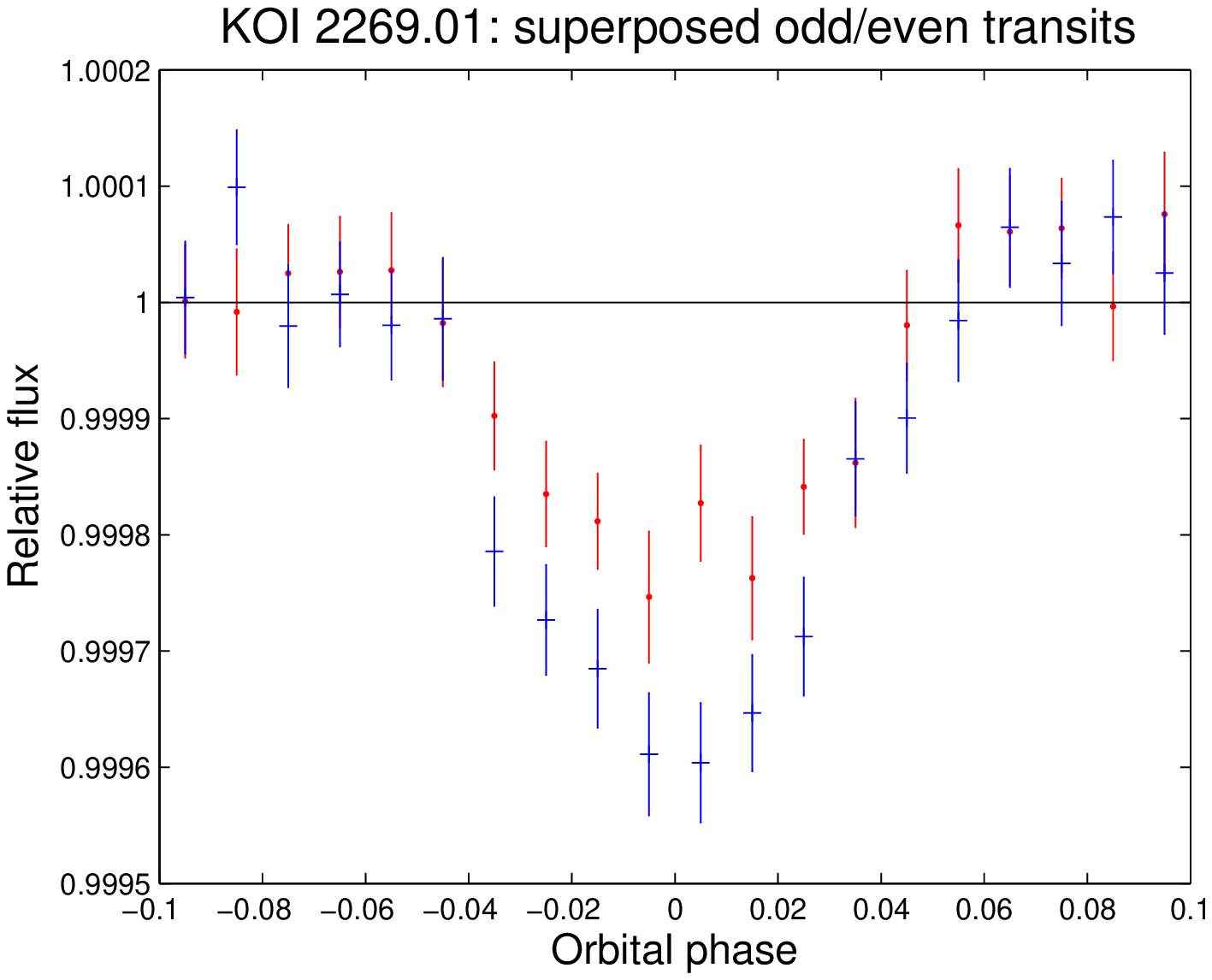}
\caption{Similar to Figure \ref{KOI225oddeven}, heavily binned to 0.01 phase.}
\label{KOI2269oddeven}
\end{figure}

\end{document}